\documentclass[pre,twocolumn,aps,floatfix,longbibliography,natbib,nofootinbib]{revtex4-1}

\usepackage{pslatex,graphicx,dcolumn,bm,natbib,amssymb,amsmath,color,mathtools,mathrsfs,eufrak}
\usepackage[utf8]{inputenc}
\usepackage[T1]{fontenc}

\usepackage[english]{babel}
\usepackage{blindtext}
\usepackage{lipsum} 

\newcommand{\be}{\begin{equation}}
\newcommand{\ee}{\end{equation}}
\newcommand{\ket}[1]{|#1\rangle}
\newcommand{\rk}[1]{\text{rank} #1}

\newcommand{\pfluid}{$p_0 > p_0^*$}
\newcommand{\psolid}{$p_0 \le p_0^*$}
\DeclarePairedDelimiter\abs{\lvert}{\rvert}%
\newcommand{\dpp}{\abs{p_0 - p_0^*} }

\newcommand{\ba}{\begin{array}}
\newcommand{\ea}{\end{array}}

\newcommand{\tr}{\mathop{\mathrm{tr}}}

\newcommand{\vm}{vertex model }

\newcommand{\mg}{\mathcal{G}}
\newcommand{\mv}{\mathcal{V}}
\newcommand{\mt}{\mathcal{T}}

\newcommand{\mi}{\mathcal{I}}

\newcommand{\mm}{\mathcal{M}}
\newcommand{\muu}{\mathcal{U}}

\newcommand{\rd}{\mathrm{d}}

\newcommand{\lp}{\left(}
\newcommand{\rp}{\right)}

\begin{document}

\title{ Multicellular rosettes drive fluid-solid transition in epithelial tissues}

\author{Le Yan}
\email{lyan@kitp.ucsb.edu}
\affiliation{%
Kavli Institute for Theoretical Physics, University of California, Santa Barbara, CA 93106\\
}
\author{Dapeng Bi }%
 \email{d.bi@northeastern.edu}
\affiliation{%
 Department of Physics, Northeastern University, MA 02115, USA\\
}

\begin{abstract}
Models for confluent biological tissues often describe the network formed by cells as a triple-junction network, similar to foams. However, higher order vertices or multicellular rosettes are prevalent in developmental and {\it in vitro} processes and have been recognized as crucial in many important aspects of  morphogenesis, disease and physiology. In this work, we study the influence of rosettes on the mechanics of a confluent tissue. We find that the existence of rosettes in a tissue can greatly influence its rigidity. Using a generalized  \vm and effective medium theory we find a fluid-to-solid transition driven by rosette density and intracellular tensions. This  transition exhibits several hallmarks of a second-order  phase transition such as a growing correlation length and a universal critical scaling in the vicinity a critical point.  Further, we elucidate the nature of rigidity transitions in  dense biological tissues  and other cellular structures  using a generalized Maxwell constraint counting approach. This answers a long-standing puzzle of the origin of solidity in these systems.
\end{abstract}
\maketitle

Multicellular organization in tissues is important to understanding many aspects of biology and medicine, such as embryonic development, disease generation and progression. Particularly in fully confluent epithelial tissues where cells are densely packed and have tight adherent junctions between them, the behavior and response of cells can be highly collective and differ from the single-cell-level behavior. Many models have been proposed to understand the emergence of this collective behavior and the biophysical properties of tissues. 
In the past two decades, a class of cell-based models known as the \vm has been developed to study epithelial tissue mechanics ~\cite{ngai_honda,Farhadifar2007,Hufnagel2007,Staple2010,manning_2010,Fletcher2014,bi_softmatter,bi_nphys_2015,Bi2016,Yang2017,Merkel2017,Noll2017,boromand_2018, teomy_2018, cellgpu,Spencer_lubensky_2017,barton_silke_rastko,bi_nphys_2015,moshe_2017,atia_nat_phys_2018,Alt_VM_review_2017,flocking_spv,C8SM00446C,staddon_plos_cb,li_pnas_2018}. In the \vm,  each cell is represented as a deformable polygonal inclusion, with edges and vertices shared by neighboring cells. 
 This class of models often assumes exactly three cells meet at any vertex in an epithelial tissue and four-fold vertices only occur as an intermediate state during a T1 rearrangement.

However, in many tissues~\cite{harding_rosette_review} higher-order vertices where four or more  cells meet can occur.  
A prominent example occurs in Drosophila embryogenesis, where a mixture of T1-junctions
~\cite{weaire_rivier_1984,Zallen2004, Bertet2004, Zallen_zallen,Rauzi_2008} (vertices where four cells meet) and multicellular rosettes~\cite{Escudero_Freeman, Blankenship2006,  Fernandez-Gonzalez09, Tamada12,Kasza2014,sun_et_al_basal_rosettes_2017_ncb} (vertices with five or more cells)  have been observed during the elongation of the body axis. 
It has been further shown that as the body length of the embryo doubles due to a series of cellular rearrangements, a majority of cells in the  developing epithelium participate in rosette formation~\cite{Blankenship2006,Kasza2014}.
The morphogenesis of the Drosophila eye is also facilitated by rosettes~\cite{robertson_2012}. 
In zebrafish lateral line development, the lateral line is composed of mechanosensory organs called neuromasts which are formed from rosettes composed of 20 or more cells~\cite{GOMPEL200169,Nechiporuk08,Lecaudey08}.
In other vertebrates, rosettes have been observed in  the neural plate of the chick~\cite{Nishimura_Takeichi_2008} and mouse~\cite{Williams14} embryos, in the development of the mouse visceral endoderm~\cite{trichas_2012}, the kidney tubule ~\cite{lienkamp_2012} and the pancreas~\cite{Villasenor_2010}. 
They are also  observed in adults and \emph{in vitro} such as adult mammalian brains and cultured neural stem cells~\cite{zhang_2001,elkabetz_2008}. 

The importance of cellular rosettes  has been widely recognized. They have been proposed as an efficient mechanism for tissue remodeling and growth during the body elongation in the Drosophila embryo~\cite{Blankenship2006,Tamada12, Kasza2014} and are thought to be crucial for the orderliness of collective cell migration in  mouse  visceral endoderm~\cite{harding_rosette_review,trichas_2012}. Cancer pathologists visually inspecting histologic samples from tumors use cellular rosettes as  strong indications of malignancy such as  medulloblastoma and retinoblastoma~\cite{Wippold2006}. 
However, compared to the large body of live-imaging and molecular studies~\cite{harding_rosette_review},  there is surprisingly few modeling~\cite{trichas_2012} and even less physical understanding of rosettes regarding how they influence tissue mechanics.

Here we develop a generalized \vm and an effective medium theory that takes into account the presence of    rosettes. 
We also make experimentally testable  predictions regarding the strong correlations and the interplay between cellular topology and mechanical tensions in a tissue. 
 We find that the tissue rigidity is precisely controlled by the density of { higher-order vertices including rosettes and T1 junctions as well as}    a single parameter that { tunes} intracellular tensions in the tissue. 
 The results show that a tissue behaving as a fluid can be rigidified by the creation of just a few  { higher order vertices}. The transition between fluid and rigid states exhibit many hallmarks of a second-order phase transition, such as a growing correlation length and a universal critical scaling in the vicinity a critical point. Furthermore, we  offer a unifying perspective on rigidity transitions in dense confluent tissues and  cellular materials by elucidating  the nature of this transition using a generalized Maxwell constraint counting.

\begin{figure*}[t]
\begin{center}
\includegraphics[width=1\textwidth]{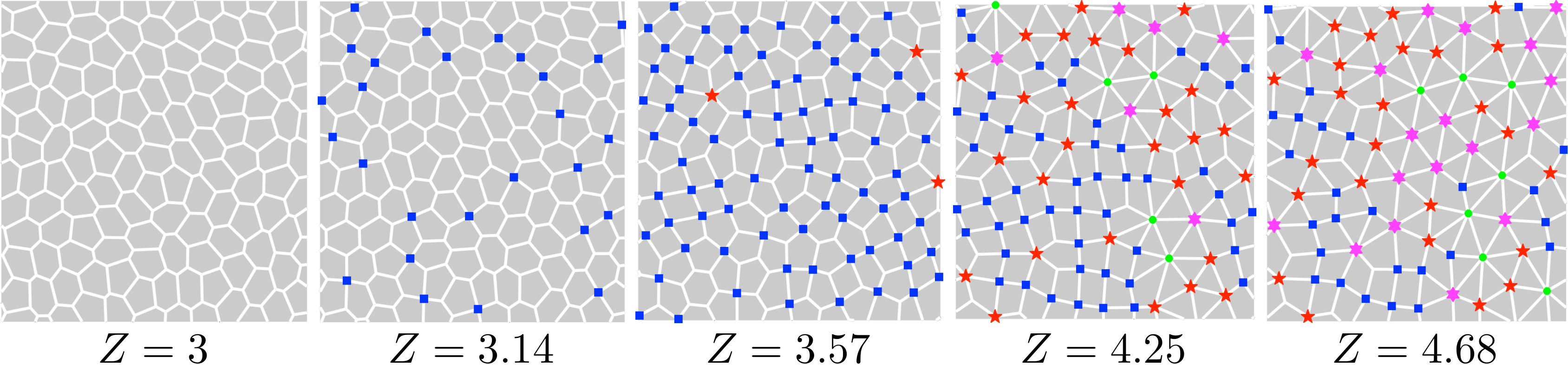}
\caption{
 {\bf Introducing higher order vertices into the \vm.} In conventional vertex models, all vertices have 3-fold coordination. Here we apply a protocol that randomly collapses cell edges to create vertices with  coordination higher than 3. 4-fold vertices are called `T1-junctions' and 5-fold or higher vertices are termed \emph{rosettes}. We characterize the state of a tissue by the average vertex coordination number, $Z$. Simulation snapshots show states with different values of $Z$.  Squares, five-pointed stars and six-pointed stars indicate vertices with coordination of $4,5$ or $6$, respectively. Disks correspond to vertices with coordination of $7$ or more.  
  }
\label{fig:snapshots}
\end{center}
\end{figure*}

\section{Generalized \vm}
We begin with the most generic form of the \vm for a homogeneous tissue~\cite{Farhadifar2007}, where the total  energy is given by a sum over the mechanical cost of deforming individual cells,
\be
\label{single_E_p0}
U  = \sum_{\alpha=1}^F \left[ K_A(A_\alpha-A_0)^2 + K_P {(P_\alpha -P_0)}^2\right].
\ee
 The first term results from a combination of three-dimensional cell incompressibility and the monolayer's resistance to height fluctuations or cell bulk elasticity~\cite{Hufnagel2007,bi_nphys_2015}, where $K_A$ is a height elasticity and $A_{\alpha}$ is the cross-sectional area (apical) of cell $\alpha$ and $A_0$ is the  preferred  area for the cell. The second term in equation~\eqref{single_E_p0} is quadratic in the cell cross-sectional perimeter $P_\alpha$ and models the active contractility of the actin-myosin subcellular cortex, with elastic constant $K_P$~\cite{Farhadifar2007}, and $P_0$ is an effective target shape index, representing an interfacial tension set by a competition between the cortical tension and the energy of cell-cell adhesion~\cite{manning_2010,bi_nphys_2015} between two contacting cells.
In Eq.~\eqref{single_E_p0}, the sum is over all $F$ number of cells in the tissue. The cell areas ($\{A_\alpha\}$) and perimeter ($\{P_\alpha\}$)   are fully determined by the positions of vertices $\{{\bf R}_i\}$. Due to a combination of cortical tension and cell-cell adhesion~\cite{Farhadifar2007, Heer_Martin_tension_review}, each cell contributes an effective line tension~\cite{Yang2017} to adjacent edges, controlled by $P_0$, which can also be interpreted as a geometrical non-dimensional shape index $p_0=P_0/\sqrt{A_0}$. 
It was  demonstrated that~\cite{Farhadifar2007, bi_nphys_2015,moshe_2017} the tissue undergoes a phase transition at  $p_0 = p_0^* \approx 3.81$, independent of the strength of area constraints $\frac{K_AA_0}{K_P}$. When \psolid, the tissue behaves like rigid solid with a finite shear modulus. Above $p_0^*$, the tissue becomes soft and fluid-like with a vanishing shear modulus.

\section{Manipulation of cellular topology and the creation of {higher-order vertices}}

\begin{figure*}[t]
\begin{center}
\includegraphics[width=1\textwidth]{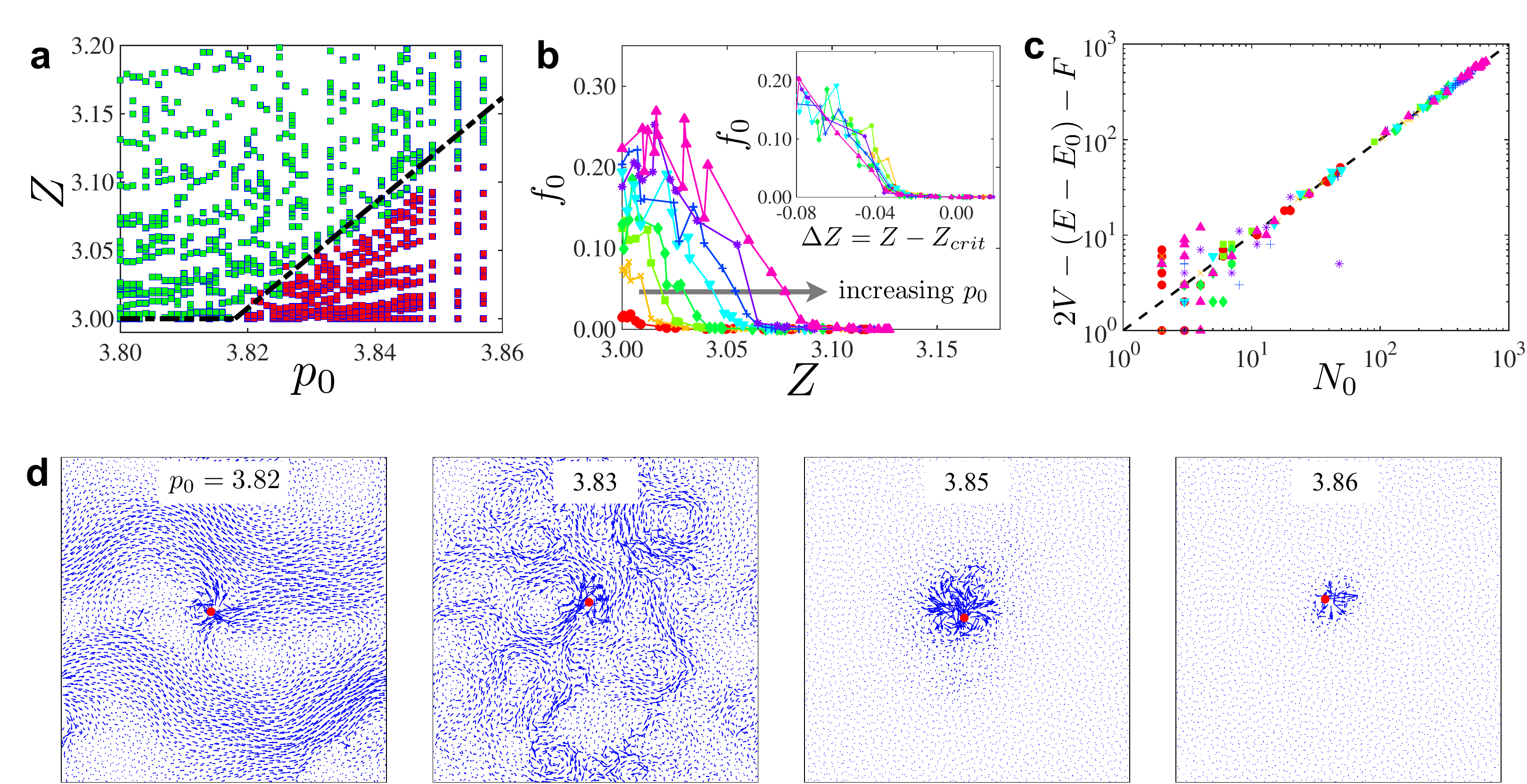}
\caption{
{\bf Tissue mechanical properties}
{\bf   (a)}
{The density of higher order vertices $Z$ and the non-dimensional} cell shape index $p_0$ together control the mechanical rigidity of the tissue.  Here green  data points indicate states with no floppy modes, i.e. mechanically rigid and purple points highlight states with a finite number of floppy modes and zero shear modulus. When only tri-junctions are allowed ($Z=3$), rigidity only occurs for tissues with \psolid. However, when a sufficient number of higher order vertices are present  ($Z>3$), tissues  can be rigidified even at \pfluid. The dashed  line indicates the phase boundary between rigid and fluid states, given by $Z_{crit}(p_0)$(eq.~\ref{zcrit}).
{\bf   (b)}
The fraction of floppy modes $f_0$  as a function of $Z$ at various values of $p_0$ (from left to right, $p0 =  3.829, 3.83,3.832, 3.834,  3.836, 3.838,   3.84, 3.842$). 
Inset:
When the same data from (b) is plotted as a function of $\Delta Z = Z - Z_{crit}(p_0)$, All data collapse onto a single universal curve indicating that the rigidity transition occurs at a line of critical points given by $Z_{crit}(p_0)$. 
{\bf   (c)} 
For the data points in (b), the actual number of floppy modes is compared against the prediction from 
the Maxwell constraint counting (eq.~\eqref{counting_2}). 
{\bf   (d)} 
Response of the tissue after a single edge collapse at $Z=3$. The displacement vector field of all vertices are shown. The red dot indicates the location of the collapsed edge near the center of the tissue. 
}
\label{fig:floppy_modes}
\end{center}
\end{figure*}

In developmental and {\it in vitro} examples, the general mechanism for rosette formation is the contraction of actomyosin networks in cells~\cite{harding_rosette_review}
This can be manifested in several ways. For example, in Drosophila body axis elongation rosettes are due to planar polarized constriction ~\cite{Zallen2004, Bertet2004, Blankenship2006, Fernandez-Gonzalez09,Rauzi_2008, Kasza2014}; while in gastrulation and neural tube closure, actomyosin structures in the apical cell surface constrict to form rosettes~\cite{GOMPEL200169,Nechiporuk08,Lecaudey08}. Also when cells delaminate or extrude from epithelia~\cite{Kocgozlu2016,Toyama2008, Marinari2012, Slattum2009}, a vertex with more than {four} edges can be left behind that becomes a center of a rosette. Similarly, a multicellular rosette can also form as the result of a wound closure~\cite{Razzell2014,Brugues2014}. %

Here instead of  focusing on  the origin of cellular rosettes   which can be varied for different processes , we assume they have been created via one of the observed mechanisms and ask how  their presence affect the mechanics at the tissue level. In practice, we create rosettes {and T1-junctions} via a simple protocol of random collapse of edges. During this process, an edge is chosen at random and its length is reduced to zero. The two vertices on the end of the edge is then merged into a single vertex.   The fractions of T1 vertices and rosettes generated in this random protocol turn out to be consistent with the fractions in the fly embryo epithelial tissue during the germ band extension~\cite{Blankenship2006}.  We carry out this process while making sure the number of cells $F$ remains constant. These moves  model  the convergence of vertices in developmental processes and cell extrusion events.  

Under periodic boundary conditions, the network will always obey Euler's characteristic formula $V-E+F=0$, where $V$ and $E$ are the number of vertices and edges, respectively.  { The density of higher order vertices is} captured by the average vertex coordination number, given by
\be
Z = 2 E / V.
\label{zdef}
\ee
Beginning  with a tissue  
 at $Z=3$ (i.e. containing only tri-junctions),  
an arbitrary value of $Z$ between $3$ and $6$ by applying a series of edge-collapse moves.
 Representative simulation snapshots are shown in Fig.~\ref{fig:snapshots}. After  $Z$ is changed, the energy is minimized using the conjugate-gradient method. For simplicity, no additional topological changes are performed (e.g. T1 transitions, cell divisions or cell apoptosis/extrusion) during the minimization.  
This corresponds to looking at when { higher-order vertices} are formed but not immediately resolved~\cite{Blankenship2006,Fernandez-Gonzalez09, Rauzi_2008, Tamada12,Kasza2014,Streichan17} or in systems where they are persistent for extended periods of time~\cite{Shook6947_persistent_rosettes,trichas_2012,Williams14,Razzell3715}.
  We also performed an alternative set of simulations in which T1 and T2 (cell apoptosis) transitions  are allowed as shown in appendix Fig.~\ref{fig:T1_T2} which do not affect our findings.  
We apply the protocol using two types of initial states at $Z=3$: (1) random tissue networks where cell shapes are obtained from a random Voronoi tessellation~\cite{Bi2016,li_pnas_2018} and (2) an ordered hexagonal tiling~\cite{Farhadifar2007,Staple2010} where every cell is a regular hexagon.

\section{Tissue mechanical rigidity transitions}
We begin by probing the mechanical response of the tissue as a function of the average coordination number $Z$ and the non-dimensional cell shape index $p_0$. In order to determine whether the tissue is mechanically rigid, we analyze the zero modes of the Hessian matrix %
\begin{equation}
M_{i \mu j \nu} = \frac{\partial^2 U}{\partial R_{i\mu}\partial R_{j\nu}},
\label{hessian}
\end{equation}
where ${\bf R}_{i}$ and   ${\bf R}_{j}$ are positions of vertices $i$ and $j$ while $\mu, \nu$ are Cartesian coordinates. The Hessian has dimensions of $2V \times 2V$ and a eigenspectrum of $2V$ eigenvalues  $\{ \lambda_k = \omega_k^2 \}$.  
The presence of zero eigenvalues  indicate the loss of rigidity, which would correspond to  floppy modes or zero modes 
that allow deformation of the system without changes in the total energy.  The number of floppy modes $N_0$ can therefore be used as a measure to distinguish between the rigid and fluid states.  Mathematically, the number of floppy modes is just the nullity of $M$, i.e. number of linearly independent solutions to $M \ket{\bf{\delta R}}=0$.

In Fig.~\ref{fig:floppy_modes}(a), the rigid states ( $N_0 = 0$) and the fluid states ($N_0 > 0$) are  shown for simulation results at  all values pairs ($p_0,Z$). At $Z=3$, the tissue undergoes a rigidity transition at $p_0 = p_0^*\approx3.818 $ ($p_0^*\approx3.722$ for the hexagonally ordered initial state  (Fig.~\ref{hex_si})). This   recapitulates previous results on tri-junction only tissues~\cite{Farhadifar2007,Staple2010,bi_nphys_2015}.   However, Fig.~\ref{fig:floppy_modes}(a) further shows that a fluidized tissue at $Z=3$ can be {rigidified} when $Z$ is increased.  The value of $Z$ where the tissue rigidifies depends on $p_0$. The boundary between rigid and non-rigid states follows a  line of  transition points $Z_{crit} (p_0)$ given by
\be
  Z_{crit}(p_0)=\begin{cases}
    3 & \text{\psolid}.\\
     3 + \mathcal{B} \left[ p_0-p_0^* \right] & \text{\pfluid},
  \end{cases}
  \label{zcrit}
\ee
where  the slope of the boundary separating solid and fluid states ($\mathcal{B}\approx 3.85$) is obtained through empirical fitting. 
 In appendix \ref{app:rigidity_line} we develop a meanfield model which provides an accurate prediction for the numerical value of $\mathcal{B}$.

  To better understand how the creation of { higher order vertices}  can rigidify a tissue, we analyze the behavior of the  fraction of zero modes $f_0 \equiv N_0/(2V)$ as a function of $Z$ for various \pfluid~  in Fig.~\ref{fig:floppy_modes}(b). Using the  relation of $Z_{crit}(p_0)$(eq.~\eqref{zcrit}), we then re-plot $f(\lambda=0)$ vs. $Z-Z_{crit}$ in  Fig.~\ref{fig:floppy_modes}(b-inset) and show that all curves can be collapsed onto a  universal  master curve. This  suggests that the nature of the rosette-driven rigidity transition may be universal across different $p_0$ values.   
Strikingly, $f_0$ decreases faster than the increasing number of topological constraints $Z$, which betrays the Maxwell counting theorem~\cite{maxwell_counting} in traditional rigidity transitions.
 
\begin{figure*}[t]
\begin{center}
\includegraphics[width=1\textwidth]{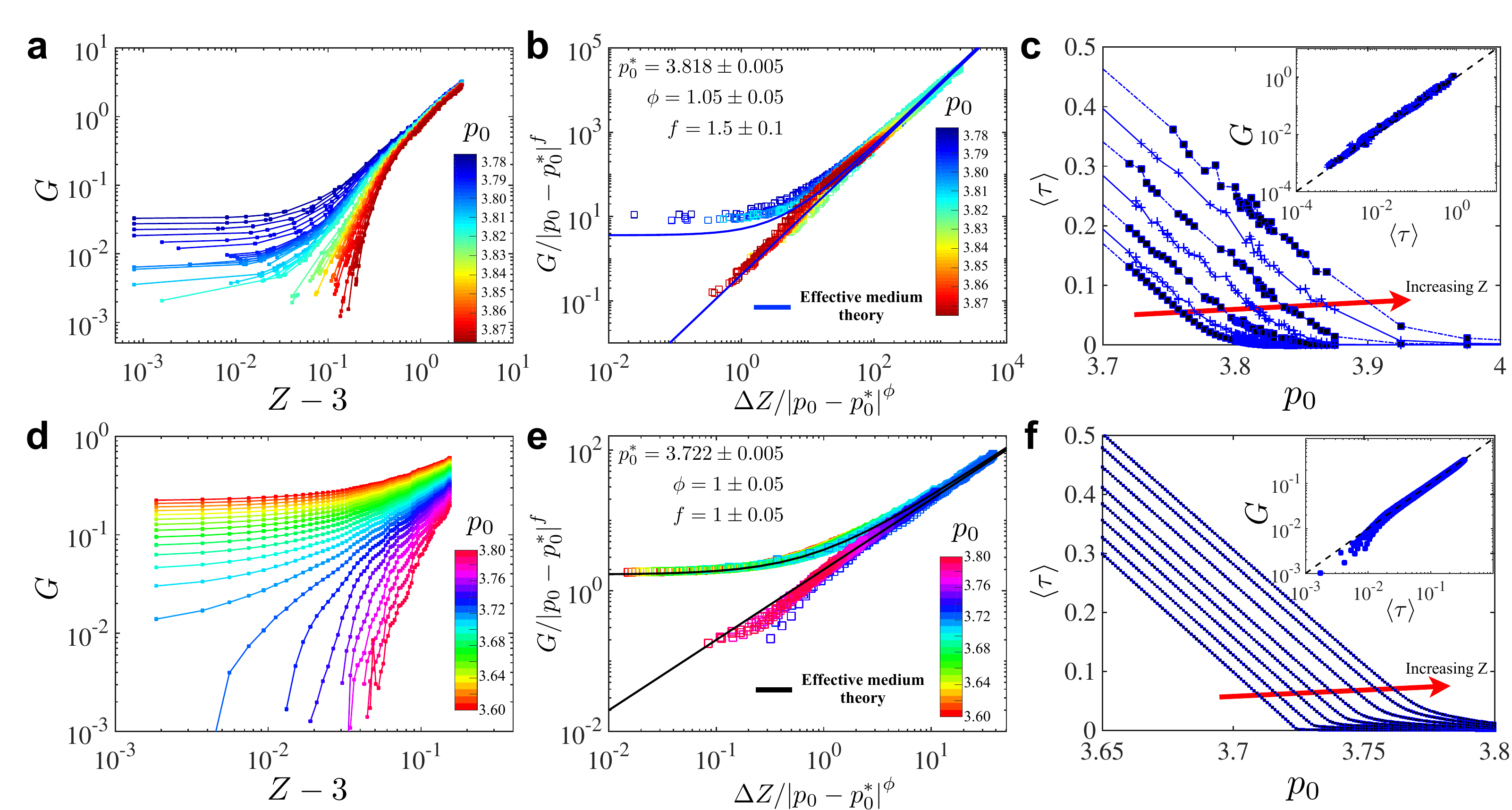}
\caption{
{\bf Shear modulus scaling in the rigid regime.} 
{\bf  (a)} 
 The shear modulus G as a function of $Z-3$ for a range of $p_0$ values. Here the initial states at $Z=3$ are \emph{disordered}. 
 {\bf  (b)} 
When  data from (a) is re-plotted according to the critical scaling ansatz (eq.~\ref{sm_scaling}), they collapse onto two distinct branches. The  solid lines are predictions from the effective medium theory (EMT) (eq.~\eqref{emt_scaling_relation}).
{\bf  (c)} 
Average tension of the tissue as function of $p_0$ for various values of $Z$ as indicated by the red arrow. Here $Z = 3, 3.2 ,3.4,3.6, 3.8, 4, 4.2$. The {inset} shows that the shear modulus is always a linear function of the tension, regardless of the value of $Z$. The dashed line indicates a linear relationship. 
{\bf  (d)} 
The shear modulus as function of $Z-3$ for a range of $p_0$ values. Here the initial states at $Z=3$ is an ordered \emph{hexagonal packing}. At $Z=3$ the rigidity transition occurs at $p_0=p_0*\approx 3.722$~\cite{Farhadifar2007}.
{\bf  (e)} 
Re-scaled shear modulus data according the scaling ansatz (eq.~\ref{sm_scaling}) show that all data collapse onto two distinct branches. The black solid lines are predictions from effective medium theory (eq.~\eqref{emt_scaling_relation}). 
{\bf  (f)} 
Average tension of the tissue as function of $p_0$ for various values of $Z$ as indicated by the red arrow. Here $Z = 3, 3.03, 3.05,     3.07,     3.1,    3.12,    3.14,    3.17$. The {inset} shows that the shear modulus is always a linear function of the tension, regardless of the value of $Z$. Here the initial states at $Z=3$ is a ordered \emph{hexagonal packing}. The dashed line indicates a linear relationship. 
}
\label{fig:sm}
\end{center}
\end{figure*}

\subsection{Nature of the  transition}
To explain the nature of the transition between the fluid and rigid states in Fig.~\ref{fig:floppy_modes}(a), we employ a generalized version of the 
Maxwell constraint counting theorem~\cite{maxwell_counting}: 
the rigidity is lost when the {number of independent constraints} no longer matches
 the {number of degrees of freedom} ($2V$ for the model) and the number of floppy modes is given by the difference. 
It's tempting to simply prescribe one constraint per cell for each area term and each perimeter term in eq.~\eqref{single_E_p0}. However, this will result in an erroneous counting of $N_0=2V-2F = (4-Z)V$, which deviates from the above numerical evidence and suggests that the floppy modes should always be present as long as $Z<4$! 
To accurately perform constraint counting~\cite{Kane_Lubensky}, we apply  the rank-nullity relation to the  Hessian matrix: 
\be
 N_0 \equiv \text{Nullity}(M) =
\overbrace{2 \ \ V}^{\text{Degrees of freedom}} \ - \ \overbrace{\rk(M). }^{\text{No. of indep. constraints}}
\label{counting_00}
 \ee
 and count the  {number of independent constraints for mechanical equilibrium} by calculating $\rk(M)$.

To  illustrate this  how this applies to the \vm , we will consider the Hessian of the energy without the area contribution, i.e. setting  $K_A=0$ in eq.~\eqref{single_E_p0}. 
 A complete calculation for $\rk(M)$ for  arbitrary value of $K_A$ is performed in appendix~\ref{app:counting}. The calculate its rank, we first rewrite the Hessian (eq.~\eqref{hessian}) 
\be
M_{i\mu,j\nu}=K_P \left[\sum_{\alpha=1}^F\frac{\partial p_\alpha}{\partial R_{i\mu}}\frac{\partial p_\alpha}{\partial R_{j\nu}}+\sum_{m=1}^E\tau_m\frac{\partial^2l_m}{\partial R_{i\mu}\partial R_{j\nu}}\right].
\label{hessian_p}
\ee 
The first term in eq.~(\ref{hessian_p}) is positive definite and contribute a total count of $F$ to $\rk(M)$. The second term sums over all $E$ edges where  where $\tau_m = (p_\alpha-p_0) + (p_\beta-p_0)$  is the mechanical line tension~\cite{Yang2017}  for an edge $m$ shared by cell $\alpha$ and $\beta$ and $l_m$ is the  edge length. The second term in eq.~(\ref{hessian_p}) can be rewritten as
\be
\sum_{m=1}^E\frac{\tau_m}{l_m}\hat{e}_m^\perp\hat{e}_m^\perp,
\label{hessian_p_2}
\ee
where
 $\hat{e}_m^\perp=\mathcal{R}(\pi/2)\hat{e}_m$ is 90-degrees rotation of the edge unit vector 
$\hat{e}_m={\bf l}_m/l_m$.
Therefore, terms in this sum (Eq.~\eqref{hessian_p_2})
are positive definite only for edges with $\tau_m>0$.  As a consequence, only edges with a positive tension contribute to  $\rk(M)$. This would result in $E-E_0$ number of independent constraints, where $E_0$ is the number of edges with zero tension $\tau_m=0$.

Together  we obtain the rank of the Hessian matrix to be  $F+E-E_0$. and the Maxwell rigidity criterion for the \vm
\be
N_0 = \overbrace{2 \ \ V}^{\text{Degrees of freedom}} \ - \ \overbrace{ \left[ (E-E_0)+F \right]}^{\text{No. of indep. constraints}}.
\label{counting_2}
\ee 
Eq.~\eqref{counting_2} means that both change topology of the network (e.g. changing $Z,E$) or changing the mechanical state ($E_0$) can influence the rigidity of the tissue. For example, at  $Z=3$, Euler formula dictates $2V=E+F$ and hence  $N_0=E_0$. This suggests that zero modes emerge due to edges with zero tension.  

Next we put this generalized Maxwell relation  to the test by directly comparing the predicted $N_0$ using Eq.~\eqref{counting_2} to the number of zero modes calculated from the Hessian for different $Z$ and $p_0$ in the fluid phase. We obtain an excellent agreement between theoretical prediction and simulation data (Fig.~\ref{fig:floppy_modes}(c)). 
The agreement between the two measures in Fig.~\ref{fig:floppy_modes}(c) implies that the entire rigidity line in Eq.~\eqref{zcrit} is isostatic, i.e. the number of degrees of freedom matches the number of constraints.
The slope of this rigidity line can be obtained through a meanfield model proposed in Appendix~\ref{app:rigidity_line},  which is plotted as the dashed line in Fig.~\ref{fig:floppy_modes}(a).
There are several special cases of constraint counting which are detailed in  Appendix~\ref{app:counting} and Table~\ref{all_counting_table}. First, there is a fixed upper limit to the number of independent constraint, i.e. that they cannot exceed the degrees of freedom (i.e. $\rk(M) \le 2V$). There is also a lower threshold on the number of constraints. When $K_A=0$, the number of constraints cannot drop below $F$. This means that when $K_A=0$, it is not possible for a tissue to have zero modes when $Z = 6$. When $K_A > 0$, the number of constraints cannot drop below $2F$. This means that a tissue will always be rigid for a state with $Z \ge 4$.

\subsection{Non-local response near the transition}
The  rigidity transition is associated with a diverging mechanical lengthscale at the transition. As higher order vertices are created using the edge collapse move,  each move results in  a topology change in the network where $V \to V-1$,  $E \to E-1$ and  $Z \to Z + (Z-2)/(V-1)$. These  changes reduce the number of zero modes  by modifying the degrees of freedom and constraints according to eq.~\eqref{counting_2}. 
Furthermore, rosettes { and T1-junctions} change the mean polygon geometry and induce tension, which has non-local effects on other cells at a distance. 
To capture this effect, we start  from a state at $Z=3$  and measure tissue response to { an edge collapse that creates a single four-fold vertex}. Fig.~\ref{fig:floppy_modes}(d) shows the displacement map of all vertices in response to a   single edge collapse . Closer to the  transition point of $p_0=3.81$, the response is highly non-local and involves a majority of vertices,  with the displacement vector forming spatially extended swirls-like patterns. At $p_0$ values  further away from the transition, the response is more localized. We quantify this growing  lengthscale closer to $p_0^*$ by a spatial correlation function of the displacement field  $C_{\bf{d}}(r) = \langle {\bf{d}}(0)\cdot{\bf{d}}({\bf{r}}) \rangle$ shown in Fig.~\ref{fig:swirls}(a). 
This growing lengthscale is highly reminiscent of the diverging dynamical lengthscale approaching a jamming or glass transition~\cite{LiuNagelReview}. 
\newline

\begin{figure*}[t]
\begin{center}
\includegraphics[width=0.7\textwidth]{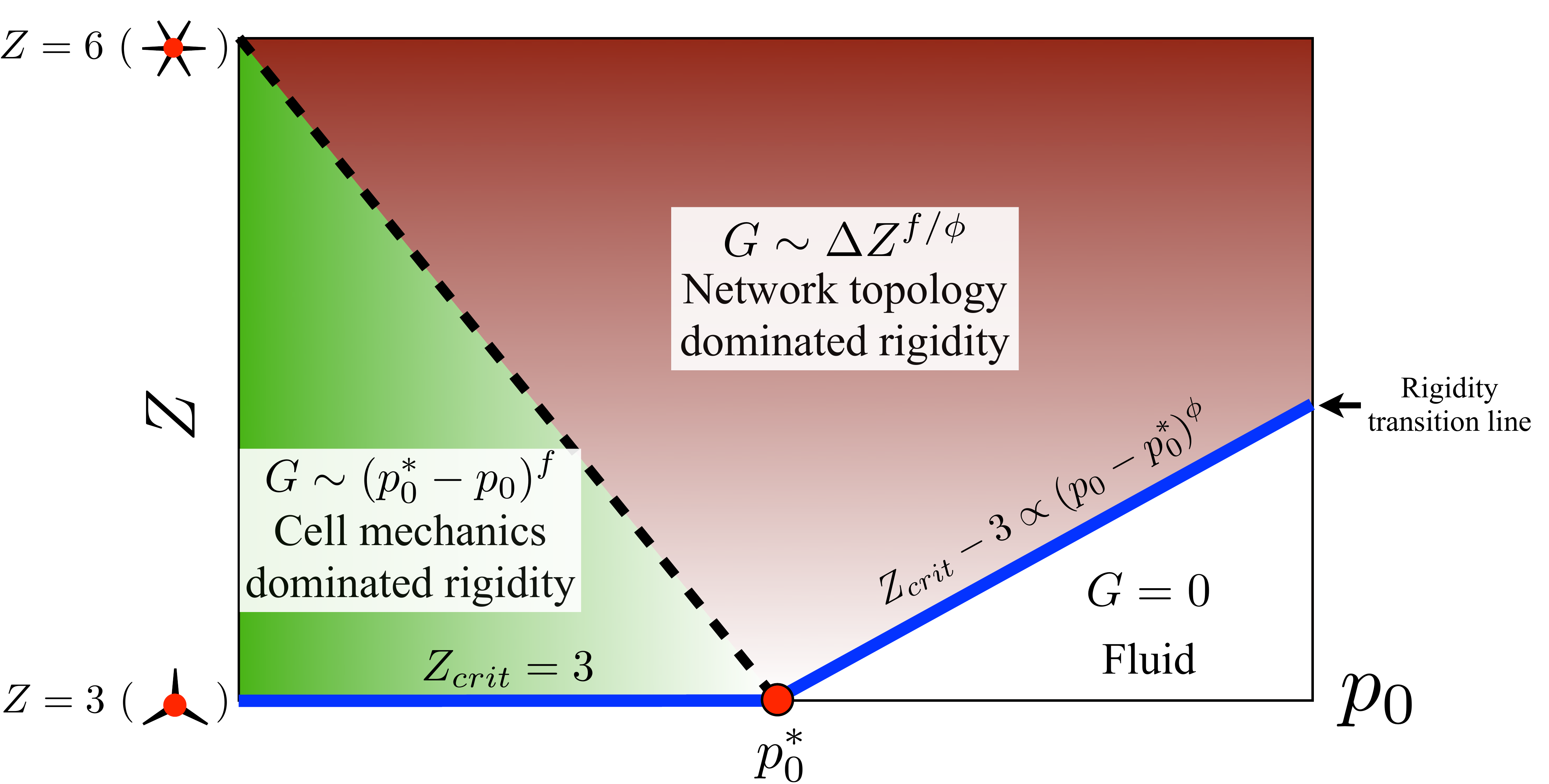}
\caption{
{\bf A phase diagram in $Z-p_0$ space.} The critical scaling of $G$ suggests three distinct mechanical regimes. When  $\Delta Z  \ll {\dpp}^\phi$, the tissue is rigid with a shear modulus that is dominated by cellular mechanics which is encoded in the shape index $p_0$. $\Delta Z  \gg {\dpp}^\phi$, the topology of the cell network dominates tissue rigidity where the shear modulus has a power law dependence on the average vertex coordination $Z$. The tissue becomes fluid like as floppy modes emerge at the line given by $Z_{crit}(p_0)$. 
}
\label{fig:phase_diagram}
\end{center}
\end{figure*}

\subsection{Critical scaling of the shear modulus}
We next probe the tissue mechanics within the rigid  phase  by analyzing the shear modulus.    
The shear modulus $G$ (defined in Appendix~\ref{app:simul_methods})  as a function of $Z$ for different values of $p_0$ is shown in Fig.~\ref{fig:sm}(a). 
The functional dependence of $G$ on $Z$ separates into two regimes based on the value $p_0$. For \psolid, $G$ is finite as $Z \to 3$ and increases with increasing $Z$. 
For \pfluid, as expected from the behavior of zero modes,   $G$  vanishes  at the rigidity transition line $Z_{crit}(p_0)$. In the limit of large $Z$,  $G$ becomes  less dependent on $Z$.  

Given the hallmarks of a critical point observed for $(p_0=p_0^*, Z=3)$~\cite{bi_nphys_2015}  and the rapidly growing correlation lengthscale near the transition,  we propose a critical-scaling ansatz for the shear modulus
\be
G = {\dpp}^f {g}_{\pm} \left( \frac{\Delta Z}{{\dpp}^\phi} \right).
\label{sm_scaling}
\ee
$g_{+}(y), g_{-}(y)$ are the branches of the crossover scaling function for \psolid \ and \pfluid. 
Here $y = \Delta Z/ {\dpp}^\phi$  serves as the crossover scaling variable with  exponent $\phi$. In Fig.~\ref{fig:sm}(b), we re-plot all data using the rescaled variables $G /{\dpp}^f$ and $ \Delta Z/ {\dpp}^\phi$.  The best collapse is obtained with  exponents $f = 1.5 \pm 0.1$ and $\phi = 1.05 \pm 0.05$. 
Furthermore, the branches of the crossover scaling function lead to two distinct mechanical regimes:  
{\bf{(I)}} When \psolid, in the limit of $y \ll 1$, $g_{-}(y) = \text{const}$ or equivalently $G \propto {\dpp}^f$. 
{\bf{(II)}} When $y \gg 1$, the two branches merge or $g_{-}(y) = g_{+}(y) \propto y^{f/\phi}$. This mean the shear modulus becomes independent of cell shapes and  depends only on the network topology or $G \propto {\Delta Z}^{f/\phi}$. 

We also perform the scaling analysis on states initialized from the hexagonal tiling. We calculate  $G$ for all data shown in Fig.~\ref{hex_si} and plot them as a function of $Z$ and $p_0$ values in Fig.~\ref{fig:sm}(d,e). 
Testing the same scaling ansatz (eq.~\ref{sm_scaling}) gives good  scaling collapse of  and yield   $f=1\pm0.05$,  $\phi=1\pm0.05$ and $p_0^*\approx 3.722$. 

\section{Effective Medium Theory}
 To better understand the scaling relations for shear modulus, 
we develop an effective medium theory (EMT)~\cite{Feng85,Mao10,Wyart10,DeGiuli14,Lubensky_review,chase_rigidity,Das_schwarz_plosone_2012} near the critical point $(Z=3,p_0^*)$.
To capture the nature of the tension-induced rigidity, we map the random tension network described by the Hessian in  eq.~\eqref{hessian_p} to a uniformly stressed  medium whose Hessian is given by
\be
\overline\mm=\mm^{\rm topo}+k_{\rm eff}\mm^{\rm ss}.
\ee 
Here $\mm^{\rm topo}$ maps to the Hessian term dependent only on the topology of the tissue network  which is given by the first term in eq.~\eqref{hessian_p} and  $\mm^{\rm ss}$ maps to the tension-dependent term (second term in eq.~\eqref{hessian_p}). 
$k_{\rm eff}$ is the effective tension, uniform on all edges. Now, replacing the tension on edge $m$ with $k_m=\tau_m/l_m$, a random variable obeying the probability distribution, 
\be
p(k)=P\delta(k-\bar{k})+(1-P)\delta(k), 
\ee
which characterizes both edges of zero internal tension and the ones of order $\bar{k}$ (a more general distribution is considered in Appendix~\ref{app:emt}), 
results in a scattering potential,
\be
\mv = \mm_m-\overline\mm=\left\{\begin{array}{cc}
(k_m-k_{\rm eff})e_{m\mu}^{\perp}e_{m\nu}^{\perp}  & \text{if}\quad i=j\in {\partial} m\\
-(k_m-k_{\rm eff})e_{m\mu}^{\perp}e_{m\nu}^{\perp}  & \text{if}\quad i\neq j\in {\partial} m\\
0 & \text{otherwise}
\end{array}\right.,
\ee
where $\partial m$ is the set of vertices defining the edge $m$.

The Green's function of the perturbed system $\mg$ can be written in terms of the Green's function of the effective medium $\overline{\mg}=(\overline\mm-\omega^2\mi)^{-1}$,
\be
\mg = (\mm-\omega^2\mi)^{-1}=\overline\mg + \overline\mg\mt\overline\mg,
\ee
where $\mt=-\mv\sum_{n\geq0}(-\overline\mg\mv)^n$ sums over all multiple-scattering contributions of the edge $m$. 
The EMT assumes the effective medium resembles the random medium if  on average the replacement does not effect the mechanical propagation, i.e. $\int p(k)\mg(k)\rd k=\overline\mg$,  
\be
P\mt(k_m=\bar{k})+(1-P)\mt(k_m=0)=0.
\label{ee}
\ee
This results in the self-consistency equation given by
\be
P\frac{\bar{k}-k_{\rm eff}}{1+(\bar{k}-k_{\rm eff})G_m}+(1-P)\frac{-k_{\rm eff}}{1-k_{\rm eff}G_m}=0,
\label{emt}
\ee
where $G_m=\langle\overline\mg\rangle_m=2\hat{e}_{m}^{\perp}\cdot\overline\mg_{ii}\cdot \hat{e}_{m}^{\perp}-2\hat{e}_{m}^{\perp}\cdot\overline\mg_{ij}\cdot\hat{e}_{m}^{\perp}$. 
The perimeters contribute $F/V=\frac{Z-2}{2}$ independent constraints on each vertex, so on vertex $i$, 
$\tr\langle\overline\mg\overline\mm\rangle_{i,\omega=0}=\frac{Z-2}{2}+k_{\rm eff}\frac{Z}{2}G_m=d$.  
So $G_m=\frac{h}{k_{\rm eff}}$, where $h=\frac{6-Z}{Z}$ (A more strict derivation is done for the honeycomb lattice in Appendix~\ref{app:emt}). 
Inserting it into the self-consistency equation eq.~\eqref{emt}, we have the effective tension 
\be
k_{\rm eff}=\left\{\begin{array}{cc}
\bar{k}\frac{P-h}{1-h} & P\geq P_c=h=\frac{6-Z}{Z}\\
0 & \text{otherwise}
\end{array}\right..
\ee
As $P\in[0,1]$, $Z^*=3$ is the minimal coordination for vertex network to be stabilized by tension. Number of zero modes $N_0=E(P_c-P)=E\frac{6-Z}{Z}-EP=2V-F-(E-E_0)$. 

The elastic modulus is related to the Fourier transform of the Green's function as $\lim_{{\bf q}\to0}C_{ijkl}q_iq_k\mg_{jl,\omega=0}=1$~\cite{Landau86}. As shown above, near the rigidity threshold, the Green's function of the effective medium is singular as $\overline\mg\sim k_{\rm eff}^{-1}q^{-2}$. So the shear modulus scales as the effective tension, $G\propto k_{\rm eff}\sim\bar{k}$.

\begin{figure*}[t]
\begin{center}
\includegraphics[width=1\textwidth]{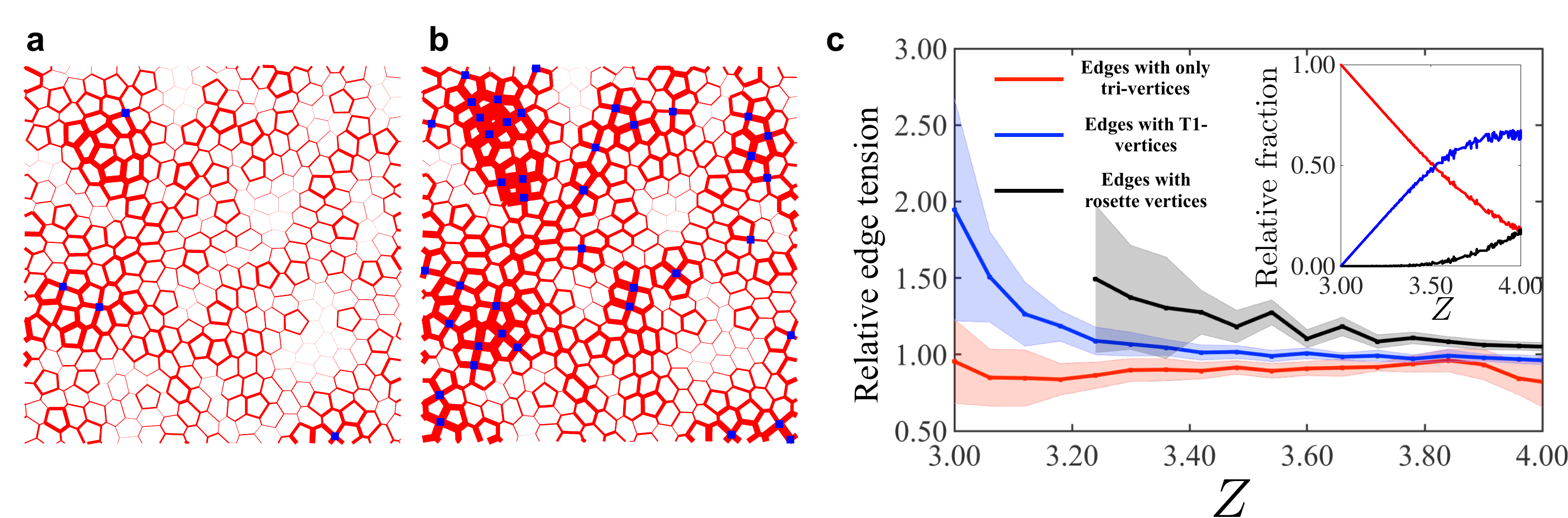}
\caption{
{\bf Spatial correlation between { higher-order vertices} and tensions.} 
{\bf  (a)} At $p_0=3.8$ and $Z\approx 3.01$, a few { higher order vertices} are present and their creation causes the tension in the tissue to increase from $Z=3$. { Here higher order vertices are indicated by blue markers and the thickness of cell edges are proportional to the tension magnitude.}
{\bf  (b)} Holding $p_0=3.8$ constant and further introducing { higher order vertices} brings the tissue to $Z \approx 3.06$. There is a significant degree of correlation between {higher-order} vertices and heightened local tensions
{
{\bf (c)} Relative edge tension due tri-junctions (red), T1-junctions (blue) and rosettes (black) plotted as function of $Z$ at fixed $p_0=3.8$. The solid lines shows the median values of each group and shaded regions indicate the standard deviation. 
The relative tension is defined as the edge tension scaled by the mean tension in the entire tissue $\langle \tau(Z) \rangle$, which scales as $ (Z-3)^{1.5}$.  
Inset: The relative populations of tri-, T1 and rosette vertices as function of $Z$. 
}
}
\label{spatial_tension}
\end{center}
\end{figure*}

\subsection*{Effective Medium Theory predicts the critical scaling observed in simulations}
As shown in the insets of Fig.~\ref{fig:sm} (c\&f), the theory predicts that the shear modulus scales as the effective tension on edges, $G\sim {k_{\rm eff}}$, which is determined by a self-consistency equation and proportional to the mean tension in the network $\bar{k}$ when the Maxwell criterion is saturated $N_0=0$.
In the limit $\bar{k}\to0$, the average tension vanishes with the geometric frustration,
\be
G\propto\bar{k}\sim|p_0-p_0^*(Z)|^f.
\ee 
Inserting the eq.~\eqref{zcrit}, the scaling relation for the  modulus
\be
\label{emt_scaling_relation}
G \propto 
\left\{\ba{lc}
\dpp^f \left[ 1+\frac{Z-3}{\mathcal{B} \dpp}\right]^f & p_0<p_0^*\\
\dpp^f \left[\frac{Z-Z_{crit}(p_0)}{\mathcal{B} \dpp}\right]^f & p_0>p_0^*
\ea\right.
\ee
This implies that $\phi=1$.  
The EMT also gives the exponent $f$. On one hand, from the definition of the  energy eq.~\eqref{single_E_p0}, we have $U/F\sim \bar{k}^2$ when $K_A=0$. 
On the other, the energy $U$ is a smooth function of $p_0$ vanishing at threshold $p_0^*$, so that we can Taylor expand it near $p_0^*$: $U/F=c_2(p_0^*-p_0)^2+c_3(p_0^*-p_0)^3+...$. 
When the vertex network does not relax, for example on the  ordered hexagonal cells, the actual perimeter $p_\alpha$ stays as $p_0^*$ as $p_0$ decreases, so $\bar{k}\propto p_0^*-p_0$, $f=1$. 
When the network does relax, 
the energy vanishes at the quadratic order, $c_2=0$, due to the floppy nature of the unstressed network. To the lowest order, 
$U/F\propto[p_0^*(Z)-p_0]^3$, which implies an exponent of $f=3/2$. This explains the different exponents found in the disordered and ordered cases, as also confirmed numerically in Fig.~\ref{fig:sm} (c\&f). The EMT prediction eq.~\eqref{emt_scaling_relation} with the only fitting parameter $p_0^*$, shown as solid lines in Fig.~\ref{fig:sm} (b) and (e), agrees well with the numerical results.

\section{discussion and conclusions}
In this work, we have  revealed that the topology of the tissue network (controlled by the { average vertex coordination} $Z$) and the intracellular tensions  (controlled by the parameter $p_0$) can greatly influence the rigidity of the tissue. The interplay of these parameters give rise to a fluid to solid transition as well as different mechanical regimes   in the solid phase. The rich set of behaviors  are summarized in a  phase diagram (Fig.~\ref{fig:phase_diagram}). 
 Until now, there has been a lack of theoretical explanation for the recently observed jamming transitions and rigidity in dense tissues~\cite{bi_nphys_2015}. Whereas conventional wisdom on constraint counting would erroneously suggest that a tissue should always be fluid-like, this work explains why $Z=3$ states can be stabilized. It further offers a unifying perspective on why a rigidity transition can be expected at all in a cellular material by using an generalized Maxwell-constraint counting approach.

This work makes several experimentally verifiable quantitative predictions for cell shape and tissue mechanics. 
First, our model provides a criterion to determine the rigidity of a tissue from direct measures of cellular geometry. 
 As the values of $Z$, cell perimeters $\{P\}$, and cell areas $\{A\}$, are experimentally accessible, Eq.~\eqref{pstar} predicts that a tissue should be  fluid-like  if 
$
p \equiv  \left\langle \frac{P}{\sqrt{A}} \right\rangle > p_0(Z=3) + \mathcal{B}^{-1} (Z-3)
$
and rigid  if
$
p \equiv  \left\langle \frac{P}{\sqrt{A}} \right\rangle \le p_0(Z=3) + \mathcal{B}^{-1} (Z-3) 
$,
where $p_0(Z=3) \approx 3.8$ and $\mathcal{B} \approx 3.85$ follow from the theoretical results of this paper. 
Second, the scaling relation of how the tension of in the tissue would grow with the creation of { higher-order vertices}, $\langle \tau \rangle \propto (Z-3)^{1.5}$ can be verified with tension measured in laser-ablation experiments~\cite{Hutson145, Rauzi_2008,Fernandez-Gonzalez09} or mechanical inference methods~\cite{Brodland22111, Chiou_Shraiman_2012,Ishihara_compare_EPJE, ISHIHARA2012201,Kasza2014,Brodland_cellfit}.

{ 
Our result provides a direct understanding of 
the observed fluidization of epithelium in Drosophila embryo development~\cite{Streichan17}. 
Preliminary work ~\cite{Bi_private} on measuring shape index (ratio between cell perimeter and $\sqrt{\text{area}}$) as well as the coordination number $Z$ in the developing Drosophila embryo suggest the tissue is  on the solid side but remains close to the   solid-liquid phase boundary before gastrulation; however it crosses  boundary and transitions into a fluid-like state when ventral furrow forms and dives far into the liquid phase during the fast germ band extension associated with rosette forming. Further,  our predictions that rigidity is lost and shear modulus vanishing in the fluid phase is consistent with elastic moduli measured in~\cite{Streichan17}.  These observations~\cite{Streichan17,Bi_private} suggest that rigidity vanishes at the formation of ventral furrow and the onset of fast germ band extension event.}

We also predict a high degree of correlation between higher-order vertices and adjacent edge tensions. A representative example is shown in Fig.~\ref{spatial_tension}(a \& b). Here T1-junctions and rosettes are marked in blue and cell edge are drawn with widths proportional to the edge tensions.  
We consistently observe higher-order vertices coincide with nearby high edge tensions.  
To quantify this correlation, edges are grouped according to the degree of their adjacent vertices ($Z_1$ and $Z_2$).  This results in three different categories of edges: (1) edges associated with only tri-junctions ($Z_1=Z_2=3$), (2) edges associated with T1-junctions (at least one adjacent vertex is a T1-junction), and (3) edges associated with rosettes (at least one adjacent vertex is a rosette i.e. $Z_1>4$ or $Z_2>4$). The relative fraction for these edge types are plotted as function of $Z$ in Fig.~\ref{spatial_tension}(c) inset.
Interestingly, the statistics show very good agreement~Fig.~\ref{fig:blankenship_compare} with the cell topologies measured during Drosophila embryo elongation~\cite{Blankenship2006}. 
Next, for edge type we compare their tensions to the mean tension of the entire tissue, which is shown in Fig.~\ref{spatial_tension}(c). This predicts as $Z$ increases, the edges near higher-order {vertices} consistent carry more tension compared to tri-junction vertices. In particular, at the onset of higher-order vertices appearing, tensions associated with T1-junctions are in the range of $1.5\pm0.3$ times the tension near tri-junctions. Again these numbers match up well with the experimental results from Drosophila embryo elongation where tensions associated with edges pointing along anterior-posterior (AP) direction is typically 1.7 times~\cite{Fernandez-Gonzalez09} compared to edges pointing in the dorsal-ventral (DV) direction. While a more systematic study is warranted  to model  the elongation processes in Drosophila development. Our minimal model  suggest that the correlations between junctional tension and cellular topologies may be more universal. For future work, we will build on these general results to specifically model the feedback mechanism that leads to rosette formation in Drosophila embryogenesis where both increased rosette count and increased tensions are predicted to emerge from the myosin positive feedback loop~\cite{Fernandez-Gonzalez09,Pouille09}.
Finally, the criticality of this rigidity transition results in non-local response to perturbations near the critical point, as shown in Fig.~\ref{fig:floppy_modes}(d). 
We thus expect the laser ablation~\cite{Pouille09} of cell edges to induce a similar long-range effect to collapsing edges near the onset of tissue rigidity. The spatial extension of the response could even be used as an indicator of how far the tissue is from the rigidity transition. 

In this work we have focused  mainly on the mechanics of tissues with stable rosettes. However, rosettes in actual tissues can vary greatly in their life-time.  This ranges from rosettes that are transient and resolve quickly to form new structures to rosettes that persist for an extended period of time or may not resolve at all. 
  For example,   the life time of higher order vertices  are {thought to be linked to the Hippo pathway and  the  force-sensitive protein Ajuba}~\cite{Razzell3715,SUN2016694,Ibarjcs214700}.
Our predictions are therefore applicable to the mechanical response of the tissue at timescales shorter than the rosette resolution time. 
In the version of the \vm here, we have not included biological feedback of cells in response to mechanical stress. However in Drosophila embryo elongation for example, myosin motors can slide more on actin in response to higher tensions and as a result,  the observed cell perimeters can change~\cite{Noll2017}. At the same time, myosin recruitment is stimulated to sustain the perimeter at high tensions. These feedback events occur at a characteristic time scale corresponding to the relaxation and resolution of rosettes. We would expect a breakdown of the static rigidity theory in these regimes. For the Drosophila embryo this corresponds to a timescale of several minutes~\cite{Doubrovinski17}.  Modeling dynamical movements and transient formation+resolution of rosette will be an interesting avenue for future research.

Another important aspect of this work is the use of a generalized Maxwell constraint counting argument to explain why rigidity transitions can occur in tissues in general. Whereas conventional counting arguments suggest~\cite{2018arXiv180901586M} that the \vm is always under-constrained, we show that this this not to be the case through an accurate counting of mechanical constraints. For example, conventional constraint counting would predict that a cellular material should only be stable for $Z>4$. However, most epithelial tissues can be stable at precisely $Z = 3$. This work explains why $Z=3$ states are stabilized and further offers a unifying perspective on why a rigidity transition can be expected at all in a cellular material by using a Maxwell-constraint counting approach. 
Our understanding of this rigidity transition is base on a generalized Maxwell counting that properly count the critical role of tension. It's a natural extension of the stress-driven rigidity transition in packings and fiber networks close to the critical topology with $Z=2d$~\cite{Sharma_nphys_2016, DeGiuli14,  feng_pre_2015, feng_2016_SM_nonlinear}. 
The appearance of the tension can also be viewed as a result of the geometric incompatibility of the material metric as studied in a continuum treatment of the vertex model by Moshe and coworkers~\cite{moshe_2017}. 
The similar insight of tension and geometric compatibility was also proposed in a recent study by Merkel and Manning~\cite{Merkel2017} on the 3D Voronoi-based cell model, a variation of the vertex model~\cite{Li2014, Bi2016, barton_silke_rastko}.

The constraint counting developed here can be easily generalized to the Voronoi-based cell models. In those  models, the degrees of freedom are cell centers (whose number is $2F$) rather than cell vertices and the cell shapes are obtained from a Voronoi tessellation of the cell centers. 
For the same energy Eq.~\eqref{single_E_p0},  the number of constraints is the same as that given by Eq.~\eqref{counting_2}, i.e. $\text{max}(2F,E-E_0+F)$. Which leads to a constraint counting that is given by $N_0-N_{ss}=2F-\text{max}(2F,E-E_0+F)$.  When edges carry tensions and the number of constraints is greater than the degrees of freedom $E-E_0+F > 2F$, we expect there to be redundant constraints which will result in $N_{ss}$ states of self-stress. Interestingly, when $2F>E-E_0+F$, the counting leads to $N_0 = 0$, indicating that the system is always marginal and thus lack of a transition to fluid in the linear response, as recently observed and studied in~\cite{C7SM02127E}.  
However,  the solid-fluid transition is still evidently shown in the Voronoi-based model with self-propelling cells~\cite{Bi2016,li_pnas_2018}. The nature of this observed transition is determined by two aspects. First, there is still a transition to self-stressed state when the number of  edges under tension, $E-E_0$ becomes larger than the number of cells $F$ at the same threshold $p_0=3.81$. Second, the marginal nature of the system when $p_0>3.81$ makes the linear response fragile and plastic nonlinear processes (T1 rearrangement, rosette formation, etc.) take over the response, so that the system behaves as a liquid whenever a finite energy is injected to it.
 This is supported by an ongoing work~\cite{Bi_private}, which appears to show that in the marginal range of  Voronoi-based cell models, 
T1 rearrangements can be easily triggered with a perturbation whose magnitude vanishes with the system size.  This could very well suggest an example where the Maxwell constraint counting and linear-response theory fails to predict the loss of the rigidity.  

\section*{Acknowledgements}
The authors wish to acknowledge  
Sebastian Streichan for providing helpful comments, suggestions  and preliminary experimental data. 
The authors would also like to thank 
Boris Shraiman, 
Matthieu Wyart,   
Cristina Marchetti, 
Mark Bowick, 
Jen Schwarz, 
Jin-Ah Park, 
Lisa Manning, 
Xiaoming Mao, 
Frederick MacKintosh, 
Eric DeGiuli,
Karen Kasza, 
and Bulbul Chakraborty 
for helpful discussions.
LY is supported by the Gordon and Betty Moore Foundation under Grant No. 2919. 
This research was supported in part by the Kavli Institute for Theoretical Physics under NSF Grant No. PHY17-48958 and NIH Grant No. R25GM067110.
The authors acknowledge the support of the Northeastern University Discovery Cluster.

\appendix

\section{Simulations Methods}\label{app:simul_methods}
We simulate tissues composed of $N = 625$ cells under periodic boundary conditions with box size $L=\sqrt{N}=25$ which is commensurate with the value of preferred cell area of unity.  We use $K_p=1$ for the data shown in this paper. The value of $p_0$ is varied between 3.6 and 4.3. 
We carry out the protocol using two different types of initial states at $Z=3$: (1) random tissue networks where cell shapes are obtained from the random Voronoi tessellations in the  Voronoi-based cell model~\cite{Bi2016} and (2) an ordered hexagonal tiling~\cite{Farhadifar2007,Staple2010}. After initialization, the vertex positions in the tissue is evolved to energy minimum using the Broyden-Fletcher-Goldfarb-Shanno method~\cite{L-BFGS} until the residual force magnitude on all vertices is less than $10^{-8}$. The simulations methods are based on~\cite{bi_nphys_2015}.

The shear modulus is calculated by considering the linear response of the tissue  to an infinitesimal affine strain $\gamma$.  It is given by~\cite{Maloney_PRE_2006}
\be
  G = G_{\text{affine}}-G_{\text{non-affine}} = \frac{\partial^{2}U}{\partial \gamma^{2}} \Bigr\vert_{\gamma=0} - \Xi_{i\mu}M^{-1}_{i\mu j\nu}\Xi_{j \nu}.
  \label{eq:sm}
\ee
Here 
$\Xi_{i\mu}$ is the derivative of the  force on vertex $i$ with respect to the strain, given by 
\be
\Xi_{i\mu} \equiv \frac{\partial^{2}U}{\partial \gamma \partial r_{i\mu}} \Bigr\vert_{\gamma=0}
\ee. 

\section{Mean-field model for determining the slope of the rigidity line}\label{app:rigidity_line}
The rigidity transition line in $(p_0,Z)$-plane can be expressed as eq.~\eqref{zcrit} or its inverse
\be
p_0^*(Z) = p_0^*(Z=3) + \mathcal{B}^{-1}(Z-3)
\label{pstar}
\ee
where $\mathcal{B}^{-1}=\partial p_0^*(Z) /\partial Z \vert_{Z=3}$ which captures how the \emph{critical preferred perimeter} changes when the number of edges per cell changes. We then take a mean-field approach and replace $p_0^*(Z)$ with the perimeter of a regular $n$-sided polygon (with area of unity) of $n=2Z/(Z-2)$ sides 
\be
P_n = 2 \sqrt{n} \ \tan (\pi/{n}).
\ee
to obtain $\mathcal{B}^{-1}=\partial P_n /\partial Z \vert_{Z=3} \approx 0.260$. This mean-field approximation serves well to predict the slope of the transition line (Fig.~\ref{fig:floppy_modes}(a)). However, the properties of regular polygons are not helpful for predicting the value of the transition point itself ($p_0^*(Z=3) = 3.81 (\text{disordered}), 3.72 (\text{hex})$) which is an emergent collective property of the system.

\section{Generalized Maxwell Constraint Counting for the \vm}\label{app:counting}

\subsection{Hessian (dynamical matrix) of the vertex model}

We begin by calculating the Hessian matrix of the total energy, which is given by Eq.~\eqref{single_E_p0}. 
Now since $\sum_\alpha A_\alpha = \text{constant}$ for a confluent tissue not undergoing cell number changes, it is equivalent to work with the simpler form of energy
\be
U  = \frac{1}{2}\sum_{\alpha=1}^F \left[  K_A {A_\alpha}^2 + K_P (P_\alpha - P_0)^2 \right].
\label{full_u}
\ee
In Eq.~\eqref{full_u}, we have also multiplied the energy by a factor of $1/2$ for convenience when taking derivatives. 

The network of cells is given by vertices (at positions $\{ {\bf R}_{i}\} $) and edges (specified by vectors edge vectors ${\bf L}_m$). The relation between edges and vertices is given by the directed adjacency matrix $g_{mi}$. Edge vectors can be calculated using 
\be
{\bf L}_m=\sum_ig_{mi}{\bf R}_i,
\ee 
where
$g_{mi}$ is nonzero if vertex $i$ is on edge $m$; it is $+1$  vertex $i$ forms the head of edge vector $m$ and $-1$ if it is the tail.  The relation between edges and facets (cells) is given by edge-facet adjacency matrices $h_{m\alpha}$, which is $+1 (-1)$ if edge $m$ goes counter-clockwise (clockwise) around facet $\alpha$ and zero otherwise. 
In this notation, both the area and the perimeter can be written in vertices and edges,
\be
\begin{split}
A_\alpha &= 
\frac{1}{4}\sum_{m,i}|g_{mi}|h_{m\alpha}({\bf R}_i\times{\bf L}_m)\cdot{\bf z}
=-\frac{1}{4}\sum_{m,i}|g_{mi}|h_{m\alpha}{\bf R}_i\cdot{\bf L}_m^{\perp} \\
P_\alpha & =\sum_m|h_{m\alpha}| L_m.
\end{split}
\label{A_P_def}
\ee
First, it is instructive to calculate the total force on each vertex, which is given by the gradient of Eq.~\eqref{full_u},
\be
f_{i\mu} \equiv  - \frac{\partial{U}}{\partial{R}_{i\mu}} = 
\sum_{\alpha=1}^F \left[  K_A {A_\alpha}  \frac{\partial{A_\alpha}}{\partial{R}_{i\mu}} + K_P (P_\alpha - P_0)  \frac{\partial{P_\alpha}}{\partial{R}_{i\mu}} \right].
\label{full_force_facet}
\ee
Based on the definitions of area and perimeter (eq~\eqref{A_P_def}), we obtain the geometric derivatives
\be
\begin{split}
\frac{\partial A_\alpha}{\partial R_{i\mu}} &=-\frac{1}{2}\sum_m|g_{mi}|h_{m\alpha}L_m e_{m\mu}^\perp 
\qquad\text{and} \\
\frac{\partial P_\alpha}{\partial R_{i\mu}} &=\sum_m|h_{m\alpha}|g_{mi}{e}_{m\mu}.
\end{split}
\label{A_P_derivatives}
\ee
Therefore using Eq.~\eqref{A_P_derivatives} it is possible to rewrite Eq.~\eqref{full_force_facet} in terms of a sum over all edges 
\be
f_{i\mu} = -\sum_{m=1}^E 
\left(
-\frac{1}{2} K_A \sigma_m L_m  |g_{mi}| e_{m \mu}^{\perp}+
K_P \tau_m g_{mi}  e_{m\mu} 
\right)
\label{full_force_edges}
\ee
where we have separated the force on each vertex into two mutually orthogonal components: a component {along the edge vector}  $\hat{\bf e}_{m} \equiv  {\bf{L_m}} /  L_m $ and a component {perpendicular to the edge vector} given by $\hat{\bf e}_{m}^\perp = \mathcal{R}(\pi/2) \hat{\bf e}_{m} $, or the edge vector rotated by $\pi/2$.  The magnitude of forces are given by
\be
\begin{split}
\tau_m &=\sum_\alpha|h_{m\alpha}|(P_\alpha-P_0)= [(P_\alpha- P_0)+(P_\beta- P_0)], \\
\sigma_m  &=\sum_\alpha h_{m\alpha}A_\alpha= A_\alpha - A_\beta.
\end{split}
\ee
 $K_P \tau_m$ is the line tension due to mismatch between the actual  perimeters of cell $\alpha$ and $\beta$ from the preferred perimeter. $K_A \sigma_m/2$ is the pressure due to the difference between the areas of cells $\alpha$ and $\beta$. 

The Hessian is a $2V\times2V$ matrix given by 
\be
\begin{split}
M_{j\nu,i\mu}  \equiv &  \frac{\partial{U}^2}{\partial{R}_{j\nu}\partial{R}_{i\mu}} \\ 
= &
\sum_m
\Bigg[
K_P \left(
\frac{\partial \tau_m}{\partial R_{j\nu}} g_{mi} e_{m \mu} +  \frac{\partial e_{m\mu}}{\partial R_{j\nu} } \tau_m g_{mi} 
\right) \\
&-\frac{1}{2}K_A 
\left(
\frac{\partial \sigma_m}{\partial R_{j\nu}}  L_{m \mu}^{\perp} |g_{m i}| +
\frac{\partial L_{m \mu}^{\perp}}{\partial R_{j\nu}} \sigma_m |g_{mi}|
\right)
\Bigg]
\end{split}
\label{hessian_0}
\ee
which can be written explicitly as 
\be
\begin{split}
M_{j\nu,i\mu} =\sum_{m=1}^E\sum_{n=1}^E
 &  \Bigg[
K_P \ e_{n \nu} g_{nj} * H_{nm} * e_{m \mu}  g_{mi}  \\
&+K_P \ e_{n \nu}^\perp g_{nj}  * T_{nm} *  e_{m \mu}^\perp g_{mi} \\
& +\frac{K_A}{4} \ e_{n \nu}^\perp g_{nj} * \tilde{H}_{nm}  * e_{m \mu}^\perp g_{mi}  \\
& +\frac{K_A}{2}  \ e_{n\nu} g_{nj}  * \Sigma_{nm} * e_{m\mu}^\perp g_{mi} \\
& -\frac{K_A}{2} e_{n\nu}^\perp g_{nj}  * \Sigma_{nm} * e_{m\mu} g_{mi}   
\Bigg].
\label{hessian_1}
\end{split}
\ee

Eq.~\eqref{hessian_1} has been written in a symmetric form where each term is a ($2V\times2V$) matrix decomposed into a product of three matrices $(2V\times E)(E\times E)(E\times 2V)$. Here the $E\times E$ matrices are defined as
\be
\begin{split}
H_{nm} &= \sum_{\alpha=1}^F \vert h_{n \alpha} h_{m \alpha} \vert \\
\tilde{H}_{nm} &= L_n L_m \sum_{\alpha=1}^F  h_{n \alpha} h_{m \alpha}  \\
T_{nm} &= \frac{\tau_m}{L_m}\delta_{nm}  \\
\Sigma_{nm} &= \sigma_m \delta_{nm}. 
\end{split}
\label{edge_mat_def}
\ee
The matrices $H, \tilde{H}$ give the relationship between edges and facets. $T, \Sigma$ are diagonal matrices which give  the tension and pressure on each edge. The geometric property of edges and relationship between edges and vertices is captured by the  $2V\times E$ matrices
\be
\begin{split}
S_{n,j,\nu}^\| = e_{n \nu} g_{nj} \\
S_{n,j,\nu}^\perp = e_{n \nu}^\perp g_{nj}
\end{split}
\label{S_mat_def}
\ee
which are mutually independent by definition. 
 In matrix form, Eq.~\eqref{hessian_1} can be written  using Eqs.~\eqref{S_mat_def} and~\eqref{edge_mat_def} in matrix form as a product of block matrices. 
\be
\begin{split}
M &= S \ C \ S^T \\
& =
\left(S^\| \quad \vert \quad S^\perp \right)
\left(
    \begin{array}{c|c}
	K_P H & -\frac{1}{2}K_A \Sigma \\
	      \hline
      \frac{1}{2} K_A \Sigma & K_P T + \frac{1}{4} K_A \tilde{H}
    \end{array}
\right)    
 {\left(S^\| \quad \vert \quad S^\perp \right)}^T.
\label{hessian_block}
\end{split}
\ee

Here we have defined the augmented matrix $S \equiv (S^\| \ \vert \ S^\perp)$. And a $2E \times 2E$ block matrix 
\be
C=
\left(
    \begin{array}{c|c}
	K_P H & -\frac{1}{2}K_A \Sigma \\
	      \hline
      \frac{1}{2} K_A \Sigma & K_P T + \frac{1}{4} K_A \tilde{H}
    \end{array}
\right). 
\label{hessian_center_mat}
\ee
$C$ is block anti-symmetric.

\subsection{Constraint counting}
We apply the rank-nullity theorem to the Hessian matrix in order to accurately count the number of constraints~\cite{Kane_Lubensky}
\be
\rk(M) + \text{nullity}(M) = 2V.
\label{rank_nullity_hessian}
\ee
The rank of $M$ give the number of independent constraints for force balance~\cite{Kane_Lubensky, Lubensky_review}. The nullity of $M$ is the dimensionality of the null space of $M$, i.e. the number of solutions to $M \ket{\delta {\bf{R}}_i} = 0$. These correspond to the number of ways to infinitesimally perturb the vertices such that the energy of the system does not change. This is exactly the number of {zero-modes} in the Hessian, which we have termed $N_0$. Therefore the constrain counting becomes
\be
\label{full_counting}
\begin{split}
\text{nullity}(M) = N_0  = \overbrace{2 \ \ V}^{\text{Degrees of freedom}} \ - \ \overbrace{\rk(M) }^{\text{No. of indep. constraints}}.
\end{split}
\ee

\begin{table*}[t]
\begin{center}
\begin{tabular}{llcc}
\centering
\renewcommand{\arraystretch}{1.3}
Model   \quad   \quad & 
$N_{\text{constraint}}=\rk(M)$ & 
Rigidity criterion ( $N_{d.o.f.} >N_{\text{constraint}}$)   \quad & 
$N_0= N_{d.o.f.} -N_{\text{constraint}}$\footnote{excluding the $d=2$ number of trivial transitional zero modes}  \\ \hline \hline 
%
$K_A = 0$   & 
$\text{min}\left[2V,\text{max}(E+F-E_0,F) \right] $  & 
$ 
\begin{array} {ll}  
Z \ge \frac{3}{1-E_0/2E} \ &\text{if} \quad (2-6/Z)E<E_0   
 \end{array}
$ &
$  
\begin{array} {ll} N_0= &  \begin{cases}
(3-Z)V+E_0 \ &\text{if} \quad (2-6/Z)E<E_0 \\
0 \ &\text{otherwise} 
\end{cases}
 \end{array}
$  
%
\\
\hline
$K_A > 0$   & 
$\text{min}\left[2V,\text{max}(E+F-E_0,2F) \right] $  & 
$ 
\begin{array} {ll}  
\begin{cases}
Z \ge \frac{3}{1-E_0/2E} \ &\text{if} \quad (2-6/Z)E<E_0 < E/2 \\
Z \ge 4 \ &\text{if} \quad E_0 > E/2
\end{cases}
 \end{array}
$ &
$  
\begin{array} {ll} N_0= &  \begin{cases}
(3-Z)V+E_0 \ &\text{if} \quad (2-6/Z)E<E_0 < E/2 \\
(4-Z)V \ &\text{if} \quad E_0 > E/2 \\
0 \ &\text{otherwise} 
\end{cases}
 \end{array}
$  
  \\
\hline \hline
\end{tabular}
\caption{
{
A overview of constraint counting in \vm. The cases for when $K_A=0$ and $K_A>0$ are presented separately. The rigidity criterion is the minimum value of $Z$ needed to ensure rigidity. Note that the count of edges (E), vertices(V) and cells(F) are related through the Euler characteristic ($V-E+F=0$) and the definition for vertex coordination $Z = 2E/V$. 
}
 }
\label{all_counting_table}
\end{center}
\vspace{-0.25 in}
\end{table*}

To calculate the rank of $M$, we first note that it has a trivial upper bound of $2V$ (in which case $N_0=0$).  Additionally since $M$ is a product of $S$ and $C$ and $S$ is full rank, it follows that   
\footnote{$\rk(AB)=min(\rk(A),\rk(B))$ }
$\rk(M)=\rk(C)$ and by application of Guttman rank additivity of a block matrix~\cite{Guttman} we can write down a general relation
\be
\begin{split}
\rk(M) = & \rk(K_P \ H) + \\
&  \rk(K_P \ T  + \frac{1}{4} K_A \tilde{H} + \frac{1}{4}K_A^2 K_P^{-1} \ \Sigma \ H^\dagger \ \Sigma).
\label{rank_M}
\end{split}
\ee
In Eq.~\eqref{rank_M}, $H^\dagger$ is the Morse-Penrose pseudo-inverse of $H$ since below we will show that it is always less than full rank and therefore not invertible. Since $H$  can be expressed as a product of matrices with lesser rank $F$( Eq.~\eqref{edge_mat_def}), it always holds that $\rk(H)=F$. The reason that the pseudo-inverse can be done without changing the actual dimension of the $C$ matrix is that one can show that the rank of $\Sigma$ is also $F$, and these $F$ directions are not independent of the $F$ directions in $H$.

In the case of $K_A=0$, $\rk(M) = \rk(H) + \rk(T)$. The rank of $T$ is precisely given by the number of edges with $\tau_m > 0$ ( Eq.~\eqref{edge_mat_def}) which we have defined as $E-E_0$ in the main text. Hence the constraint counting for when $K_A=0$ is given by
\be
N_0 = 2V-\text{min}\left[2V,(E-E_0+F)\right].
\label{counting_0}
\ee
{ The minimum function is used in Eq.~\eqref{counting_0} because the rank of a matrix can not exceed the its largest dimension}

\begin{figure}[htpb]
\begin{center}
\includegraphics[width=.75\columnwidth]{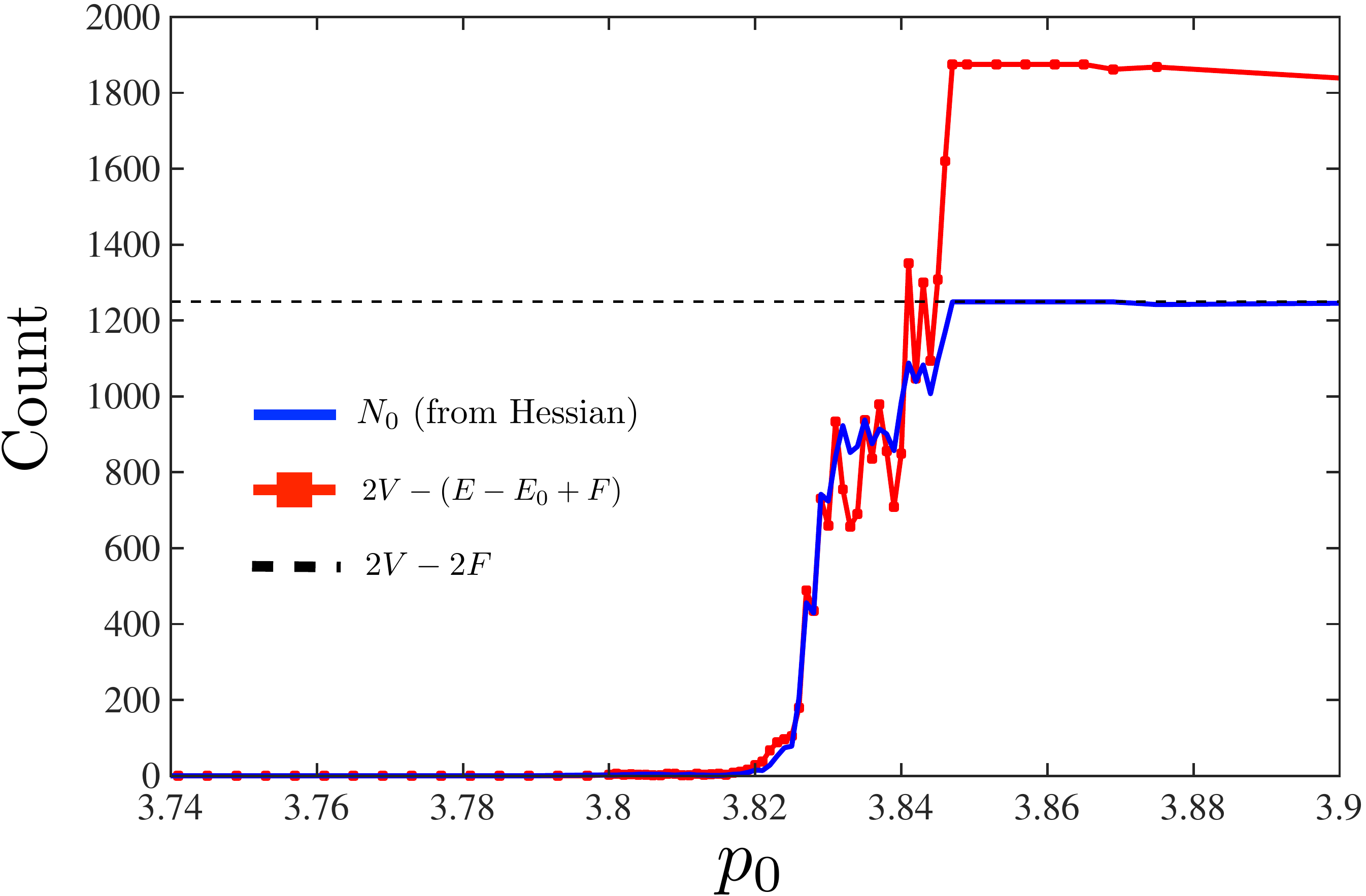}
\caption{
Maxwell counting in the vertex model is given by $N_0 = 2V-\text{max}(2F, E-E_0+F)$. 
}
\label{z_3_counting}
\end{center}
\end{figure}

When $K_A>0$ the rank $M$ can take on a continuum of values, however it is still  possible to give bounds for the rank of $M$.  First even when all tensions and pressures vanish, i.e. $\tau_m=0, \ \sigma_m = 0$ on all edges, the ranks of both $H$ and $\tilde{H}$ would still remain at $F$.  This would provide the lower bound of $\rk(M) \ge 2F$. On the other hand, when all even when all tensions and pressures are finite, the second term in Eq.~\eqref{rank_M} cannot exceed $E+F$. We therefore obtain the  limits for the constraint counting Eq.~\eqref{full_counting}
\be
2V - (E+F) \le N_0 \le 2V-2F.
\label{counting_1}
\ee
When zero tension edges are taken into account, we obtain the more general result
\be
N_0 = \overbrace{2 \ \ V}^{\text{Degrees of freedom}} \ - \ \overbrace{\text{min}\left[2V,\text{max}(2F, E-E_0+F) \right]}^{\text{No. of indep. constraints}}.
\label{counting_2_appendix}
\ee
We explicitly test this in a system of $F=625$ cells at fixed $Z=3$. In Fig.~\ref{z_3_counting}, the number of zero modes are first  calculated directed from the Hessian matrix and plotted for states at various values of $p_0$. The quantity $2V-(E-E_0+F)$ is also plotted as function of $p_0$ for the same states and they closely track the behavior of $N_0$ until $2V-(E-E_0+F)$ becomes larger than the upper limit of zero modes, $2V-2F$.  

Eq.~\eqref{counting_1} also allows us to make interesting predictions. Using the fixed topological relations $E= Z V/2$ and $F =(Z-2) V/2$, Eq.~\eqref{counting_1} becomes a condition for the fraction of zero modes in the system $f_0$, which is given by
\be
f_0 \equiv \frac{N_0}{2V} \in [0,\frac{4-Z}{2}].
\label{counting_3}
\ee
Eq.~\eqref{counting_1} suggests that when $Z>4$, no zero modes should be present in the system. 
At $Z=3$, the fraction of zero modes is at most $1/2$. 

{
In Table~\ref{all_counting_table}, we list the Maxwell counting for both $K_A=0$ and $K_A>0$ cases, with the rigidity criterion as well as the number of zero modes in the system for all possible cases. 
}

\section{Effective Medium Theory Details}\label{app:emt}

We derive the effective medium theory for the case with only perimeter constraints, $K_A=0$ and $K_P>0$. To approximate the scattering behavior of the Hessian 
$
M=K_PS^\parallel HS^{\parallel T}+K_PS^\perp TS^{\perp T}
$, 
we propose the following effective medium,
\be
\overline\mm=\mm^{\rm topo}+k_{\rm eff}\mm^{ss},
\ee
with $\mm^{\rm topo}$ mapping to the cell topology term $K_PS^\parallel HS^{\parallel T}$ and $k_{\rm eff}\mm^{ss}$ corresponding to the stress contribution $K_PS^\perp TS^{\perp T}$.

\subsection{General stress distribution}
We consider a general distribution of random stresses on edges $k_m=\frac{\tau_m}{L_m}$,
\be
p(k)=P\rho(k)+(1-P)\delta(k),
\ee
where function $\rho(k)$ is normalized in $k>0$ with a single stress scale $\bar{k}$,
\be
\rho(k) =\frac{1}{\bar{k}}\hat{\rho}(\frac{k}{\bar{k}}).
\ee 

Following the standard coherent potential approximation procedure~\cite{Lubensky_review}, the self-consistent equation of the effective medium reads, 

\be
P\int_0^{\infty}\rd k\rho(k)\frac{k-k_{\rm eff}}{1+(k-k_{\rm eff})G_m}+(1-P)\frac{-k_{\rm eff}}{1-k_{\rm eff}G_m}=0,
\label{ee_2}
\ee
where $G_m=\langle\overline\mg\rangle_m=\sum_{i\mu,j\nu}g_{mi}e_{i\mu}^\perp\overline\mg g_{mj}e_{j\nu}^\perp$ is singular as $G_m=\frac{h}{k_{\rm eff}}$ with $h=\frac{6-Z}{Z}$. 
Expanding near vanishing stress to the first order of $k_{\rm eff}$, we get 
\be
k_{\rm eff} = \bar{k}\frac{h}{a P}\frac{P-h}{1-h},\qquad a=\int_0^\infty\rd x\frac{1}{x}\hat{\rho}(x).
\ee

\begin{figure}[h!]
\centering
\includegraphics[width=.5\linewidth]{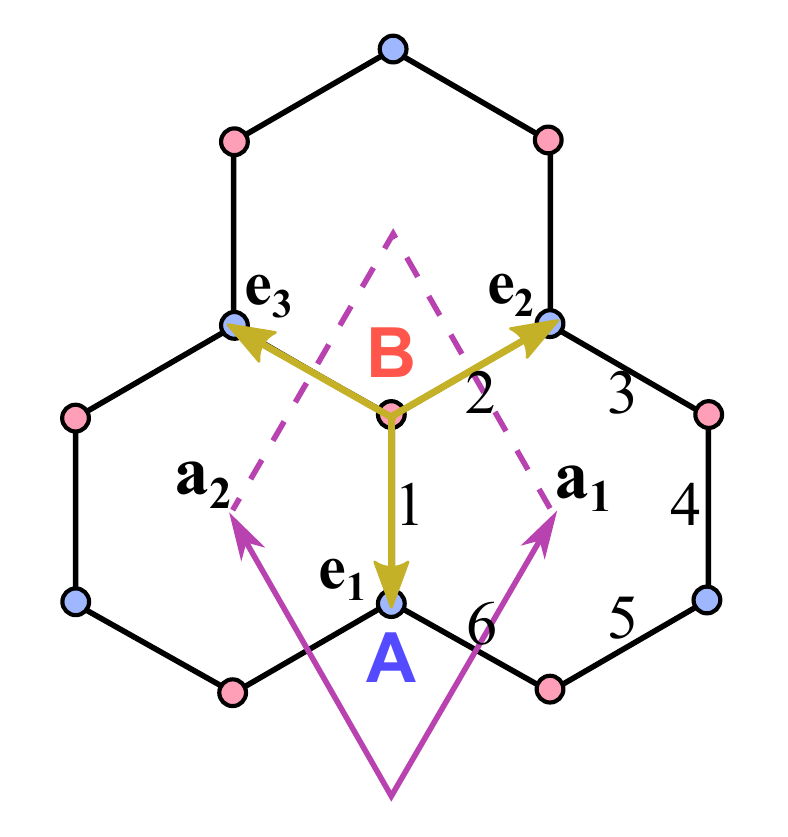}
\caption{\small{Primitive cells of the honeycomb lattice.}}\label{honeycomb}
\end{figure}

\subsection{$G_m$ on honeycomb lattice}
We derive $G_m$ on honeycomb lattice, where $Z=3$.
The honeycomb lattice contain two particles per unit cell, as shown in Fig.~\ref{honeycomb}, so the displacement is described by a four-dimensional vector,
\be
{\bf u}_l = (u_{l,A,x},u_{l,A,y},u_{l,B,x},u_{l,B,y}),
\ee
for each unit cell $l$ at ${\bf R}_l = {\bf R}_{(l_1,l_2)}=l_1{\bf a}_1+l_2{\bf a}_2$, where ${\bf a}$ are primitive translation vectors. Setting the edge length of hexagons to unity, ${\bf a}_1 = \sqrt{3}(\frac{1}{2},\frac{\sqrt{3}}{2})$, ${\bf a}_2=\sqrt{3}(-\frac{1}{2},\frac{\sqrt{3}}{2})$.

In Fourier space,
\be\ba{c}
{\bf u}_{\bf q} = \sum_l{\bf u}_l e^{-i{\bf q}\cdot{\bf R}_l},\\
{\bf u}_{l} = \frac{1}{N}\sum_{\bf q}{\bf u}_{\bf q} e^{i{\bf q}\cdot{\bf R}_l},
\ea\ee
the elastic energy change becomes,
\be
\Delta U = \frac{1}{2N^2}\sum_{{\bf q},{\bf q}'}{\bf u}_{\bf q}\cdot\overline\mm_{-{\bf q},{\bf q}'}\cdot{\bf u}_{{\bf q}'},
\ee
where $N$ is the number of unit cells. For homogeneous honeycomb lattice, 
\be
\overline\mm_{{\bf q},{\bf q}'} = N\delta_{{\bf q},{\bf q}'}\mm_{\bf q}.
\ee

We first need to compute the band structure for the first term in the Hessian of the cell perimeter constraints when $k_{\rm eff}=0$. 
Each unit cell contains a hexagonal cell on lattice. The six edges of a cell, labeled in Fig.~\ref{honeycomb}, are
\be\ba{c}
{\bf B}_{1,{\bf q}} = (0,1,0,-1), \\ 
{\bf B}_{2,{\bf q}} = (-\frac{\sqrt{3}}{2}e^{-i(\frac{\sqrt{3}}{2}q_x+\frac{3}{2}q_y)},-\frac{1}{2}e^{-i(\frac{\sqrt{3}}{2}q_x+\frac{3}{2}q_y)},\frac{\sqrt{3}}{2},\frac{1}{2}), \\ 
{\bf B}_{3,{\bf q}} = (\frac{\sqrt{3}}{2}e^{-i(\frac{\sqrt{3}}{2}q_x+\frac{3}{2}q_y)},-\frac{1}{2}e^{-i(\frac{\sqrt{3}}{2}q_x+\frac{3}{2}q_y)},-\frac{\sqrt{3}}{2}e^{-i\sqrt{3}q_x},\frac{1}{2}e^{-i\sqrt{3}q_x}),\\
{\bf B}_{4,{\bf q}} = (0,e^{-i\sqrt{3}q_x},0,-e^{-i\sqrt{3}q_x}),\\
{\bf B}_{5,{\bf q}} = (-\frac{\sqrt{3}}{2}e^{-i\sqrt{3}q_x},-\frac{1}{2}e^{-i\sqrt{3}q_x},\frac{\sqrt{3}}{2}e^{-i(\frac{\sqrt{3}}{2}q_x-\frac{3}{2}q_y)},\frac{1}{2}e^{-i(\frac{\sqrt{3}}{2}q_x-\frac{3}{2}q_y)}), \\
{\bf B}_{6,{\bf q}} = (\frac{\sqrt{3}}{2},-\frac{1}{2},-\frac{\sqrt{3}}{2}e^{-i(\frac{\sqrt{3}}{2}q_x-\frac{3}{2}q_y)},\frac{1}{2}e^{-i(\frac{\sqrt{3}}{2}q_x-\frac{3}{2}q_y)}).\\
\ea\ee
The corresponding Hessian is,
\be
\mm_{\bf q}^{\rm topo}=\sum_{m,l=1}^6{\bf B}_{m,{\bf q}}\otimes{\bf B}_{l,-{\bf q}}
=\lp\ba{cc}
\mm_{1}^{\rm topo} & \mm_{12}^{\rm topo}\\
\mm_{21}^{\rm topo} & \mm_{2}^{\rm topo}
\ea\rp,
\ee
where
\begin{widetext}
\be
\mm_{1}^{\rm topo}= \mm_2^{{\rm topo}*}=\lp\ba{cc}
\frac{3}{2}(1-\cos\sqrt{3}q_x) & \sqrt{3}i(\frac{1}{2}\sin\sqrt{3}q_x-\sin\frac{\sqrt{3}}{2}q_x e^{i\frac{3}{2}q_y})\\
-\sqrt{3}i(\frac{1}{2}\sin\sqrt{3}q_x-\sin\frac{\sqrt{3}}{2}q_x e^{-i\frac{3}{2}q_y}) & \frac{3}{2}+\frac{1}{2}\cos\sqrt{3}q_x-2\cos\frac{\sqrt{3}}{2}q_x\cos\frac{3}{2}q_y
\ea\rp,
\ee
\be
\mm_{12}^{\rm topo}= \mm_{21}^{{\rm topo}*}=\lp\ba{cc}
\frac{3}{2}(1-\cos\sqrt{3}q_x) & -\sqrt{3}i(\frac{1}{2}\sin\sqrt{3}q_x-\sin\frac{\sqrt{3}}{2}q_xe^{-i\frac{3}{2}q_y})\\
-\sqrt{3}i(\frac{1}{2}\sin\sqrt{3}q_x-\sin\frac{\sqrt{3}}{2}q_xe^{-i\frac{3}{2}q_y}) & -\cos^2\frac{\sqrt{3}}{2}q_x+2\cos\frac{\sqrt{3}}{2}q_xe^{-i\frac{3}{2}q_y}-e^{-i3y}
\ea\rp.
\ee
\end{widetext}
There are three zero bands and one acoustic branch for each unit cell. Besides a normalization prefactor, they are
\be
\begin{split}
\tilde\psi_{0,1}&=\lp\ba{c}{\bf g}\\e^{-i{\bf q}\cdot\hat{\bf e}_1}c({\bf q}){\bf g}^*\ea\rp,\quad  \\
\tilde\psi_{0,2}&=\lp\ba{c}{\bf g}\\-e^{-i{\bf q}\cdot\hat{\bf e}_1}c({\bf q}){\bf g}^*\ea\rp,\quad \\
\tilde\psi_{0,3}&=\lp\ba{c}{\bf f}\\e^{-i{\bf q}\cdot\hat{\bf e}_1}{\bf f}^*\ea\rp, \\
\tilde\psi_a&=\lp\ba{c}{\bf f}\\-e^{-i{\bf q}\cdot\hat{\bf e}_1}{\bf f}^*\ea\rp,
\end{split}
\ee
where ${\bf f}=\sum_{j=1}^3\hat{\bf e}_je^{i{\bf q}\cdot\hat{\bf e}_j}$, ${\bf g}=\sum_{j=1}^3\hat{\bf e}_j^\perp e^{-i{\bf q}\cdot\hat{\bf e}_j}$, and $c({\bf q})=-\frac{1}{\sqrt{3+2\eta({\bf q})}}\sum_{j=1}^3e^{-i{\bf q}\cdot\hat{\bf e}_j}$ with $\eta({\bf q})=\frac{1}{2}\sum_{j\neq k}e^{i{\bf q}\cdot(\hat{\bf e}_j-\hat{\bf e}_k)}$. The acoustic branch corresponds to dilation modes of cells, obeying the dispersion relation $\lambda_a^{\rm topo}=2(3-\eta({\bf q}))$.  

Include the internal tension $k_{\rm eff}>0$ contribution to the Hessian matrix,
\be
k_{\rm eff}\mm^{ss}_{\bf q}=k_{\rm eff}\sum_{j=1}^3{\bf D}_{j,{\bf q}}\otimes {\bf D}_{j,-{\bf q}}=k_{\rm eff}\lp\ba{cc}
\mm_1^{ss} & \mm_{12}^{ss}\\
\mm_{21}^{ss} & \mm_2^{ss}
\ea\rp,
\ee
where vectors ${\bf D}$ correspond to the perpendicular directions, 
\be
\begin{split}
{\bf D}_{1,{\bf q}} = &(-1,0,1,0);\\
{\bf D}_{2,{\bf q}} = & \\
& (\frac{1}{2}e^{-i(\frac{\sqrt{3}}{2}q_x+\frac{3}{2}q_y)},-\frac{\sqrt{3}}{2}e^{-i(\frac{\sqrt{3}}{2}q_x+\frac{3}{2}q_y)},-\frac{1}{2},\frac{\sqrt{3}}{2});\\
{\bf D}_{3,{\bf q}}= & \\
& (\frac{1}{2}e^{-i(-\frac{\sqrt{3}}{2}q_x+\frac{3}{2}q_y)},\frac{\sqrt{3}}{2}e^{-i(-\frac{\sqrt{3}}{2}q_x+\frac{3}{2}q_y)},-\frac{1}{2},-\frac{\sqrt{3}}{2}).
\end{split}
\ee
\be
\mm_{1}^{ss}= \mm_2^{ss}=\frac{3}{2}\mi,
\ee
\be
\begin{split}
\mm_{12}^{ss}= & \mm_{21}^{ss*} \\
= &\lp\ba{cc} 
-1-\frac{1}{2}e^{-i\frac{3}{2}q_y}\cos\frac{\sqrt{3}}{2}q_x & -i\frac{\sqrt{3}}{2}e^{-i\frac{3}{2}q_y}\sin\frac{\sqrt{3}}{2}q_x\\
-i\frac{\sqrt{3}}{2}e^{-i\frac{3}{2}q_y}\sin\frac{\sqrt{3}}{2}q_x & -\frac{3}{2}e^{-i\frac{3}{2}q_y}\cos\frac{\sqrt{3}}{2}q_x
\ea\rp.
\end{split}
\ee
Band structure of $\mm^{ss}$ includes one zero band, one acoustic branch and two optical branches, 
\be
\begin{split}
\tilde\phi_0 =& \lp\ba{c}{\bf f}\\-e^{-i{\bf q}\cdot\hat{\bf e}_1}{\bf f}^*\ea\rp,\\
\tilde\phi_{a}=&\lp\ba{c}{\bf g}\\e^{-i{\bf q}\cdot\hat{\bf e}_1}c({\bf q}){\bf g}^*\ea\rp,\\
\tilde\phi_{o,1}=&\lp\ba{c}{\bf g}\\-e^{-i{\bf q}\cdot\hat{\bf e}_1}c({\bf q}){\bf g}^*\ea\rp, \\
\tilde\phi_{o,2}=&\lp\ba{c}{\bf f}\\e^{-i{\bf q}\cdot\hat{\bf e}_1}{\bf f}^*\ea\rp
\end{split}
\ee
corresponding to $\lambda_0^{ss}=0$, $\lambda_a^{ss}=\frac{3}{2}-\frac{\sqrt{3+2\eta({\bf q})}}{2}$, $\lambda_{o,1}^{ss}=\frac{3}{2}+\frac{\sqrt{3+2\eta({\bf q})}}{2}$, and $\lambda_{o,2}^{ss}=3$. 

Notice that the only acoustic branch of $\mm_{\bf q}^{\rm topo}$ is the only zero band of $\mm_{\bf q}^{ss}$. So when there is internal tension $k_{\rm eff}>0$, the full Hessian matrix 
\be
\overline\mm_{\bf q}(k_{\rm eff})=\mm_{\bf q}^{\rm topo}+k_{\rm eff}\mm_{\bf q}^{ss}
\ee
is invertible, 
\be
G_m=\int_{1BZ}\frac{\rd^2{\bf q}}{v_1}{\bf D}_{1,-{\bf q}}\cdot\muu_{\bf q}\muu_{\bf q}^\dagger\left[\overline\mm_{\bf q}(k_{\rm eff})-\omega^2\mi\right]^{-1}\muu_{\bf q}\muu_{\bf q}^\dagger\cdot{\bf D}_{1,{\bf q}},
\ee
where 
\begin{widetext}
\be
\begin{split}
\muu_{\bf q}&=(\phi_0,\phi_a,\phi_{o,1},\phi_{o,2}) \\
&=\frac{1}{2\sqrt{3-\eta({\bf q})}}\lp\ba{cccc}
{\bf f} & {\bf g} & {\bf g} & {\bf f}\\
-e^{-i{\bf q}\cdot\hat{\bf e}_1}{\bf f}^* &  e^{-i{\bf q}\cdot\hat{\bf e}_1}c({\bf q}){\bf g}^* &
-e^{-i{\bf q}\cdot\hat{\bf e}_1}c({\bf q}){\bf g}^* & e^{-i{\bf q}\cdot\hat{\bf e}_1}{\bf f}^*
\ea\rp.
\end{split}
\ee
\end{widetext}
So $G_m|_{\omega=0}=\frac{1}{k_{\rm eff}}$.


\subsection{Phase diagram far from $Z=3$}

\begin{figure}[htpb]
\centering
\includegraphics[width=.47\linewidth]{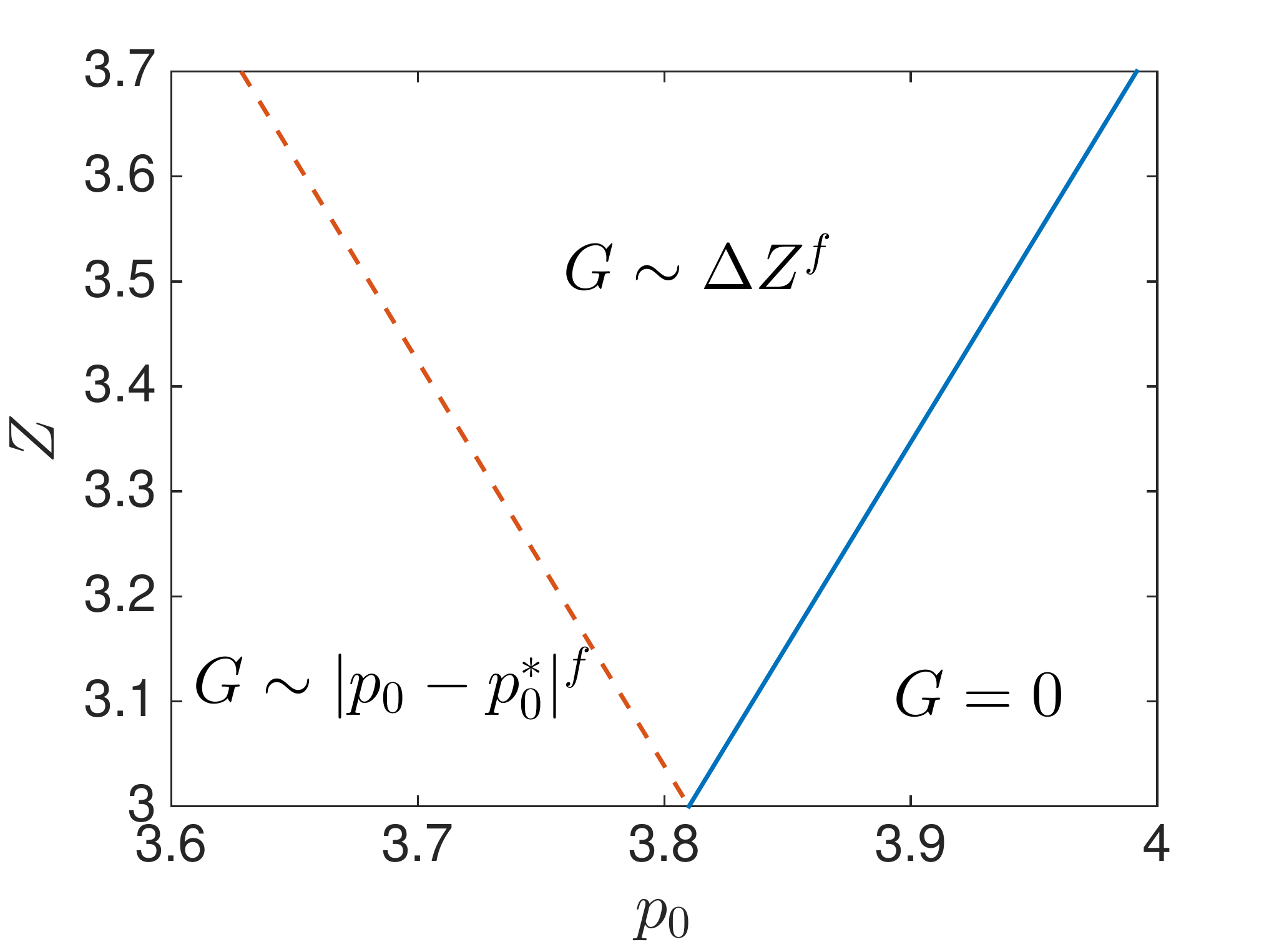}
\includegraphics[width=.47\linewidth]{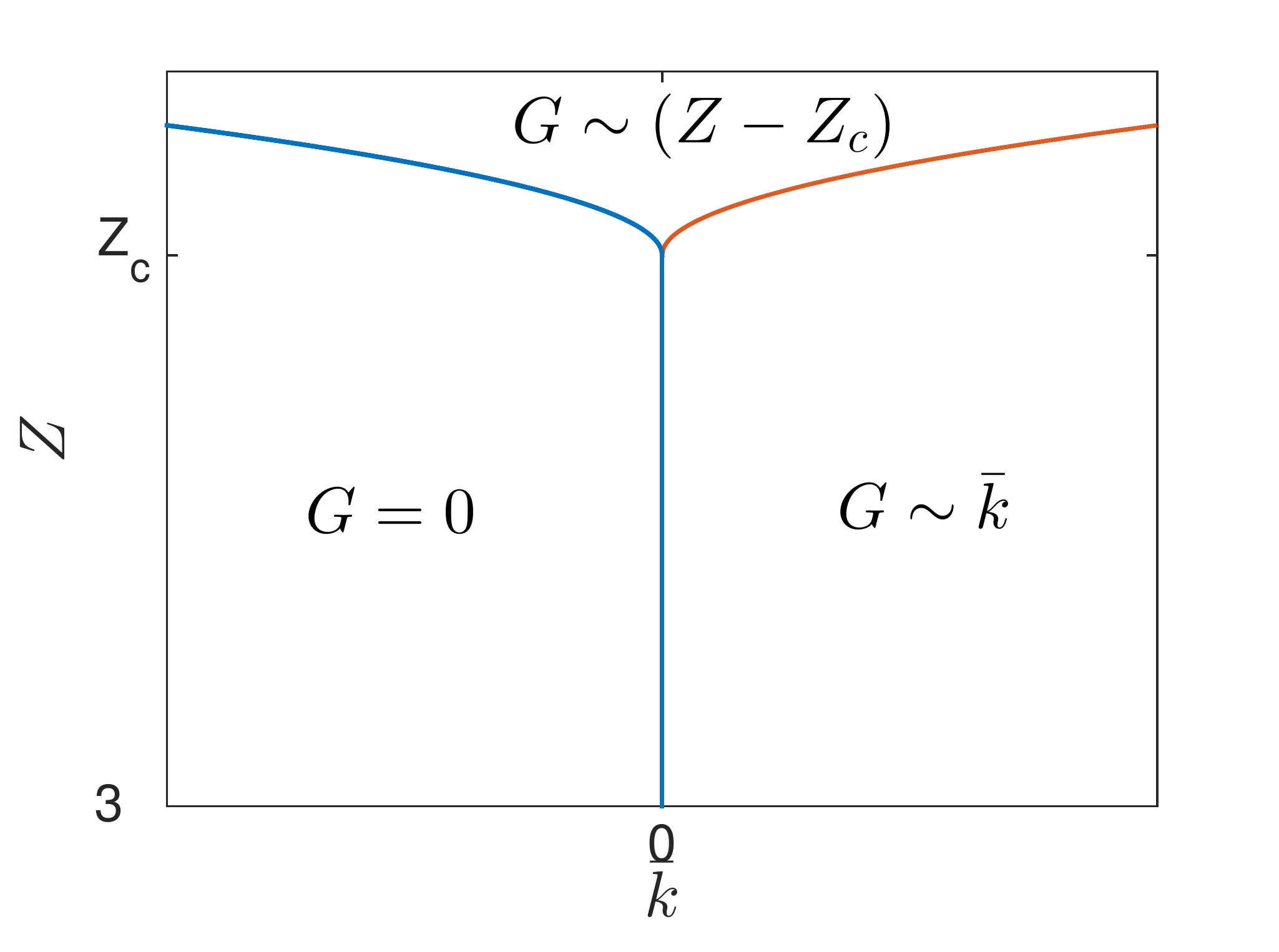}
\caption{\small{Left: Phase diagram $Z$-$p_0$. Right: Phase diagram $Z$-$k$, where $Z_c=4$ for $K_A>0$ and $Z_c=6$ for $K_A=0$. The solid-fluid boundary is shown by the blue line. The red lines separate the regions where mechanical moduli are dominated by different control parameters. }}\label{phase}
\end{figure}

In the main text, we have derived the phase diagram near the critical point $(Z=3, p_0^*)$, illustrated by the left panel in Fig.~\ref{phase}. Though we do see a cross-over from the perimeter dominated region to the topology dominated region, the nature of the rigidity does not change in the vicinity of tri-junction tissues $Z=3$. 
As long as the number of constraints $2F<2V$ and $Z<Z_c=4$ (or when $K_A=0$, $F< 2V$ and $Z<Z_c=6$), internal tension $k>0$ is necessary to stabilize the structure, as illustrated in the phase diagram with internal tension as a control parameter in the right panel of Fig.~\ref{phase}. Above $Z_c$, the topology alone can rigidify the structure, and the solid-liquid transition happens along the line, $Z-Z_c\sim \sqrt{-k}$ as predicted in \cite{DeGiuli14}, shown in  Fig.~\ref{phase}. In this new topology dominated region, the shear modulus crossovers to a linear dependence on the junction connectivity $Z-Z_c$.

\begin{figure*}[htpb]
\begin{center}
\includegraphics[width=2\columnwidth]{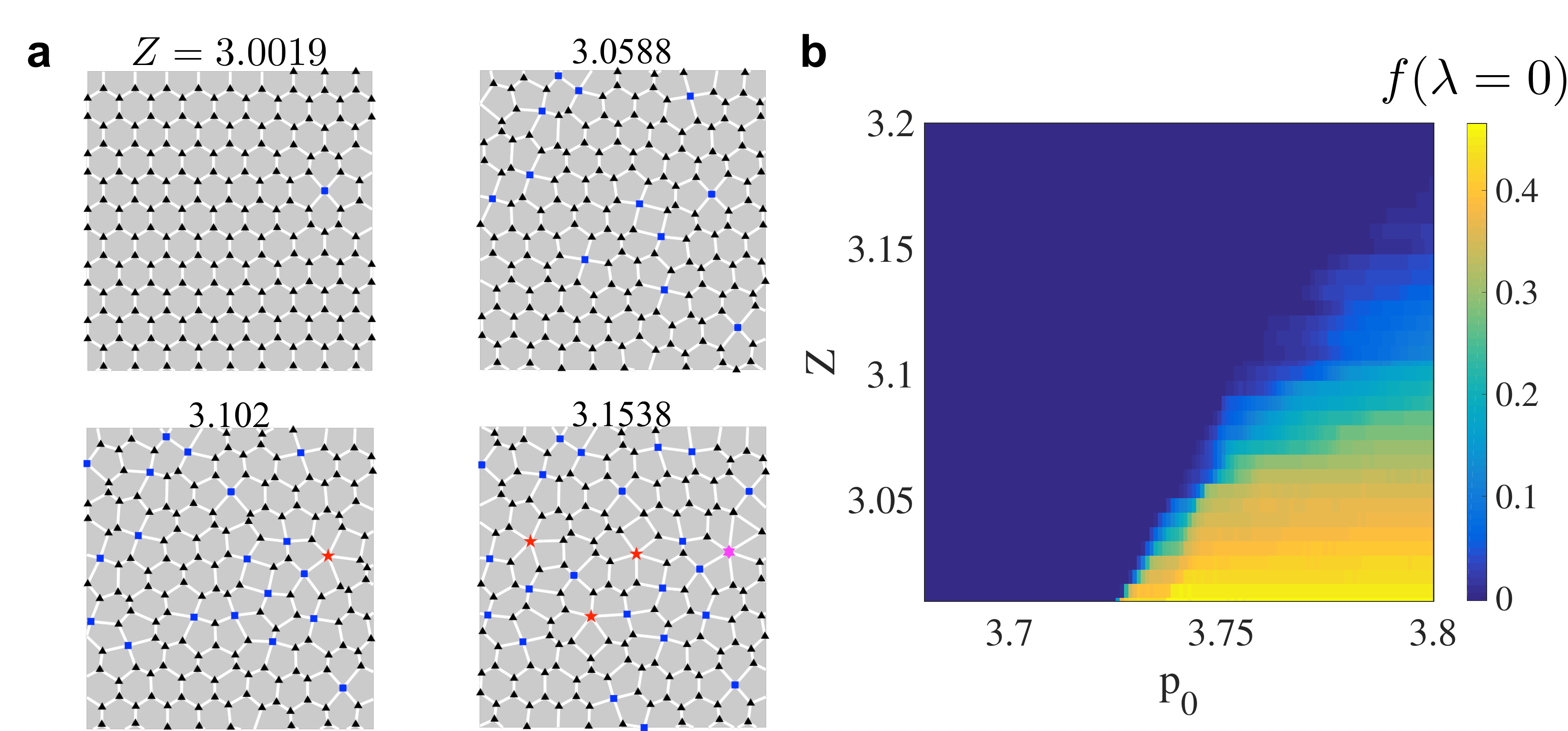}
\caption{
{\bf (a)} 
Application of the edge pinch to a initially ordered hexagonal lattice. 
{\bf  (b)} 
The corresponding $Z-p_0$ phase diagram. Color bar gives the fraction of floppy modes in the tissue. 
}
\label{hex_si}
\end{center}
\end{figure*}

\begin{figure}[htpb]
\begin{center}
\includegraphics[width=1\columnwidth]{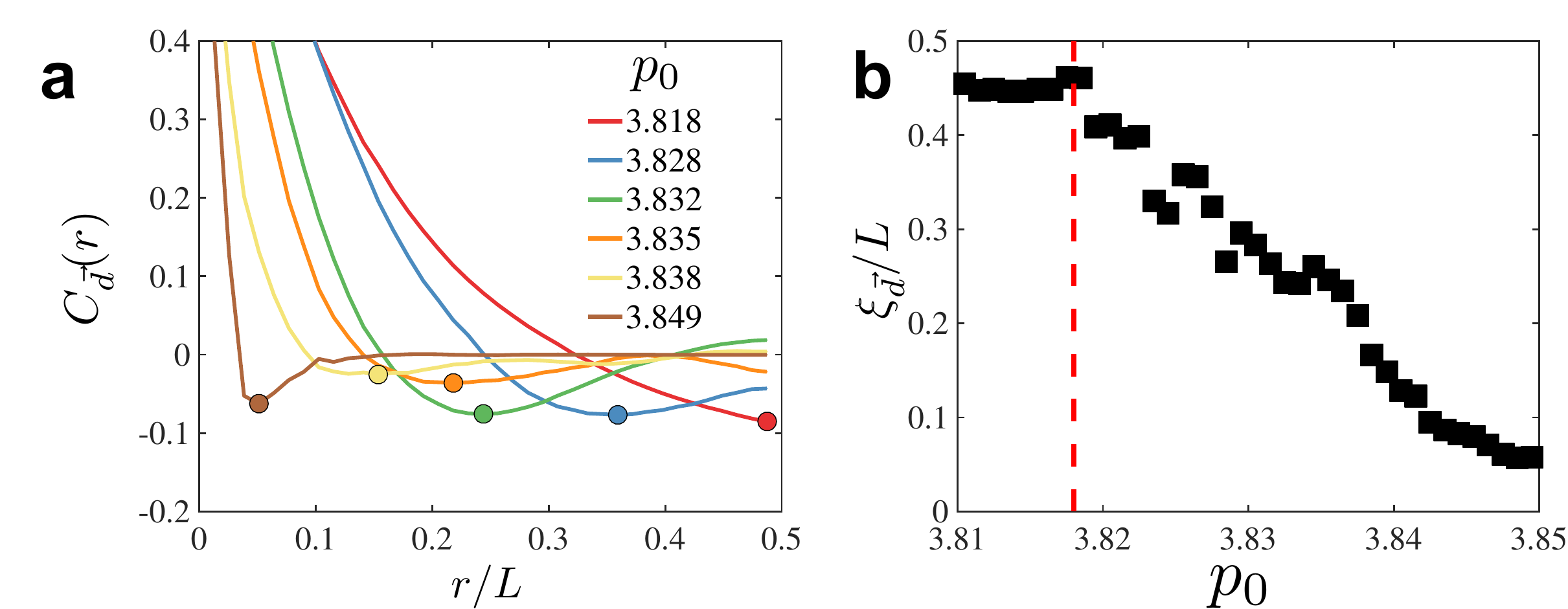}
\caption{
{\bf  (a)}
Spatial autocorrelation of the vertex displacement field after a single edge is pinched at $Z=3$. The dot on each curve indicates the location of the negative dip in $C_{\vec{d}}(r)$ which defines the swirl size $\xi_d$. 
{\bf   (b)}
The swirl size $\xi_d$ associated with a single edge pinch at $Z=3$. In the rigid tissue, the swirl size spans the  on the order of the tissue size but decreases in the fluid tissue. This suggests that near the  fluid solid transition, a single pinched edge will have long range effects where as deeper in the fluid state the effect of pinching is much more local. 
}
\label{fig:swirls}
\end{center}
\end{figure}

\begin{figure}[htpb]
\begin{center}
\includegraphics[width=0.8\columnwidth]{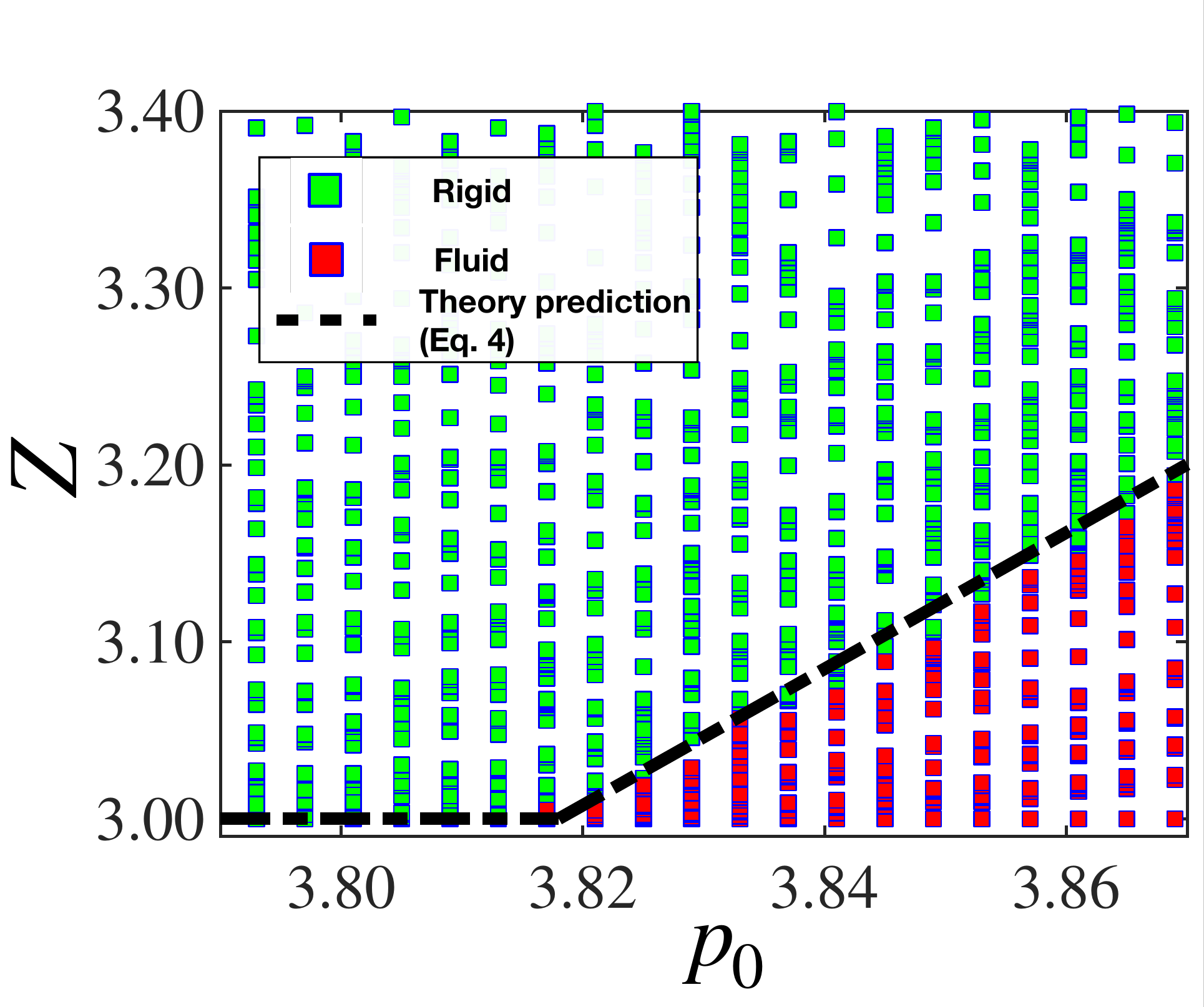}
\caption{
{ We also performed an alternative set of simulations in which T1 and T2 (cell apoptosis) transitions  are allowed. Specifically, T1 transitions are attempted on an edge when its length is shorter than 16\% the average edge lengths in the system while T2 transition are carried out on cells with area less than 10\% of the average cell area. The phase diagram for tissue mechanical states remained the same with this protocol. 
}
}
\label{fig:T1_T2}
\end{center}
\end{figure}

\begin{figure}[htpb]
\begin{center}
\includegraphics[width=1\columnwidth]{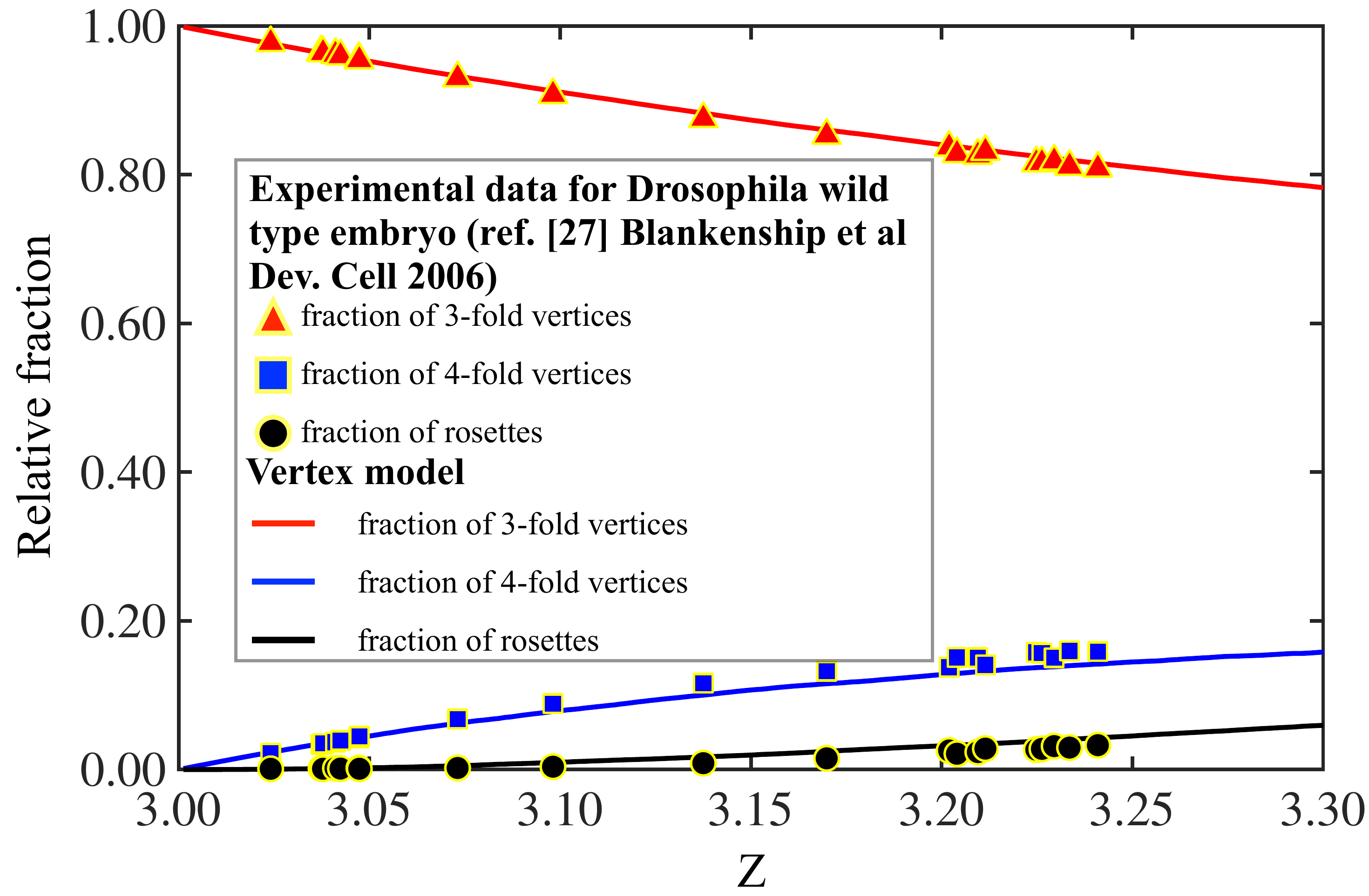}
\caption{
{ 
A direct comparison between the topologies generated in \vm and experimental observations in wild type Drosophila embryo~\cite{Blankenship2006}. The relative fractions for (1) three-fold vertices, (2) four-fold vertices and (3) rosettes are plotted as function of $Z$. }
}
\label{fig:blankenship_compare}
\end{center}
\end{figure}

\clearpage

\begin{thebibliography}{0}%
\makeatletter
\providecommand \@ifxundefined [1]{%
 \@ifx{#1\undefined}
}%
\providecommand \@ifnum [1]{%
 \ifnum #1\expandafter \@firstoftwo
 \else \expandafter \@secondoftwo
 \fi
}%
\providecommand \@ifx [1]{%
 \ifx #1\expandafter \@firstoftwo
 \else \expandafter \@secondoftwo
 \fi
}%
\providecommand \natexlab [1]{#1}%
\providecommand \enquote  [1]{``#1''}%
\providecommand \bibnamefont  [1]{#1}%
\providecommand \bibfnamefont [1]{#1}%
\providecommand \citenamefont [1]{#1}%
\providecommand \href@noop [0]{\@secondoftwo}%
\providecommand \href [0]{\begingroup \@sanitize@url \@href}%
\providecommand \@href[1]{\@@startlink{#1}\@@href}%
\providecommand \@@href[1]{\endgroup#1\@@endlink}%
\providecommand \@sanitize@url [0]{\catcode `\\12\catcode `\$12\catcode
  `\&12\catcode `\#12\catcode `\^12\catcode `\_12\catcode `\%12\relax}%
\providecommand \@@startlink[1]{}%
\providecommand \@@endlink[0]{}%
\providecommand \url  [0]{\begingroup\@sanitize@url \@url }%
\providecommand \@url [1]{\endgroup\@href {#1}{\urlprefix }}%
\providecommand \urlprefix  [0]{URL }%
\providecommand \Eprint [0]{\href }%
\providecommand \doibase [0]{http://dx.doi.org/}%
\providecommand \selectlanguage [0]{\@gobble}%
\providecommand \bibinfo  [0]{\@secondoftwo}%
\providecommand \bibfield  [0]{\@secondoftwo}%
\providecommand \translation [1]{[#1]}%
\providecommand \BibitemOpen [0]{}%
\providecommand \bibitemStop [0]{}%
\providecommand \bibitemNoStop [0]{.\EOS\space}%
\providecommand \EOS [0]{\spacefactor3000\relax}%
\providecommand \BibitemShut  [1]{\csname bibitem#1\endcsname}%
\let\auto@bib@innerbib\@empty
\end{thebibliography}%


\begin{thebibliography}{89}%
\makeatletter
\providecommand \@ifxundefined [1]{%
 \@ifx{#1\undefined}
}%
\providecommand \@ifnum [1]{%
 \ifnum #1\expandafter \@firstoftwo
 \else \expandafter \@secondoftwo
 \fi
}%
\providecommand \@ifx [1]{%
 \ifx #1\expandafter \@firstoftwo
 \else \expandafter \@secondoftwo
 \fi
}%
\providecommand \natexlab [1]{#1}%
\providecommand \enquote  [1]{``#1''}%
\providecommand \bibnamefont  [1]{#1}%
\providecommand \bibfnamefont [1]{#1}%
\providecommand \citenamefont [1]{#1}%
\providecommand \href@noop [0]{\@secondoftwo}%
\providecommand \href [0]{\begingroup \@sanitize@url \@href}%
\providecommand \@href[1]{\@@startlink{#1}\@@href}%
\providecommand \@@href[1]{\endgroup#1\@@endlink}%
\providecommand \@sanitize@url [0]{\catcode `\\12\catcode `\$12\catcode
  `\&12\catcode `\#12\catcode `\^12\catcode `\_12\catcode `\%12\relax}%
\providecommand \@@startlink[1]{}%
\providecommand \@@endlink[0]{}%
\providecommand \url  [0]{\begingroup\@sanitize@url \@url }%
\providecommand \@url [1]{\endgroup\@href {#1}{\urlprefix }}%
\providecommand \urlprefix  [0]{URL }%
\providecommand \Eprint [0]{\href }%
\providecommand \doibase [0]{http://dx.doi.org/}%
\providecommand \selectlanguage [0]{\@gobble}%
\providecommand \bibinfo  [0]{\@secondoftwo}%
\providecommand \bibfield  [0]{\@secondoftwo}%
\providecommand \translation [1]{[#1]}%
\providecommand \BibitemOpen [0]{}%
\providecommand \bibitemStop [0]{}%
\providecommand \bibitemNoStop [0]{.\EOS\space}%
\providecommand \EOS [0]{\spacefactor3000\relax}%
\providecommand \BibitemShut  [1]{\csname bibitem#1\endcsname}%
\let\auto@bib@innerbib\@empty
\bibitem [{\citenamefont {Nagai}\ and\ \citenamefont
  {Honda}(2001)}]{ngai_honda}%
  \BibitemOpen
  \bibfield  {author} {\bibinfo {author} {\bibfnamefont {Tatsuzo}\ \bibnamefont
  {Nagai}}\ and\ \bibinfo {author} {\bibfnamefont {Hisao}\ \bibnamefont
  {Honda}},\ }\bibfield  {title} {\enquote {\bibinfo {title} {A dynamic cell
  model for the formation of epithelial tissues},}\ }\href {\doibase
  10.1080/13642810108205772} {\bibfield  {journal} {\bibinfo  {journal}
  {Philosophical Magazine Part B}\ }\textbf {\bibinfo {volume} {81}},\ \bibinfo
  {pages} {699--719} (\bibinfo {year} {2001})}\BibitemShut {NoStop}%
\bibitem [{\citenamefont {Farhadifar}\ \emph {et~al.}(2007)\citenamefont
  {Farhadifar}, \citenamefont {Roeper}, \citenamefont {Aigouy}, \citenamefont
  {Eaton},\ and\ \citenamefont {Julicher}}]{Farhadifar2007}%
  \BibitemOpen
  \bibfield  {author} {\bibinfo {author} {\bibfnamefont {Reza}\ \bibnamefont
  {Farhadifar}}, \bibinfo {author} {\bibfnamefont {Jens-Christian}\
  \bibnamefont {Roeper}}, \bibinfo {author} {\bibfnamefont {Benoit}\
  \bibnamefont {Aigouy}}, \bibinfo {author} {\bibfnamefont {Suzanne}\
  \bibnamefont {Eaton}}, \ and\ \bibinfo {author} {\bibfnamefont {Frank}\
  \bibnamefont {Julicher}},\ }\bibfield  {title} {\enquote {\bibinfo {title}
  {The influence of cell mechanics, cell-cell interactions, and proliferation
  on epithelial packing},}\ }\href
  {http://www.sciencedirect.com/science/article/pii/S0960982207023342}
  {\bibfield  {journal} {\bibinfo  {journal} {Current Biology}\ }\textbf
  {\bibinfo {volume} {17}},\ \bibinfo {pages} {2095 -- 2104} (\bibinfo {year}
  {2007})}\BibitemShut {NoStop}%
\bibitem [{\citenamefont {Hufnagel}\ \emph {et~al.}(2007)\citenamefont
  {Hufnagel}, \citenamefont {Teleman}, \citenamefont {Rouault}, \citenamefont
  {Cohen},\ and\ \citenamefont {Shraiman}}]{Hufnagel2007}%
  \BibitemOpen
  \bibfield  {author} {\bibinfo {author} {\bibfnamefont {Lars}\ \bibnamefont
  {Hufnagel}}, \bibinfo {author} {\bibfnamefont {Aurelio~A.}\ \bibnamefont
  {Teleman}}, \bibinfo {author} {\bibfnamefont {H.}~\bibnamefont {Rouault}},
  \bibinfo {author} {\bibfnamefont {Stephen~M.}\ \bibnamefont {Cohen}}, \ and\
  \bibinfo {author} {\bibfnamefont {Boris~I.}\ \bibnamefont {Shraiman}},\
  }\bibfield  {title} {\enquote {\bibinfo {title} {On the mechanism of wing
  size determination in fly development},}\ }\href {10.1073/pnas.0607134104}
  {\bibfield  {journal} {\bibinfo  {journal} {Proceedings of the National
  Academy of Sciences}\ }\textbf {\bibinfo {volume} {104}},\ \bibinfo {pages}
  {3835--3840} (\bibinfo {year} {2007})}\BibitemShut {NoStop}%
\bibitem [{\citenamefont {Staple}\ \emph {et~al.}(2010)\citenamefont {Staple},
  \citenamefont {Farhadifar}, \citenamefont {Roeper}, \citenamefont {Aigouy},
  \citenamefont {Eaton},\ and\ \citenamefont {Julicher}}]{Staple2010}%
  \BibitemOpen
  \bibfield  {author} {\bibinfo {author} {\bibfnamefont {D.~B}\ \bibnamefont
  {Staple}}, \bibinfo {author} {\bibfnamefont {R}~\bibnamefont {Farhadifar}},
  \bibinfo {author} {\bibfnamefont {J.~C}\ \bibnamefont {Roeper}}, \bibinfo
  {author} {\bibfnamefont {B}~\bibnamefont {Aigouy}}, \bibinfo {author}
  {\bibfnamefont {S}~\bibnamefont {Eaton}}, \ and\ \bibinfo {author}
  {\bibfnamefont {F}~\bibnamefont {Julicher}},\ }\bibfield  {title} {\enquote
  {\bibinfo {title} {Mechanics and remodelling of cell packings in
  epithelia},}\ }\href {http://www.ncbi.nlm.nih.gov/pubmed/21082210} {\bibfield
   {journal} {\bibinfo  {journal} {Eur. Phys. J. E}\ }\textbf {\bibinfo
  {volume} {33}},\ \bibinfo {pages} {117--127} (\bibinfo {year}
  {2010})}\BibitemShut {NoStop}%
\bibitem [{\citenamefont {Manning}\ \emph {et~al.}(2010)\citenamefont
  {Manning}, \citenamefont {Foty}, \citenamefont {Steinberg},\ and\
  \citenamefont {Schoetz}}]{manning_2010}%
  \BibitemOpen
  \bibfield  {author} {\bibinfo {author} {\bibfnamefont {M.~Lisa}\ \bibnamefont
  {Manning}}, \bibinfo {author} {\bibfnamefont {Ramsey~A.}\ \bibnamefont
  {Foty}}, \bibinfo {author} {\bibfnamefont {Malcolm~S.}\ \bibnamefont
  {Steinberg}}, \ and\ \bibinfo {author} {\bibfnamefont {Eva-Maria}\
  \bibnamefont {Schoetz}},\ }\bibfield  {title} {\enquote {\bibinfo {title}
  {Coaction of intercellular adhesion and cortical tension specifies tissue
  surface tension},}\ }\href@noop {} {\bibfield  {journal} {\bibinfo  {journal}
  {Proceedings of the National Academy of Sciences}\ }\textbf {\bibinfo
  {volume} {107}},\ \bibinfo {pages} {12517--12522} (\bibinfo {year}
  {2010})}\BibitemShut {NoStop}%
\bibitem [{\citenamefont {Fletcher}\ \emph {et~al.}(2014)\citenamefont
  {Fletcher}, \citenamefont {Osterfield}, \citenamefont {Baker},\ and\
  \citenamefont {Shvartsman}}]{Fletcher2014}%
  \BibitemOpen
  \bibfield  {author} {\bibinfo {author} {\bibfnamefont {Alexander~G}\
  \bibnamefont {Fletcher}}, \bibinfo {author} {\bibfnamefont {Miriam}\
  \bibnamefont {Osterfield}}, \bibinfo {author} {\bibfnamefont {Ruth~E}\
  \bibnamefont {Baker}}, \ and\ \bibinfo {author} {\bibfnamefont {Stanislav~Y}\
  \bibnamefont {Shvartsman}},\ }\bibfield  {title} {\enquote {\bibinfo {title}
  {{Vertex models of epithelial morphogenesis.}}}\ }\href {\doibase
  10.1016/j.bpj.2013.11.4498} {\bibfield  {journal} {\bibinfo  {journal}
  {Biophysical journal}\ }\textbf {\bibinfo {volume} {106}},\ \bibinfo {pages}
  {2291--304} (\bibinfo {year} {2014})}\BibitemShut {NoStop}%
\bibitem [{\citenamefont {Bi}\ \emph {et~al.}(2014)\citenamefont {Bi},
  \citenamefont {Lopez}, \citenamefont {Schwarz},\ and\ \citenamefont
  {Manning}}]{bi_softmatter}%
  \BibitemOpen
  \bibfield  {author} {\bibinfo {author} {\bibfnamefont {Dapeng}\ \bibnamefont
  {Bi}}, \bibinfo {author} {\bibfnamefont {Jorge~H.}\ \bibnamefont {Lopez}},
  \bibinfo {author} {\bibfnamefont {J.~M.}\ \bibnamefont {Schwarz}}, \ and\
  \bibinfo {author} {\bibfnamefont {M.~Lisa}\ \bibnamefont {Manning}},\
  }\bibfield  {title} {\enquote {\bibinfo {title} {Energy barriers and cell
  migration in densely packed tissues},}\ }\href {\doibase 10.1039/C3SM52893F}
  {\bibfield  {journal} {\bibinfo  {journal} {Soft Matter}\ }\textbf {\bibinfo
  {volume} {10}},\ \bibinfo {pages} {1885--1890} (\bibinfo {year}
  {2014})}\BibitemShut {NoStop}%
\bibitem [{\citenamefont {Bi}\ \emph {et~al.}(2015)\citenamefont {Bi},
  \citenamefont {Lopez}, \citenamefont {Schwarz},\ and\ \citenamefont
  {Manning}}]{bi_nphys_2015}%
  \BibitemOpen
  \bibfield  {author} {\bibinfo {author} {\bibfnamefont {Dapeng}\ \bibnamefont
  {Bi}}, \bibinfo {author} {\bibfnamefont {J.~H.}\ \bibnamefont {Lopez}},
  \bibinfo {author} {\bibfnamefont {J.~M.}\ \bibnamefont {Schwarz}}, \ and\
  \bibinfo {author} {\bibfnamefont {M.~Lisa}\ \bibnamefont {Manning}},\
  }\bibfield  {title} {\enquote {\bibinfo {title} {{A density-independent
  rigidity transition in biological tissues}},}\ }\href {\doibase
  10.1038/nphys3471} {\bibfield  {journal} {\bibinfo  {journal} {Nature
  Physics}\ }\textbf {\bibinfo {volume} {11}},\ \bibinfo {pages} {1074--1079}
  (\bibinfo {year} {2015})}\BibitemShut {NoStop}%
\bibitem [{\citenamefont {Bi}\ \emph {et~al.}(2016)\citenamefont {Bi},
  \citenamefont {Yang}, \citenamefont {Marchetti},\ and\ \citenamefont
  {Manning}}]{Bi2016}%
  \BibitemOpen
  \bibfield  {author} {\bibinfo {author} {\bibfnamefont {Dapeng}\ \bibnamefont
  {Bi}}, \bibinfo {author} {\bibfnamefont {Xingbo}\ \bibnamefont {Yang}},
  \bibinfo {author} {\bibfnamefont {M.~Cristina}\ \bibnamefont {Marchetti}}, \
  and\ \bibinfo {author} {\bibfnamefont {M.~Lisa}\ \bibnamefont {Manning}},\
  }\bibfield  {title} {\enquote {\bibinfo {title} {{Motility-driven glass and
  jamming transitions in biological tissues}},}\ }\href {\doibase
  10.1103/PhysRevX.6.021011} {\bibfield  {journal} {\bibinfo  {journal}
  {Physical Review X}\ }\textbf {\bibinfo {volume} {6}},\ \bibinfo {pages}
  {1--13} (\bibinfo {year} {2016})}\BibitemShut {NoStop}%
\bibitem [{\citenamefont {Yang}\ \emph {et~al.}(2017)\citenamefont {Yang},
  \citenamefont {Bi}, \citenamefont {Czajkowski}, \citenamefont {Merkel},
  \citenamefont {Manning},\ and\ \citenamefont {Marchetti}}]{Yang2017}%
  \BibitemOpen
  \bibfield  {author} {\bibinfo {author} {\bibfnamefont {Xingbo}\ \bibnamefont
  {Yang}}, \bibinfo {author} {\bibfnamefont {Dapeng}\ \bibnamefont {Bi}},
  \bibinfo {author} {\bibfnamefont {Michael}\ \bibnamefont {Czajkowski}},
  \bibinfo {author} {\bibfnamefont {Matthias}\ \bibnamefont {Merkel}}, \bibinfo
  {author} {\bibfnamefont {M.~Lisa}\ \bibnamefont {Manning}}, \ and\ \bibinfo
  {author} {\bibfnamefont {M.~Cristina}\ \bibnamefont {Marchetti}},\ }\bibfield
   {title} {\enquote {\bibinfo {title} {Correlating cell shape and cellular
  stress in motile confluent tissues},}\ }\href {\doibase
  10.1073/pnas.1705921114} {\bibfield  {journal} {\bibinfo  {journal}
  {Proceedings of the National Academy of Sciences}\ }\textbf {\bibinfo
  {volume} {114}},\ \bibinfo {pages} {12663--12668} (\bibinfo {year}
  {2017})}\BibitemShut {NoStop}%
\bibitem [{\citenamefont {Merkel}\ and\ \citenamefont
  {Manning}(2018)}]{Merkel2017}%
  \BibitemOpen
  \bibfield  {author} {\bibinfo {author} {\bibfnamefont {Matthias}\
  \bibnamefont {Merkel}}\ and\ \bibinfo {author} {\bibfnamefont {M~Lisa}\
  \bibnamefont {Manning}},\ }\bibfield  {title} {\enquote {\bibinfo {title} {A
  geometrically controlled rigidity transition in a model for confluent 3d
  tissues},}\ }\href {http://stacks.iop.org/1367-2630/20/i=2/a=022002}
  {\bibfield  {journal} {\bibinfo  {journal} {New Journal of Physics}\ }\textbf
  {\bibinfo {volume} {20}},\ \bibinfo {pages} {022002} (\bibinfo {year}
  {2018})}\BibitemShut {NoStop}%
\bibitem [{\citenamefont {Noll}\ \emph {et~al.}(2017)\citenamefont {Noll},
  \citenamefont {Mani}, \citenamefont {Heemskerk}, \citenamefont {Streichan},\
  and\ \citenamefont {Shraiman}}]{Noll2017}%
  \BibitemOpen
  \bibfield  {author} {\bibinfo {author} {\bibfnamefont {Nicholas}\
  \bibnamefont {Noll}}, \bibinfo {author} {\bibfnamefont {Madhav}\ \bibnamefont
  {Mani}}, \bibinfo {author} {\bibfnamefont {Idse}\ \bibnamefont {Heemskerk}},
  \bibinfo {author} {\bibfnamefont {Sebastian~J}\ \bibnamefont {Streichan}}, \
  and\ \bibinfo {author} {\bibfnamefont {Boris~I}\ \bibnamefont {Shraiman}},\
  }\bibfield  {title} {\enquote {\bibinfo {title} {{Active tension network
  model suggests an exotic mechanical state realized in epithelial tissues}},}\
  }\href {\doibase 10.1038/nphys4219} {\bibfield  {journal} {\bibinfo
  {journal} {Nature Physics}\ } (\bibinfo {year} {2017}),\
  10.1038/nphys4219}\BibitemShut {NoStop}%
\bibitem [{\citenamefont {Boromand}\ \emph {et~al.}(2018)\citenamefont
  {Boromand}, \citenamefont {Signoriello}, \citenamefont {Ye}, \citenamefont
  {O'Hern},\ and\ \citenamefont {Shattuck}}]{boromand_2018}%
  \BibitemOpen
  \bibfield  {author} {\bibinfo {author} {\bibfnamefont {Arman}\ \bibnamefont
  {Boromand}}, \bibinfo {author} {\bibfnamefont {Alexandra}\ \bibnamefont
  {Signoriello}}, \bibinfo {author} {\bibfnamefont {Fangfu}\ \bibnamefont
  {Ye}}, \bibinfo {author} {\bibfnamefont {Corey~S.}\ \bibnamefont {O'Hern}}, \
  and\ \bibinfo {author} {\bibfnamefont {Mark~D.}\ \bibnamefont {Shattuck}},\
  }\bibfield  {title} {\enquote {\bibinfo {title} {Jamming of deformable
  polygons},}\ }\href {\doibase 10.1103/PhysRevLett.121.248003} {\bibfield
  {journal} {\bibinfo  {journal} {Phys. Rev. Lett.}\ }\textbf {\bibinfo
  {volume} {121}},\ \bibinfo {pages} {248003} (\bibinfo {year}
  {2018})}\BibitemShut {NoStop}%
\bibitem [{\citenamefont {Teomy}\ \emph {et~al.}(2018)\citenamefont {Teomy},
  \citenamefont {Kessler},\ and\ \citenamefont {Levine}}]{teomy_2018}%
  \BibitemOpen
  \bibfield  {author} {\bibinfo {author} {\bibfnamefont {Eial}\ \bibnamefont
  {Teomy}}, \bibinfo {author} {\bibfnamefont {David~A.}\ \bibnamefont
  {Kessler}}, \ and\ \bibinfo {author} {\bibfnamefont {Herbert}\ \bibnamefont
  {Levine}},\ }\bibfield  {title} {\enquote {\bibinfo {title} {Confluent and
  nonconfluent phases in a model of cell tissue},}\ }\href {\doibase
  10.1103/PhysRevE.98.042418} {\bibfield  {journal} {\bibinfo  {journal} {Phys.
  Rev. E}\ }\textbf {\bibinfo {volume} {98}},\ \bibinfo {pages} {042418}
  (\bibinfo {year} {2018})}\BibitemShut {NoStop}%
\bibitem [{\citenamefont {Sussman}(2017)}]{cellgpu}%
  \BibitemOpen
  \bibfield  {author} {\bibinfo {author} {\bibfnamefont {Daniel~M.}\
  \bibnamefont {Sussman}},\ }\bibfield  {title} {\enquote {\bibinfo {title}
  {cellgpu: Massively parallel simulations of dynamic vertex models},}\
  }\href@noop {} {\bibfield  {journal} {\bibinfo  {journal} {Computer Physics
  Communications}\ }\textbf {\bibinfo {volume} {219}},\ \bibinfo {pages} {400
  -- 406} (\bibinfo {year} {2017})}\BibitemShut {NoStop}%
\bibitem [{\citenamefont {Spencer}\ \emph {et~al.}(2017)\citenamefont
  {Spencer}, \citenamefont {Jabeen},\ and\ \citenamefont
  {Lubensky}}]{Spencer_lubensky_2017}%
  \BibitemOpen
  \bibfield  {author} {\bibinfo {author} {\bibfnamefont {Meryl~A.}\
  \bibnamefont {Spencer}}, \bibinfo {author} {\bibfnamefont {Zahera}\
  \bibnamefont {Jabeen}}, \ and\ \bibinfo {author} {\bibfnamefont {David~K.}\
  \bibnamefont {Lubensky}},\ }\bibfield  {title} {\enquote {\bibinfo {title}
  {Vertex stability and topological transitions in vertex models of foams and
  epithelia},}\ }\href {\doibase 10.1140/epje/i2017-11489-4} {\bibfield
  {journal} {\bibinfo  {journal} {The European Physical Journal E}\ }\textbf
  {\bibinfo {volume} {40}},\ \bibinfo {pages} {2} (\bibinfo {year}
  {2017})}\BibitemShut {NoStop}%
\bibitem [{\citenamefont {Barton}\ \emph {et~al.}(2017)\citenamefont {Barton},
  \citenamefont {Henkes}, \citenamefont {Weijer},\ and\ \citenamefont
  {Sknepnek}}]{barton_silke_rastko}%
  \BibitemOpen
  \bibfield  {author} {\bibinfo {author} {\bibfnamefont {Daniel~L.}\
  \bibnamefont {Barton}}, \bibinfo {author} {\bibfnamefont {Silke}\
  \bibnamefont {Henkes}}, \bibinfo {author} {\bibfnamefont {Cornelis~J.}\
  \bibnamefont {Weijer}}, \ and\ \bibinfo {author} {\bibfnamefont {Rastko}\
  \bibnamefont {Sknepnek}},\ }\bibfield  {title} {\enquote {\bibinfo {title}
  {Active vertex model for cell-resolution description of epithelial tissue
  mechanics},}\ }\href {\doibase 10.1371/journal.pcbi.1005569} {\bibfield
  {journal} {\bibinfo  {journal} {PLOS Computational Biology}\ }\textbf
  {\bibinfo {volume} {13}},\ \bibinfo {pages} {1--34} (\bibinfo {year}
  {2017})}\BibitemShut {NoStop}%
\bibitem [{\citenamefont {Moshe}\ \emph {et~al.}(2018)\citenamefont {Moshe},
  \citenamefont {Bowick},\ and\ \citenamefont {Marchetti}}]{moshe_2017}%
  \BibitemOpen
  \bibfield  {author} {\bibinfo {author} {\bibfnamefont {Michael}\ \bibnamefont
  {Moshe}}, \bibinfo {author} {\bibfnamefont {Mark~J.}\ \bibnamefont {Bowick}},
  \ and\ \bibinfo {author} {\bibfnamefont {M.~Cristina}\ \bibnamefont
  {Marchetti}},\ }\bibfield  {title} {\enquote {\bibinfo {title} {Geometric
  frustration and solid-solid transitions in model 2d tissue},}\ }\href
  {\doibase 10.1103/PhysRevLett.120.268105} {\bibfield  {journal} {\bibinfo
  {journal} {Phys. Rev. Lett.}\ }\textbf {\bibinfo {volume} {120}},\ \bibinfo
  {pages} {268105} (\bibinfo {year} {2018})}\BibitemShut {NoStop}%
\bibitem [{\citenamefont {Atia}\ \emph {et~al.}(2018)\citenamefont {Atia},
  \citenamefont {Bi}, \citenamefont {Sharma}, \citenamefont {Mitchel},
  \citenamefont {Gweon}, \citenamefont {A.~Koehler}, \citenamefont {DeCamp},
  \citenamefont {Lan}, \citenamefont {Kim}, \citenamefont {Hirsch},
  \citenamefont {Pegoraro}, \citenamefont {Lee}, \citenamefont {Starr},
  \citenamefont {Weitz}, \citenamefont {Martin}, \citenamefont {Park},
  \citenamefont {Butler},\ and\ \citenamefont {Fredberg}}]{atia_nat_phys_2018}%
  \BibitemOpen
  \bibfield  {author} {\bibinfo {author} {\bibfnamefont {Lior}\ \bibnamefont
  {Atia}}, \bibinfo {author} {\bibfnamefont {Dapeng}\ \bibnamefont {Bi}},
  \bibinfo {author} {\bibfnamefont {Yasha}\ \bibnamefont {Sharma}}, \bibinfo
  {author} {\bibfnamefont {Jennifer~A.}\ \bibnamefont {Mitchel}}, \bibinfo
  {author} {\bibfnamefont {Bomi}\ \bibnamefont {Gweon}}, \bibinfo {author}
  {\bibfnamefont {Stephan}\ \bibnamefont {A.~Koehler}}, \bibinfo {author}
  {\bibfnamefont {Stephen~J.}\ \bibnamefont {DeCamp}}, \bibinfo {author}
  {\bibfnamefont {Bo}~\bibnamefont {Lan}}, \bibinfo {author} {\bibfnamefont
  {Jae~Hun}\ \bibnamefont {Kim}}, \bibinfo {author} {\bibfnamefont {Rebecca}\
  \bibnamefont {Hirsch}}, \bibinfo {author} {\bibfnamefont {Adrian~F.}\
  \bibnamefont {Pegoraro}}, \bibinfo {author} {\bibfnamefont {Kyu~Ha}\
  \bibnamefont {Lee}}, \bibinfo {author} {\bibfnamefont {Jacqueline~R.}\
  \bibnamefont {Starr}}, \bibinfo {author} {\bibfnamefont {David~A.}\
  \bibnamefont {Weitz}}, \bibinfo {author} {\bibfnamefont {Adam~C.}\
  \bibnamefont {Martin}}, \bibinfo {author} {\bibfnamefont {Jin-Ah}\
  \bibnamefont {Park}}, \bibinfo {author} {\bibfnamefont {James~P.}\
  \bibnamefont {Butler}}, \ and\ \bibinfo {author} {\bibfnamefont {Jeffrey~J.}\
  \bibnamefont {Fredberg}},\ }\bibfield  {title} {\enquote {\bibinfo {title}
  {Geometric constraints during epithelial jamming},}\ }\href@noop {}
  {\bibfield  {journal} {\bibinfo  {journal} {Nature Physics}\ }\textbf
  {\bibinfo {volume} {14}},\ \bibinfo {pages} {613--620} (\bibinfo {year}
  {2018})}\BibitemShut {NoStop}%
\bibitem [{\citenamefont {Alt}\ \emph {et~al.}(2017)\citenamefont {Alt},
  \citenamefont {Ganguly},\ and\ \citenamefont
  {Salbreux}}]{Alt_VM_review_2017}%
  \BibitemOpen
  \bibfield  {author} {\bibinfo {author} {\bibfnamefont {Silvanus}\
  \bibnamefont {Alt}}, \bibinfo {author} {\bibfnamefont {Poulami}\ \bibnamefont
  {Ganguly}}, \ and\ \bibinfo {author} {\bibfnamefont {Guillaume}\ \bibnamefont
  {Salbreux}},\ }\bibfield  {title} {\enquote {\bibinfo {title} {Vertex models:
  from cell mechanics to tissue morphogenesis},}\ }\href {\doibase
  10.1098/rstb.2015.0520} {\bibfield  {journal} {\bibinfo  {journal}
  {Philosophical Transactions of the Royal Society of London B: Biological
  Sciences}\ }\textbf {\bibinfo {volume} {372}} (\bibinfo {year} {2017}),\
  10.1098/rstb.2015.0520}\BibitemShut {NoStop}%
\bibitem [{\citenamefont {Giavazzi}\ \emph {et~al.}(2018)\citenamefont
  {Giavazzi}, \citenamefont {Paoluzzi}, \citenamefont {Macchi}, \citenamefont
  {Bi}, \citenamefont {Scita}, \citenamefont {Manning}, \citenamefont
  {Cerbino},\ and\ \citenamefont {Marchetti}}]{flocking_spv}%
  \BibitemOpen
  \bibfield  {author} {\bibinfo {author} {\bibfnamefont {Fabio}\ \bibnamefont
  {Giavazzi}}, \bibinfo {author} {\bibfnamefont {Matteo}\ \bibnamefont
  {Paoluzzi}}, \bibinfo {author} {\bibfnamefont {Marta}\ \bibnamefont
  {Macchi}}, \bibinfo {author} {\bibfnamefont {Dapeng}\ \bibnamefont {Bi}},
  \bibinfo {author} {\bibfnamefont {Giorgio}\ \bibnamefont {Scita}}, \bibinfo
  {author} {\bibfnamefont {M.~Lisa}\ \bibnamefont {Manning}}, \bibinfo {author}
  {\bibfnamefont {Roberto}\ \bibnamefont {Cerbino}}, \ and\ \bibinfo {author}
  {\bibfnamefont {M.~Cristina}\ \bibnamefont {Marchetti}},\ }\bibfield  {title}
  {\enquote {\bibinfo {title} {Flocking transitions in confluent tissues},}\
  }\href {\doibase 10.1039/C8SM00126J} {\bibfield  {journal} {\bibinfo
  {journal} {Soft Matter}\ }\textbf {\bibinfo {volume} {14}},\ \bibinfo {pages}
  {3471--3477} (\bibinfo {year} {2018})}\BibitemShut {NoStop}%
\bibitem [{\citenamefont {Czajkowski}\ \emph {et~al.}(2018)\citenamefont
  {Czajkowski}, \citenamefont {Bi}, \citenamefont {Manning},\ and\
  \citenamefont {Marchetti}}]{C8SM00446C}%
  \BibitemOpen
  \bibfield  {author} {\bibinfo {author} {\bibfnamefont {Michael}\ \bibnamefont
  {Czajkowski}}, \bibinfo {author} {\bibfnamefont {Dapeng}\ \bibnamefont {Bi}},
  \bibinfo {author} {\bibfnamefont {M.~Lisa}\ \bibnamefont {Manning}}, \ and\
  \bibinfo {author} {\bibfnamefont {M.~Cristina}\ \bibnamefont {Marchetti}},\
  }\bibfield  {title} {\enquote {\bibinfo {title} {Hydrodynamics of
  shape-driven rigidity transitions in motile tissues},}\ }\href {\doibase
  10.1039/C8SM00446C} {\bibfield  {journal} {\bibinfo  {journal} {Soft Matter}\
  }\textbf {\bibinfo {volume} {14}},\ \bibinfo {pages} {5628--5642} (\bibinfo
  {year} {2018})}\BibitemShut {NoStop}%
\bibitem [{\citenamefont {Staddon}\ \emph {et~al.}(2018)\citenamefont
  {Staddon}, \citenamefont {Bi}, \citenamefont {Tabatabai}, \citenamefont
  {Ajeti}, \citenamefont {Murrell},\ and\ \citenamefont
  {Banerjee}}]{staddon_plos_cb}%
  \BibitemOpen
  \bibfield  {author} {\bibinfo {author} {\bibfnamefont {Michael~F.}\
  \bibnamefont {Staddon}}, \bibinfo {author} {\bibfnamefont {Dapeng}\
  \bibnamefont {Bi}}, \bibinfo {author} {\bibfnamefont {A.~Pasha}\ \bibnamefont
  {Tabatabai}}, \bibinfo {author} {\bibfnamefont {Visar}\ \bibnamefont
  {Ajeti}}, \bibinfo {author} {\bibfnamefont {Michael~P.}\ \bibnamefont
  {Murrell}}, \ and\ \bibinfo {author} {\bibfnamefont {Shiladitya}\
  \bibnamefont {Banerjee}},\ }\bibfield  {title} {\enquote {\bibinfo {title}
  {Cooperation of dual modes of cell motility promotes epithelial stress
  relaxation to accelerate wound healing},}\ }\href {\doibase
  10.1371/journal.pcbi.1006502} {\bibfield  {journal} {\bibinfo  {journal}
  {PLOS Computational Biology}\ }\textbf {\bibinfo {volume} {14}},\ \bibinfo
  {pages} {1--23} (\bibinfo {year} {2018})}\BibitemShut {NoStop}%
\bibitem [{\citenamefont {Li}\ \emph {et~al.}(2018)\citenamefont {Li},
  \citenamefont {Das},\ and\ \citenamefont {Bi}}]{li_pnas_2018}%
  \BibitemOpen
  \bibfield  {author} {\bibinfo {author} {\bibfnamefont {Xinzhi}\ \bibnamefont
  {Li}}, \bibinfo {author} {\bibfnamefont {Amit}\ \bibnamefont {Das}}, \ and\
  \bibinfo {author} {\bibfnamefont {Dapeng}\ \bibnamefont {Bi}},\ }\bibfield
  {title} {\enquote {\bibinfo {title} {Biological tissue-inspired tunable
  photonic fluid},}\ }\href {\doibase 10.1073/pnas.1715810115} {\bibfield
  {journal} {\bibinfo  {journal} {Proceedings of the National Academy of
  Sciences}\ }\textbf {\bibinfo {volume} {115}},\ \bibinfo {pages} {6650--6655}
  (\bibinfo {year} {2018})},\ \Eprint
  {http://arxiv.org/abs/http://www.pnas.org/content/115/26/6650.full.pdf}
  {http://www.pnas.org/content/115/26/6650.full.pdf} \BibitemShut {NoStop}%
\bibitem [{\citenamefont {Harding}\ \emph {et~al.}(2014)\citenamefont
  {Harding}, \citenamefont {McGraw},\ and\ \citenamefont
  {Nechiporuk}}]{harding_rosette_review}%
  \BibitemOpen
  \bibfield  {author} {\bibinfo {author} {\bibfnamefont {Molly~J.}\
  \bibnamefont {Harding}}, \bibinfo {author} {\bibfnamefont {Hillary~F.}\
  \bibnamefont {McGraw}}, \ and\ \bibinfo {author} {\bibfnamefont {Alex}\
  \bibnamefont {Nechiporuk}},\ }\bibfield  {title} {\enquote {\bibinfo {title}
  {The roles and regulation of multicellular rosette structures during
  morphogenesis},}\ }\href {\doibase 10.1242/dev.101444} {\bibfield  {journal}
  {\bibinfo  {journal} {Development}\ }\textbf {\bibinfo {volume} {141}},\
  \bibinfo {pages} {2549--2558} (\bibinfo {year} {2014})}\BibitemShut {NoStop}%
\bibitem [{\citenamefont {Weaire}\ and\ \citenamefont
  {Rivier}(1984)}]{weaire_rivier_1984}%
  \BibitemOpen
  \bibfield  {author} {\bibinfo {author} {\bibfnamefont {D.}~\bibnamefont
  {Weaire}}\ and\ \bibinfo {author} {\bibfnamefont {N.}~\bibnamefont
  {Rivier}},\ }\bibfield  {title} {\enquote {\bibinfo {title} {Soap, cells and
  statistics?random patterns in two dimensions},}\ }\href {\doibase
  10.1080/00107518408210979} {\bibfield  {journal} {\bibinfo  {journal}
  {Contemporary Physics}\ }\textbf {\bibinfo {volume} {25}},\ \bibinfo {pages}
  {59--99} (\bibinfo {year} {1984})}\BibitemShut {NoStop}%
\bibitem [{\citenamefont {Zallen}\ and\ \citenamefont
  {Wieschaus}(2004)}]{Zallen2004}%
  \BibitemOpen
  \bibfield  {author} {\bibinfo {author} {\bibfnamefont {Jennifer~A}\
  \bibnamefont {Zallen}}\ and\ \bibinfo {author} {\bibfnamefont {Eric}\
  \bibnamefont {Wieschaus}},\ }\bibfield  {title} {\enquote {\bibinfo {title}
  {{Patterned Gene Expression Directs Bipolar Planar Polarity in
  Drosophila}},}\ }\href@noop {} {\bibfield  {journal} {\bibinfo  {journal}
  {Developmental cell}\ }\textbf {\bibinfo {volume} {6}},\ \bibinfo {pages}
  {343--355} (\bibinfo {year} {2004})}\BibitemShut {NoStop}%
\bibitem [{\citenamefont {Bertet}\ \emph {et~al.}(2004)\citenamefont {Bertet},
  \citenamefont {Sulak},\ and\ \citenamefont {Lecuit}}]{Bertet2004}%
  \BibitemOpen
  \bibfield  {author} {\bibinfo {author} {\bibfnamefont {Claire}\ \bibnamefont
  {Bertet}}, \bibinfo {author} {\bibfnamefont {Lawrence}\ \bibnamefont
  {Sulak}}, \ and\ \bibinfo {author} {\bibfnamefont {Thomas}\ \bibnamefont
  {Lecuit}},\ }\bibfield  {title} {\enquote {\bibinfo {title}
  {{Myosin-dependent junction remodelling controls planar cell intercalation
  and axis elongation}},}\ }\href@noop {} {\bibfield  {journal} {\bibinfo
  {journal} {Nature}\ }\textbf {\bibinfo {volume} {429}},\ \bibinfo {pages}
  {667--671} (\bibinfo {year} {2004})}\BibitemShut {NoStop}%
\bibitem [{\citenamefont {Zallen}\ and\ \citenamefont
  {Zallen}(2004)}]{Zallen_zallen}%
  \BibitemOpen
  \bibfield  {author} {\bibinfo {author} {\bibfnamefont {Jennifer~A}\
  \bibnamefont {Zallen}}\ and\ \bibinfo {author} {\bibfnamefont {Richard}\
  \bibnamefont {Zallen}},\ }\bibfield  {title} {\enquote {\bibinfo {title}
  {Cell-pattern disordering during convergent extension in drosophila},}\
  }\href {http://stacks.iop.org/0953-8984/16/i=44/a=005} {\bibfield  {journal}
  {\bibinfo  {journal} {Journal of Physics: Condensed Matter}\ }\textbf
  {\bibinfo {volume} {16}},\ \bibinfo {pages} {S5073} (\bibinfo {year}
  {2004})}\BibitemShut {NoStop}%
\bibitem [{\citenamefont {Rauzi}\ \emph {et~al.}(2008)\citenamefont {Rauzi},
  \citenamefont {Verant}, \citenamefont {Lecuit},\ and\ \citenamefont
  {Lenne}}]{Rauzi_2008}%
  \BibitemOpen
  \bibfield  {author} {\bibinfo {author} {\bibfnamefont {Matteo}\ \bibnamefont
  {Rauzi}}, \bibinfo {author} {\bibfnamefont {Pascale}\ \bibnamefont {Verant}},
  \bibinfo {author} {\bibfnamefont {Thomas}\ \bibnamefont {Lecuit}}, \ and\
  \bibinfo {author} {\bibfnamefont {Pierre-Fran{\c c}ois}\ \bibnamefont
  {Lenne}},\ }\bibfield  {title} {\enquote {\bibinfo {title} {Nature and
  anisotropy of cortical forces orienting drosophila tissue morphogenesis},}\
  }\href@noop {} {\bibfield  {journal} {\bibinfo  {journal} {Nature Cell
  Biology}\ }\textbf {\bibinfo {volume} {10}},\ \bibinfo {pages} {1401 EP --}
  (\bibinfo {year} {2008})}\BibitemShut {NoStop}%
\bibitem [{\citenamefont {Escudero}\ \emph {et~al.}(2007)\citenamefont
  {Escudero}, \citenamefont {Bischoff},\ and\ \citenamefont
  {Freeman}}]{Escudero_Freeman}%
  \BibitemOpen
  \bibfield  {author} {\bibinfo {author} {\bibfnamefont {Luis~M.}\ \bibnamefont
  {Escudero}}, \bibinfo {author} {\bibfnamefont {Marcus}\ \bibnamefont
  {Bischoff}}, \ and\ \bibinfo {author} {\bibfnamefont {Matthew}\ \bibnamefont
  {Freeman}},\ }\bibfield  {title} {\enquote {\bibinfo {title} {Myosin ii
  regulates complex cellular arrangement and epithelial architecture in
  drosophila},}\ }\bibfield  {booktitle} {\emph {\bibinfo {booktitle}
  {Developmental Cell}},\ }\href {\doibase 10.1016/j.devcel.2007.09.002}
  {\bibfield  {journal} {\bibinfo  {journal} {Developmental Cell}\ }\textbf
  {\bibinfo {volume} {13}},\ \bibinfo {pages} {717--729} (\bibinfo {year}
  {2007})}\BibitemShut {NoStop}%
\bibitem [{\citenamefont {Blankenship}\ \emph {et~al.}(2006)\citenamefont
  {Blankenship}, \citenamefont {Backovic}, \citenamefont {Sanny}, \citenamefont
  {Weitz},\ and\ \citenamefont {Zallen}}]{Blankenship2006}%
  \BibitemOpen
  \bibfield  {author} {\bibinfo {author} {\bibfnamefont {J~Todd}\ \bibnamefont
  {Blankenship}}, \bibinfo {author} {\bibfnamefont {Stephanie~T}\ \bibnamefont
  {Backovic}}, \bibinfo {author} {\bibfnamefont {Justina S~P}\ \bibnamefont
  {Sanny}}, \bibinfo {author} {\bibfnamefont {Ori}\ \bibnamefont {Weitz}}, \
  and\ \bibinfo {author} {\bibfnamefont {Jennifer~A}\ \bibnamefont {Zallen}},\
  }\bibfield  {title} {\enquote {\bibinfo {title} {{Multicellular rosette
  formation links planar cell polarity to tissue morphogenesis.}}}\ }\href
  {\doibase 10.1016/j.devcel.2006.09.007} {\bibfield  {journal} {\bibinfo
  {journal} {Developmental cell}\ }\textbf {\bibinfo {volume} {11}},\ \bibinfo
  {pages} {459--70} (\bibinfo {year} {2006})}\BibitemShut {NoStop}%
\bibitem [{\citenamefont {Fernandez-Gonzalez}\ \emph
  {et~al.}(2009)\citenamefont {Fernandez-Gonzalez}, \citenamefont
  {de~Matos~Simoes}, \citenamefont {R{\"o}per}, \citenamefont {Eaton},\ and\
  \citenamefont {Zallen}}]{Fernandez-Gonzalez09}%
  \BibitemOpen
  \bibfield  {author} {\bibinfo {author} {\bibfnamefont {Rodrigo}\ \bibnamefont
  {Fernandez-Gonzalez}}, \bibinfo {author} {\bibfnamefont {S{\'e}rgio}\
  \bibnamefont {de~Matos~Simoes}}, \bibinfo {author} {\bibfnamefont
  {Jens-Christian}\ \bibnamefont {R{\"o}per}}, \bibinfo {author} {\bibfnamefont
  {Suzanne}\ \bibnamefont {Eaton}}, \ and\ \bibinfo {author} {\bibfnamefont
  {Jennifer~A}\ \bibnamefont {Zallen}},\ }\bibfield  {title} {\enquote
  {\bibinfo {title} {Myosin ii dynamics are regulated by tension in
  intercalating cells},}\ }\href@noop {} {\bibfield  {journal} {\bibinfo
  {journal} {Developmental cell}\ }\textbf {\bibinfo {volume} {17}},\ \bibinfo
  {pages} {736--743} (\bibinfo {year} {2009})}\BibitemShut {NoStop}%
\bibitem [{\citenamefont {Tamada}\ \emph {et~al.}(2012)\citenamefont {Tamada},
  \citenamefont {Farrell},\ and\ \citenamefont {Zallen}}]{Tamada12}%
  \BibitemOpen
  \bibfield  {author} {\bibinfo {author} {\bibfnamefont {Masako}\ \bibnamefont
  {Tamada}}, \bibinfo {author} {\bibfnamefont {Dene~L}\ \bibnamefont
  {Farrell}}, \ and\ \bibinfo {author} {\bibfnamefont {Jennifer~A}\
  \bibnamefont {Zallen}},\ }\bibfield  {title} {\enquote {\bibinfo {title} {Abl
  regulates planar polarized junctional dynamics through $\beta$-catenin
  tyrosine phosphorylation},}\ }\href@noop {} {\bibfield  {journal} {\bibinfo
  {journal} {Developmental cell}\ }\textbf {\bibinfo {volume} {22}},\ \bibinfo
  {pages} {309--319} (\bibinfo {year} {2012})}\BibitemShut {NoStop}%
\bibitem [{\citenamefont {Kasza}\ \emph {et~al.}(2014)\citenamefont {Kasza},
  \citenamefont {Farrell},\ and\ \citenamefont {Zallen}}]{Kasza2014}%
  \BibitemOpen
  \bibfield  {author} {\bibinfo {author} {\bibfnamefont {Karen~E}\ \bibnamefont
  {Kasza}}, \bibinfo {author} {\bibfnamefont {Dene~L}\ \bibnamefont {Farrell}},
  \ and\ \bibinfo {author} {\bibfnamefont {Jennifer~A}\ \bibnamefont
  {Zallen}},\ }\bibfield  {title} {\enquote {\bibinfo {title} {{Spatiotemporal
  control of epithelial remodeling by regulated myosin phosphorylation.}}}\
  }\href {\doibase 10.1073/pnas.1400520111} {\bibfield  {journal} {\bibinfo
  {journal} {Proceedings of the National Academy of Sciences of the United
  States of America}\ }\textbf {\bibinfo {volume} {111}},\ \bibinfo {pages}
  {11732--7} (\bibinfo {year} {2014})}\BibitemShut {NoStop}%
\bibitem [{\citenamefont {Sun}\ \emph {et~al.}(2017)\citenamefont {Sun},
  \citenamefont {Amourda}, \citenamefont {Shagirov}, \citenamefont {Hara},
  \citenamefont {Saunders},\ and\ \citenamefont
  {Toyama}}]{sun_et_al_basal_rosettes_2017_ncb}%
  \BibitemOpen
  \bibfield  {author} {\bibinfo {author} {\bibfnamefont {Zijun}\ \bibnamefont
  {Sun}}, \bibinfo {author} {\bibfnamefont {Christopher}\ \bibnamefont
  {Amourda}}, \bibinfo {author} {\bibfnamefont {Murat}\ \bibnamefont
  {Shagirov}}, \bibinfo {author} {\bibfnamefont {Yusuke}\ \bibnamefont {Hara}},
  \bibinfo {author} {\bibfnamefont {Timothy~E.}\ \bibnamefont {Saunders}}, \
  and\ \bibinfo {author} {\bibfnamefont {Yusuke}\ \bibnamefont {Toyama}},\
  }\bibfield  {title} {\enquote {\bibinfo {title} {Basolateral protrusion and
  apical contraction cooperatively drive drosophila germ-band extension},}\
  }\href@noop {} {\bibfield  {journal} {\bibinfo  {journal} {Nature Cell
  Biology}\ }\textbf {\bibinfo {volume} {19}},\ \bibinfo {pages} {375 EP --}
  (\bibinfo {year} {2017})}\BibitemShut {NoStop}%
\bibitem [{\citenamefont {Robertson}\ \emph {et~al.}(2012)\citenamefont
  {Robertson}, \citenamefont {Pinal}, \citenamefont {Fichelson},\ and\
  \citenamefont {Pichaud}}]{robertson_2012}%
  \BibitemOpen
  \bibfield  {author} {\bibinfo {author} {\bibfnamefont {Francesca}\
  \bibnamefont {Robertson}}, \bibinfo {author} {\bibfnamefont {Noelia}\
  \bibnamefont {Pinal}}, \bibinfo {author} {\bibfnamefont {Pierre}\
  \bibnamefont {Fichelson}}, \ and\ \bibinfo {author} {\bibfnamefont {Franck}\
  \bibnamefont {Pichaud}},\ }\bibfield  {title} {\enquote {\bibinfo {title}
  {Atonal and egfr signalling orchestrate rok- and drak-dependent adherens
  junction remodelling during ommatidia morphogenesis},}\ }\href {\doibase
  10.1242/dev.080762} {\bibfield  {journal} {\bibinfo  {journal} {Development}\
  }\textbf {\bibinfo {volume} {139}},\ \bibinfo {pages} {3432--3441} (\bibinfo
  {year} {2012})}\BibitemShut {NoStop}%
\bibitem [{\citenamefont {Gompel}\ \emph {et~al.}(2001)\citenamefont {Gompel},
  \citenamefont {Cubedo}, \citenamefont {Thisse}, \citenamefont {Thisse},
  \citenamefont {Dambly-Chaudi{\`e}re},\ and\ \citenamefont
  {Ghysen}}]{GOMPEL200169}%
  \BibitemOpen
  \bibfield  {author} {\bibinfo {author} {\bibfnamefont {Nicolas}\ \bibnamefont
  {Gompel}}, \bibinfo {author} {\bibfnamefont {Nicolas}\ \bibnamefont
  {Cubedo}}, \bibinfo {author} {\bibfnamefont {Christine}\ \bibnamefont
  {Thisse}}, \bibinfo {author} {\bibfnamefont {Bernard}\ \bibnamefont
  {Thisse}}, \bibinfo {author} {\bibfnamefont {Christine}\ \bibnamefont
  {Dambly-Chaudi{\`e}re}}, \ and\ \bibinfo {author} {\bibfnamefont {Alain}\
  \bibnamefont {Ghysen}},\ }\bibfield  {title} {\enquote {\bibinfo {title}
  {Pattern formation in the lateral line of zebrafish},}\ }\href@noop {}
  {\bibfield  {journal} {\bibinfo  {journal} {Mechanisms of Development}\
  }\textbf {\bibinfo {volume} {105}},\ \bibinfo {pages} {69 -- 77} (\bibinfo
  {year} {2001})},\ \bibinfo {note} {molecular Mechanisms of Cell Migration
  during Embryogenesis}\BibitemShut {NoStop}%
\bibitem [{\citenamefont {Nechiporuk}\ and\ \citenamefont
  {Raible}(2008)}]{Nechiporuk08}%
  \BibitemOpen
  \bibfield  {author} {\bibinfo {author} {\bibfnamefont {Alex}\ \bibnamefont
  {Nechiporuk}}\ and\ \bibinfo {author} {\bibfnamefont {David~W}\ \bibnamefont
  {Raible}},\ }\bibfield  {title} {\enquote {\bibinfo {title} {Fgf-dependent
  mechanosensory organ patterning in zebrafish},}\ }\href@noop {} {\bibfield
  {journal} {\bibinfo  {journal} {Science}\ }\textbf {\bibinfo {volume}
  {320}},\ \bibinfo {pages} {1774--1777} (\bibinfo {year} {2008})}\BibitemShut
  {NoStop}%
\bibitem [{\citenamefont {Lecaudey}\ \emph {et~al.}(2008)\citenamefont
  {Lecaudey}, \citenamefont {Cakan-Akdogan}, \citenamefont {Norton},\ and\
  \citenamefont {Gilmour}}]{Lecaudey08}%
  \BibitemOpen
  \bibfield  {author} {\bibinfo {author} {\bibfnamefont {Virginie}\
  \bibnamefont {Lecaudey}}, \bibinfo {author} {\bibfnamefont {Gulcin}\
  \bibnamefont {Cakan-Akdogan}}, \bibinfo {author} {\bibfnamefont {William~HJ}\
  \bibnamefont {Norton}}, \ and\ \bibinfo {author} {\bibfnamefont {Darren}\
  \bibnamefont {Gilmour}},\ }\bibfield  {title} {\enquote {\bibinfo {title}
  {Dynamic fgf signaling couples morphogenesis and migration in the zebrafish
  lateral line primordium},}\ }\href@noop {} {\bibfield  {journal} {\bibinfo
  {journal} {Development}\ }\textbf {\bibinfo {volume} {135}},\ \bibinfo
  {pages} {2695--2705} (\bibinfo {year} {2008})}\BibitemShut {NoStop}%
\bibitem [{\citenamefont {Nishimura}\ and\ \citenamefont
  {Takeichi}(2008)}]{Nishimura_Takeichi_2008}%
  \BibitemOpen
  \bibfield  {author} {\bibinfo {author} {\bibfnamefont {Tamako}\ \bibnamefont
  {Nishimura}}\ and\ \bibinfo {author} {\bibfnamefont {Masatoshi}\ \bibnamefont
  {Takeichi}},\ }\bibfield  {title} {\enquote {\bibinfo {title}
  {Shroom3-mediated recruitment of rho kinases to the apical cell junctions
  regulates epithelial and neuroepithelial planar remodeling},}\ }\href
  {\doibase 10.1242/dev.019646} {\bibfield  {journal} {\bibinfo  {journal}
  {Development}\ }\textbf {\bibinfo {volume} {135}},\ \bibinfo {pages}
  {1493--1502} (\bibinfo {year} {2008})}\BibitemShut {NoStop}%
\bibitem [{\citenamefont {Williams}\ \emph {et~al.}(2014)\citenamefont
  {Williams}, \citenamefont {Yen}, \citenamefont {Lu},\ and\ \citenamefont
  {Sutherland}}]{Williams14}%
  \BibitemOpen
  \bibfield  {author} {\bibinfo {author} {\bibfnamefont {Margot}\ \bibnamefont
  {Williams}}, \bibinfo {author} {\bibfnamefont {Weiwei}\ \bibnamefont {Yen}},
  \bibinfo {author} {\bibfnamefont {Xiaowei}\ \bibnamefont {Lu}}, \ and\
  \bibinfo {author} {\bibfnamefont {Ann}\ \bibnamefont {Sutherland}},\
  }\bibfield  {title} {\enquote {\bibinfo {title} {Distinct apical and
  basolateral mechanisms drive planar cell polarity-dependent convergent
  extension of the mouse neural plate},}\ }\href@noop {} {\bibfield  {journal}
  {\bibinfo  {journal} {Developmental cell}\ }\textbf {\bibinfo {volume}
  {29}},\ \bibinfo {pages} {34--46} (\bibinfo {year} {2014})}\BibitemShut
  {NoStop}%
\bibitem [{\citenamefont {Trichas}\ \emph {et~al.}(2012)\citenamefont
  {Trichas}, \citenamefont {Smith}, \citenamefont {White}, \citenamefont
  {Wilkins}, \citenamefont {Watanabe}, \citenamefont {Moore}, \citenamefont
  {Joyce}, \citenamefont {Sugnaseelan}, \citenamefont {Rodriguez},
  \citenamefont {Kay}, \citenamefont {Baker}, \citenamefont {Maini},\ and\
  \citenamefont {Srinivas}}]{trichas_2012}%
  \BibitemOpen
  \bibfield  {author} {\bibinfo {author} {\bibfnamefont {Georgios}\
  \bibnamefont {Trichas}}, \bibinfo {author} {\bibfnamefont {Aaron~M.}\
  \bibnamefont {Smith}}, \bibinfo {author} {\bibfnamefont {Natalia}\
  \bibnamefont {White}}, \bibinfo {author} {\bibfnamefont {Vivienne}\
  \bibnamefont {Wilkins}}, \bibinfo {author} {\bibfnamefont {Tomoko}\
  \bibnamefont {Watanabe}}, \bibinfo {author} {\bibfnamefont {Abigail}\
  \bibnamefont {Moore}}, \bibinfo {author} {\bibfnamefont {Bradley}\
  \bibnamefont {Joyce}}, \bibinfo {author} {\bibfnamefont {Jacintha}\
  \bibnamefont {Sugnaseelan}}, \bibinfo {author} {\bibfnamefont {Tristan~A.}\
  \bibnamefont {Rodriguez}}, \bibinfo {author} {\bibfnamefont {David}\
  \bibnamefont {Kay}}, \bibinfo {author} {\bibfnamefont {Ruth~E.}\ \bibnamefont
  {Baker}}, \bibinfo {author} {\bibfnamefont {Philip~K.}\ \bibnamefont
  {Maini}}, \ and\ \bibinfo {author} {\bibfnamefont {Shankar}\ \bibnamefont
  {Srinivas}},\ }\bibfield  {title} {\enquote {\bibinfo {title} {Multi-cellular
  rosettes in the mouse visceral endoderm facilitate the ordered migration of
  anterior visceral endoderm cells},}\ }\href {\doibase
  10.1371/journal.pbio.1001256} {\bibfield  {journal} {\bibinfo  {journal}
  {PLOS Biology}\ }\textbf {\bibinfo {volume} {10}},\ \bibinfo {pages} {1--14}
  (\bibinfo {year} {2012})}\BibitemShut {NoStop}%
\bibitem [{\citenamefont {Lienkamp}\ \emph {et~al.}(2012)\citenamefont
  {Lienkamp}, \citenamefont {Liu}, \citenamefont {Karner}, \citenamefont
  {Carroll}, \citenamefont {Ronneberger}, \citenamefont {Wallingford},\ and\
  \citenamefont {Walz}}]{lienkamp_2012}%
  \BibitemOpen
  \bibfield  {author} {\bibinfo {author} {\bibfnamefont {Soeren~S}\
  \bibnamefont {Lienkamp}}, \bibinfo {author} {\bibfnamefont {Kun}\
  \bibnamefont {Liu}}, \bibinfo {author} {\bibfnamefont {Courtney~M}\
  \bibnamefont {Karner}}, \bibinfo {author} {\bibfnamefont {Thomas~J}\
  \bibnamefont {Carroll}}, \bibinfo {author} {\bibfnamefont {Olaf}\
  \bibnamefont {Ronneberger}}, \bibinfo {author} {\bibfnamefont {John~B}\
  \bibnamefont {Wallingford}}, \ and\ \bibinfo {author} {\bibfnamefont {Gerd}\
  \bibnamefont {Walz}},\ }\bibfield  {title} {\enquote {\bibinfo {title}
  {Vertebrate kidney tubules elongate using a planar cell polarity--dependent,
  rosette-based mechanism of convergent extension},}\ }\href@noop {} {\bibfield
   {journal} {\bibinfo  {journal} {Nature Genetics}\ }\textbf {\bibinfo
  {volume} {44}},\ \bibinfo {pages} {1382 EP --} (\bibinfo {year}
  {2012})}\BibitemShut {NoStop}%
\bibitem [{\citenamefont {Villasenor}\ \emph {et~al.}(2010)\citenamefont
  {Villasenor}, \citenamefont {Chong}, \citenamefont {Henkemeyer},\ and\
  \citenamefont {Cleaver}}]{Villasenor_2010}%
  \BibitemOpen
  \bibfield  {author} {\bibinfo {author} {\bibfnamefont {Alethia}\ \bibnamefont
  {Villasenor}}, \bibinfo {author} {\bibfnamefont {Diana~C.}\ \bibnamefont
  {Chong}}, \bibinfo {author} {\bibfnamefont {Mark}\ \bibnamefont
  {Henkemeyer}}, \ and\ \bibinfo {author} {\bibfnamefont {Ondine}\ \bibnamefont
  {Cleaver}},\ }\bibfield  {title} {\enquote {\bibinfo {title} {Epithelial
  dynamics of pancreatic branching morphogenesis},}\ }\href {\doibase
  10.1242/dev.052993} {\bibfield  {journal} {\bibinfo  {journal} {Development}\
  }\textbf {\bibinfo {volume} {137}},\ \bibinfo {pages} {4295--4305} (\bibinfo
  {year} {2010})}\BibitemShut {NoStop}%
\bibitem [{\citenamefont {Zhang}\ \emph {et~al.}(2001)\citenamefont {Zhang},
  \citenamefont {Wernig}, \citenamefont {Duncan}, \citenamefont {Br{\"u}stle},\
  and\ \citenamefont {Thomson}}]{zhang_2001}%
  \BibitemOpen
  \bibfield  {author} {\bibinfo {author} {\bibfnamefont {Su-Chun}\ \bibnamefont
  {Zhang}}, \bibinfo {author} {\bibfnamefont {Marius}\ \bibnamefont {Wernig}},
  \bibinfo {author} {\bibfnamefont {Ian~D.}\ \bibnamefont {Duncan}}, \bibinfo
  {author} {\bibfnamefont {Oliver}\ \bibnamefont {Br{\"u}stle}}, \ and\
  \bibinfo {author} {\bibfnamefont {James~A.}\ \bibnamefont {Thomson}},\
  }\bibfield  {title} {\enquote {\bibinfo {title} {In vitro differentiation of
  transplantable neural precursors from human embryonic stem cells},}\
  }\href@noop {} {\bibfield  {journal} {\bibinfo  {journal} {Nature
  Biotechnology}\ }\textbf {\bibinfo {volume} {19}},\ \bibinfo {pages} {1129 EP
  --} (\bibinfo {year} {2001})}\BibitemShut {NoStop}%
\bibitem [{\citenamefont {Elkabetz}\ and\ \citenamefont
  {Studer}(2008)}]{elkabetz_2008}%
  \BibitemOpen
  \bibfield  {author} {\bibinfo {author} {\bibfnamefont {Y.}~\bibnamefont
  {Elkabetz}}\ and\ \bibinfo {author} {\bibfnamefont {L.}~\bibnamefont
  {Studer}},\ }\bibfield  {title} {\enquote {\bibinfo {title} {Human
  esc-derived neural rosettes and neural stem cell progression},}\ }\href
  {\doibase 10.1101/sqb.2008.73.052} {\bibfield  {journal} {\bibinfo  {journal}
  {Cold Spring Harbor Symposia on Quantitative Biology}\ }\textbf {\bibinfo
  {volume} {73}},\ \bibinfo {pages} {377--387} (\bibinfo {year}
  {2008})}\BibitemShut {NoStop}%
\bibitem [{\citenamefont {Wippold}\ and\ \citenamefont
  {Perry}(2006)}]{Wippold2006}%
  \BibitemOpen
  \bibfield  {author} {\bibinfo {author} {\bibfnamefont {Franz~J.}\
  \bibnamefont {Wippold}}\ and\ \bibinfo {author} {\bibfnamefont
  {A.}~\bibnamefont {Perry}},\ }\bibfield  {title} {\enquote {\bibinfo {title}
  {{Neuropathology for the neuroradiologist: Rosettes and Pseudorosettes}},}\
  }\href {\doibase 10.3174/ajnr.A0781} {\bibfield  {journal} {\bibinfo
  {journal} {American Journal of Neuroradiology}\ }\textbf {\bibinfo {volume}
  {27}},\ \bibinfo {pages} {488--492} (\bibinfo {year} {2006})}\BibitemShut
  {NoStop}%
\bibitem [{\citenamefont {Heer}\ and\ \citenamefont
  {Martin}(2017)}]{Heer_Martin_tension_review}%
  \BibitemOpen
  \bibfield  {author} {\bibinfo {author} {\bibfnamefont {Natalie~C.}\
  \bibnamefont {Heer}}\ and\ \bibinfo {author} {\bibfnamefont {Adam~C.}\
  \bibnamefont {Martin}},\ }\bibfield  {title} {\enquote {\bibinfo {title}
  {Tension, contraction and tissue morphogenesis},}\ }\href {\doibase
  10.1242/dev.151282} {\bibfield  {journal} {\bibinfo  {journal} {Development}\
  }\textbf {\bibinfo {volume} {144}},\ \bibinfo {pages} {4249--4260} (\bibinfo
  {year} {2017})}\BibitemShut {NoStop}%
\bibitem [{\citenamefont {Kocgozlu}\ \emph {et~al.}(2016)\citenamefont
  {Kocgozlu}, \citenamefont {Saw}, \citenamefont {Le}, \citenamefont {Yow},
  \citenamefont {Shagirov}, \citenamefont {Wong}, \citenamefont {M{\`{e}}ge},
  \citenamefont {Lim}, \citenamefont {Toyama},\ and\ \citenamefont
  {Ladoux}}]{Kocgozlu2016}%
  \BibitemOpen
  \bibfield  {author} {\bibinfo {author} {\bibfnamefont {Leyla}\ \bibnamefont
  {Kocgozlu}}, \bibinfo {author} {\bibfnamefont {Thuan~Beng}\ \bibnamefont
  {Saw}}, \bibinfo {author} {\bibfnamefont {Anh~Phuong}\ \bibnamefont {Le}},
  \bibinfo {author} {\bibfnamefont {Ivan}\ \bibnamefont {Yow}}, \bibinfo
  {author} {\bibfnamefont {Murat}\ \bibnamefont {Shagirov}}, \bibinfo {author}
  {\bibfnamefont {Eunice}\ \bibnamefont {Wong}}, \bibinfo {author}
  {\bibfnamefont {Ren{\'{e}}~Marc}\ \bibnamefont {M{\`{e}}ge}}, \bibinfo
  {author} {\bibfnamefont {Chwee~Teck}\ \bibnamefont {Lim}}, \bibinfo {author}
  {\bibfnamefont {Yusuke}\ \bibnamefont {Toyama}}, \ and\ \bibinfo {author}
  {\bibfnamefont {Benoit}\ \bibnamefont {Ladoux}},\ }\bibfield  {title}
  {\enquote {\bibinfo {title} {{Epithelial Cell Packing Induces Distinct Modes
  of Cell Extrusions}},}\ }\href {\doibase 10.1016/j.cub.2016.08.057}
  {\bibfield  {journal} {\bibinfo  {journal} {Current Biology}\ }\textbf
  {\bibinfo {volume} {26}},\ \bibinfo {pages} {2942--2950} (\bibinfo {year}
  {2016})}\BibitemShut {NoStop}%
\bibitem [{\citenamefont {Toyama}\ \emph {et~al.}(2008)\citenamefont {Toyama},
  \citenamefont {Peralta}, \citenamefont {Wells}, \citenamefont {Kiehart},\
  and\ \citenamefont {Edwards}}]{Toyama2008}%
  \BibitemOpen
  \bibfield  {author} {\bibinfo {author} {\bibfnamefont {Yusuke}\ \bibnamefont
  {Toyama}}, \bibinfo {author} {\bibfnamefont {Xomalin~G}\ \bibnamefont
  {Peralta}}, \bibinfo {author} {\bibfnamefont {Adrienne~R}\ \bibnamefont
  {Wells}}, \bibinfo {author} {\bibfnamefont {Daniel~P}\ \bibnamefont
  {Kiehart}}, \ and\ \bibinfo {author} {\bibfnamefont {Glenn~S}\ \bibnamefont
  {Edwards}},\ }\bibfield  {title} {\enquote {\bibinfo {title} {{Apoptotic
  force and tissue dynamics during Drosophila embryogenesis.}}}\ }\href
  {\doibase 10.1126/science.1157052} {\bibfield  {journal} {\bibinfo  {journal}
  {Science (New York, N.Y.)}\ }\textbf {\bibinfo {volume} {321}},\ \bibinfo
  {pages} {1683--6} (\bibinfo {year} {2008})}\BibitemShut {NoStop}%
\bibitem [{\citenamefont {Marinari}\ \emph {et~al.}(2012)\citenamefont
  {Marinari}, \citenamefont {Mehonic}, \citenamefont {Curran}, \citenamefont
  {Gale}, \citenamefont {Duke},\ and\ \citenamefont {Baum}}]{Marinari2012}%
  \BibitemOpen
  \bibfield  {author} {\bibinfo {author} {\bibfnamefont {Eliana}\ \bibnamefont
  {Marinari}}, \bibinfo {author} {\bibfnamefont {Aida}\ \bibnamefont
  {Mehonic}}, \bibinfo {author} {\bibfnamefont {Scott}\ \bibnamefont {Curran}},
  \bibinfo {author} {\bibfnamefont {Jonathan}\ \bibnamefont {Gale}}, \bibinfo
  {author} {\bibfnamefont {Thomas}\ \bibnamefont {Duke}}, \ and\ \bibinfo
  {author} {\bibfnamefont {Buzz}\ \bibnamefont {Baum}},\ }\bibfield  {title}
  {\enquote {\bibinfo {title} {{Live-cell delamination counterbalances
  epithelial growth to limit tissue overcrowding.}}}\ }\href {\doibase
  10.1038/nature10984} {\bibfield  {journal} {\bibinfo  {journal} {Nature}\
  }\textbf {\bibinfo {volume} {484}},\ \bibinfo {pages} {542--545} (\bibinfo
  {year} {2012})}\BibitemShut {NoStop}%
\bibitem [{\citenamefont {Slattum}\ \emph {et~al.}(2009)\citenamefont
  {Slattum}, \citenamefont {McGee},\ and\ \citenamefont
  {Rosenblatt}}]{Slattum2009}%
  \BibitemOpen
  \bibfield  {author} {\bibinfo {author} {\bibfnamefont {Gloria}\ \bibnamefont
  {Slattum}}, \bibinfo {author} {\bibfnamefont {Karen~M.}\ \bibnamefont
  {McGee}}, \ and\ \bibinfo {author} {\bibfnamefont {Jody}\ \bibnamefont
  {Rosenblatt}},\ }\bibfield  {title} {\enquote {\bibinfo {title} {{P115 RhoGEF
  and microtubules decide the direction apoptotic cells extrude from an
  epithelium}},}\ }\href {\doibase 10.1083/jcb.200903079} {\bibfield  {journal}
  {\bibinfo  {journal} {Journal of Cell Biology}\ }\textbf {\bibinfo {volume}
  {186}},\ \bibinfo {pages} {693--702} (\bibinfo {year} {2009})}\BibitemShut
  {NoStop}%
\bibitem [{\citenamefont {Razzell}\ \emph {et~al.}(2014)\citenamefont
  {Razzell}, \citenamefont {Wood},\ and\ \citenamefont {Martin}}]{Razzell2014}%
  \BibitemOpen
  \bibfield  {author} {\bibinfo {author} {\bibfnamefont {William}\ \bibnamefont
  {Razzell}}, \bibinfo {author} {\bibfnamefont {Will}\ \bibnamefont {Wood}}, \
  and\ \bibinfo {author} {\bibfnamefont {Paul}\ \bibnamefont {Martin}},\
  }\bibfield  {title} {\enquote {\bibinfo {title} {{Recapitulation of
  morphogenetic cell shape changes enables wound re-epithelialisation.}}}\
  }\href {\doibase 10.1242/dev.107045} {\bibfield  {journal} {\bibinfo
  {journal} {Development (Cambridge, England)}\ }\textbf {\bibinfo {volume}
  {141}},\ \bibinfo {pages} {1814--1820} (\bibinfo {year} {2014})}\BibitemShut
  {NoStop}%
\bibitem [{\citenamefont {Brugu{\'{e}}s}\ \emph {et~al.}(2014)\citenamefont
  {Brugu{\'{e}}s}, \citenamefont {Anon}, \citenamefont {Conte}, \citenamefont
  {Veldhuis}, \citenamefont {Gupta}, \citenamefont {Colombelli}, \citenamefont
  {Mu{\~{n}}oz}, \citenamefont {Brodland}, \citenamefont {Ladoux},\ and\
  \citenamefont {Trepat}}]{Brugues2014}%
  \BibitemOpen
  \bibfield  {author} {\bibinfo {author} {\bibfnamefont {Agust{\'{i}}}\
  \bibnamefont {Brugu{\'{e}}s}}, \bibinfo {author} {\bibfnamefont {Ester}\
  \bibnamefont {Anon}}, \bibinfo {author} {\bibfnamefont {Vito}\ \bibnamefont
  {Conte}}, \bibinfo {author} {\bibfnamefont {Jim~H}\ \bibnamefont {Veldhuis}},
  \bibinfo {author} {\bibfnamefont {Mukund}\ \bibnamefont {Gupta}}, \bibinfo
  {author} {\bibfnamefont {Julien}\ \bibnamefont {Colombelli}}, \bibinfo
  {author} {\bibfnamefont {Jos{\'{e}}~J}\ \bibnamefont {Mu{\~{n}}oz}}, \bibinfo
  {author} {\bibfnamefont {G~Wayne}\ \bibnamefont {Brodland}}, \bibinfo
  {author} {\bibfnamefont {Benoit}\ \bibnamefont {Ladoux}}, \ and\ \bibinfo
  {author} {\bibfnamefont {Xavier}\ \bibnamefont {Trepat}},\ }\bibfield
  {title} {\enquote {\bibinfo {title} {{Forces driving epithelial wound
  healing}},}\ }\href {\doibase 10.1038/NPHYS3040} {\bibfield  {journal}
  {\bibinfo  {journal} {Nature Physics}\ } (\bibinfo {year} {2014}),\
  10.1038/NPHYS3040}\BibitemShut {NoStop}%
\bibitem [{\citenamefont {Streichan}\ \emph {et~al.}(2018)\citenamefont
  {Streichan}, \citenamefont {Lefebvre}, \citenamefont {Noll}, \citenamefont
  {Wieschaus},\ and\ \citenamefont {Shraiman}}]{Streichan17}%
  \BibitemOpen
  \bibfield  {author} {\bibinfo {author} {\bibfnamefont {Sebastian~J}\
  \bibnamefont {Streichan}}, \bibinfo {author} {\bibfnamefont {Matthew~F}\
  \bibnamefont {Lefebvre}}, \bibinfo {author} {\bibfnamefont {Nicholas}\
  \bibnamefont {Noll}}, \bibinfo {author} {\bibfnamefont {Eric~F}\ \bibnamefont
  {Wieschaus}}, \ and\ \bibinfo {author} {\bibfnamefont {Boris~I}\ \bibnamefont
  {Shraiman}},\ }\bibfield  {title} {\enquote {\bibinfo {title} {Global
  morphogenetic flow is accurately predicted by the spatial distribution of
  myosin motors},}\ }\href {\doibase 10.7554/eLife.27454} {\bibfield  {journal}
  {\bibinfo  {journal} {eLife}\ }\textbf {\bibinfo {volume} {7}},\ \bibinfo
  {pages} {e27454} (\bibinfo {year} {2018})}\BibitemShut {NoStop}%
\bibitem [{\citenamefont {Shook}\ \emph {et~al.}(2012)\citenamefont {Shook},
  \citenamefont {Manz}, \citenamefont {Peters}, \citenamefont {Kang},\ and\
  \citenamefont {Conover}}]{Shook6947_persistent_rosettes}%
  \BibitemOpen
  \bibfield  {author} {\bibinfo {author} {\bibfnamefont {Brett~A.}\
  \bibnamefont {Shook}}, \bibinfo {author} {\bibfnamefont {David~H.}\
  \bibnamefont {Manz}}, \bibinfo {author} {\bibfnamefont {John~J.}\
  \bibnamefont {Peters}}, \bibinfo {author} {\bibfnamefont {Sangwook}\
  \bibnamefont {Kang}}, \ and\ \bibinfo {author} {\bibfnamefont {Joanne~C.}\
  \bibnamefont {Conover}},\ }\bibfield  {title} {\enquote {\bibinfo {title}
  {Spatiotemporal changes to the subventricular zone stem cell pool through
  aging},}\ }\href {\doibase 10.1523/JNEUROSCI.5987-11.2012} {\bibfield
  {journal} {\bibinfo  {journal} {Journal of Neuroscience}\ }\textbf {\bibinfo
  {volume} {32}},\ \bibinfo {pages} {6947--6956} (\bibinfo {year}
  {2012})}\BibitemShut {NoStop}%
\bibitem [{\citenamefont {Razzell}\ \emph {et~al.}(2018)\citenamefont
  {Razzell}, \citenamefont {Bustillo},\ and\ \citenamefont
  {Zallen}}]{Razzell3715}%
  \BibitemOpen
  \bibfield  {author} {\bibinfo {author} {\bibfnamefont {William}\ \bibnamefont
  {Razzell}}, \bibinfo {author} {\bibfnamefont {Maria~E.}\ \bibnamefont
  {Bustillo}}, \ and\ \bibinfo {author} {\bibfnamefont {Jennifer~A.}\
  \bibnamefont {Zallen}},\ }\bibfield  {title} {\enquote {\bibinfo {title} {The
  force-sensitive protein ajuba regulates cell adhesion during epithelial
  morphogenesis},}\ }\href {\doibase 10.1083/jcb.201801171} {\bibfield
  {journal} {\bibinfo  {journal} {The Journal of Cell Biology}\ }\textbf
  {\bibinfo {volume} {217}},\ \bibinfo {pages} {3715--3730} (\bibinfo {year}
  {2018})}\BibitemShut {NoStop}%
\bibitem [{\citenamefont {F.R.S.}(1864)}]{maxwell_counting}%
  \BibitemOpen
  \bibfield  {author} {\bibinfo {author} {\bibfnamefont {J.~Clerk~Maxwell}\
  \bibnamefont {F.R.S.}},\ }\bibfield  {title} {\enquote {\bibinfo {title} {L.
  on the calculation of the equilibrium and stiffness of frames},}\ }\href
  {\doibase 10.1080/14786446408643668} {\bibfield  {journal} {\bibinfo
  {journal} {Philosophical Magazine}\ }\textbf {\bibinfo {volume} {27}},\
  \bibinfo {pages} {294--299} (\bibinfo {year} {1864})}\BibitemShut {NoStop}%
\bibitem [{\citenamefont {Kane}\ and\ \citenamefont
  {Lubensky}(2014)}]{Kane_Lubensky}%
  \BibitemOpen
  \bibfield  {author} {\bibinfo {author} {\bibfnamefont {C~L}\ \bibnamefont
  {Kane}}\ and\ \bibinfo {author} {\bibfnamefont {T~C}\ \bibnamefont
  {Lubensky}},\ }\bibfield  {title} {\enquote {\bibinfo {title} {{Topological
  boundary modes in isostatic lattices}},}\ }\href
  {http://dx.doi.org/10.1038/nphys2835 http://10.0.4.14/nphys2835
  http://www.nature.com/nphys/journal/v10/n1/abs/nphys2835.html{\#}supplementary-information}
  {\bibfield  {journal} {\bibinfo  {journal} {Nat Phys}\ }\textbf {\bibinfo
  {volume} {10}},\ \bibinfo {pages} {39--45} (\bibinfo {year}
  {2014})}\BibitemShut {NoStop}%
\bibitem [{\citenamefont {Liu}\ and\ \citenamefont
  {Nagel}(2010)}]{LiuNagelReview}%
  \BibitemOpen
  \bibfield  {author} {\bibinfo {author} {\bibfnamefont {Andrea~J.}\
  \bibnamefont {Liu}}\ and\ \bibinfo {author} {\bibfnamefont {Sidney~R.}\
  \bibnamefont {Nagel}},\ }\bibfield  {title} {\enquote {\bibinfo {title} {The
  jamming transition and the marginally jammed solid},}\ }\href {\doibase
  10.1146/annurev-conmatphys-070909-104045} {\bibfield  {journal} {\bibinfo
  {journal} {Annual Review of Condensed Matter Physics}\ }\textbf {\bibinfo
  {volume} {1}},\ \bibinfo {pages} {347--369} (\bibinfo {year}
  {2010})}\BibitemShut {NoStop}%
\bibitem [{\citenamefont {Feng}\ \emph {et~al.}(1985)\citenamefont {Feng},
  \citenamefont {Thorpe},\ and\ \citenamefont {Garboczi}}]{Feng85}%
  \BibitemOpen
  \bibfield  {author} {\bibinfo {author} {\bibfnamefont {Shechao}\ \bibnamefont
  {Feng}}, \bibinfo {author} {\bibfnamefont {M.~F.}\ \bibnamefont {Thorpe}}, \
  and\ \bibinfo {author} {\bibfnamefont {E.}~\bibnamefont {Garboczi}},\
  }\bibfield  {title} {\enquote {\bibinfo {title} {Effective-medium theory of
  percolation on central-force elastic networks},}\ }\href {\doibase
  10.1103/PhysRevB.31.276} {\bibfield  {journal} {\bibinfo  {journal} {Phys.
  Rev. B}\ }\textbf {\bibinfo {volume} {31}},\ \bibinfo {pages} {276--280}
  (\bibinfo {year} {1985})}\BibitemShut {NoStop}%
\bibitem [{\citenamefont {Mao}\ \emph {et~al.}(2010)\citenamefont {Mao},
  \citenamefont {Xu},\ and\ \citenamefont {Lubensky}}]{Mao10}%
  \BibitemOpen
  \bibfield  {author} {\bibinfo {author} {\bibfnamefont {Xiaoming}\
  \bibnamefont {Mao}}, \bibinfo {author} {\bibfnamefont {Ning}\ \bibnamefont
  {Xu}}, \ and\ \bibinfo {author} {\bibfnamefont {T.~C.}\ \bibnamefont
  {Lubensky}},\ }\bibfield  {title} {\enquote {\bibinfo {title} {Soft modes and
  elasticity of nearly isostatic lattices: Randomness and dissipation},}\
  }\href {\doibase 10.1103/PhysRevLett.104.085504} {\bibfield  {journal}
  {\bibinfo  {journal} {Phys. Rev. Lett.}\ }\textbf {\bibinfo {volume} {104}},\
  \bibinfo {pages} {085504} (\bibinfo {year} {2010})}\BibitemShut {NoStop}%
\bibitem [{\citenamefont {Wyart}(2010)}]{Wyart10}%
  \BibitemOpen
  \bibfield  {author} {\bibinfo {author} {\bibfnamefont {Matthieu}\
  \bibnamefont {Wyart}},\ }\bibfield  {title} {\enquote {\bibinfo {title}
  {Scaling of phononic transport with connectivity in amorphous solids},}\
  }\href@noop {} {\bibfield  {journal} {\bibinfo  {journal} {EPL (Europhysics
  Letters)}\ }\textbf {\bibinfo {volume} {89}},\ \bibinfo {pages} {64001}
  (\bibinfo {year} {2010})}\BibitemShut {NoStop}%
\bibitem [{\citenamefont {DeGiuli}\ \emph {et~al.}(2014)\citenamefont
  {DeGiuli}, \citenamefont {Laversanne-Finot}, \citenamefont {D\"uring},
  \citenamefont {Lerner},\ and\ \citenamefont {Wyart}}]{DeGiuli14}%
  \BibitemOpen
  \bibfield  {author} {\bibinfo {author} {\bibfnamefont {Eric}\ \bibnamefont
  {DeGiuli}}, \bibinfo {author} {\bibfnamefont {Adrien}\ \bibnamefont
  {Laversanne-Finot}}, \bibinfo {author} {\bibfnamefont {Gustavo~Alberto}\
  \bibnamefont {D\"uring}}, \bibinfo {author} {\bibfnamefont {Edan}\
  \bibnamefont {Lerner}}, \ and\ \bibinfo {author} {\bibfnamefont {Matthieu}\
  \bibnamefont {Wyart}},\ }\bibfield  {title} {\enquote {\bibinfo {title}
  {Effects of coordination and pressure on sound attenuation, boson peak and
  elasticity in amorphous solids},}\ }\href@noop {} {\bibfield  {journal}
  {\bibinfo  {journal} {Soft Matter}\ }\textbf {\bibinfo {volume} {10}},\
  \bibinfo {pages} {5628--5644} (\bibinfo {year} {2014})}\BibitemShut {NoStop}%
\bibitem [{\citenamefont {Lubensky}\ \emph {et~al.}(2015)\citenamefont
  {Lubensky}, \citenamefont {Kane}, \citenamefont {Mao}, \citenamefont
  {Souslov},\ and\ \citenamefont {Sun}}]{Lubensky_review}%
  \BibitemOpen
  \bibfield  {author} {\bibinfo {author} {\bibfnamefont {T~C}\ \bibnamefont
  {Lubensky}}, \bibinfo {author} {\bibfnamefont {C~L}\ \bibnamefont {Kane}},
  \bibinfo {author} {\bibfnamefont {Xiaoming}\ \bibnamefont {Mao}}, \bibinfo
  {author} {\bibfnamefont {A}~\bibnamefont {Souslov}}, \ and\ \bibinfo {author}
  {\bibfnamefont {Kai}\ \bibnamefont {Sun}},\ }\bibfield  {title} {\enquote
  {\bibinfo {title} {Phonons and elasticity in critically coordinated
  lattices},}\ }\href {http://stacks.iop.org/0034-4885/78/i=7/a=073901}
  {\bibfield  {journal} {\bibinfo  {journal} {Reports on Progress in Physics}\
  }\textbf {\bibinfo {volume} {78}},\ \bibinfo {pages} {073901} (\bibinfo
  {year} {2015})}\BibitemShut {NoStop}%
\bibitem [{\citenamefont {Broedersz}\ \emph {et~al.}(2011)\citenamefont
  {Broedersz}, \citenamefont {Mao}, \citenamefont {Lubensky},\ and\
  \citenamefont {MacKintosh}}]{chase_rigidity}%
  \BibitemOpen
  \bibfield  {author} {\bibinfo {author} {\bibfnamefont {Chase~P.}\
  \bibnamefont {Broedersz}}, \bibinfo {author} {\bibfnamefont {Xiaoming}\
  \bibnamefont {Mao}}, \bibinfo {author} {\bibfnamefont {Tom~C.}\ \bibnamefont
  {Lubensky}}, \ and\ \bibinfo {author} {\bibfnamefont {Frederick~C.}\
  \bibnamefont {MacKintosh}},\ }\bibfield  {title} {\enquote {\bibinfo {title}
  {{Criticality and isostaticity in fibre networks}},}\ }\href {\doibase
  10.1038/nphys2127} {\bibfield  {journal} {\bibinfo  {journal} {Nature
  Physics}\ }\textbf {\bibinfo {volume} {7}},\ \bibinfo {pages} {983--988}
  (\bibinfo {year} {2011})}\BibitemShut {NoStop}%
\bibitem [{\citenamefont {Das}\ \emph {et~al.}(2012)\citenamefont {Das},
  \citenamefont {Quint},\ and\ \citenamefont
  {Schwarz}}]{Das_schwarz_plosone_2012}%
  \BibitemOpen
  \bibfield  {author} {\bibinfo {author} {\bibfnamefont {Moumita}\ \bibnamefont
  {Das}}, \bibinfo {author} {\bibfnamefont {D.~A.}\ \bibnamefont {Quint}}, \
  and\ \bibinfo {author} {\bibfnamefont {J.~M.}\ \bibnamefont {Schwarz}},\
  }\bibfield  {title} {\enquote {\bibinfo {title} {Redundancy and cooperativity
  in the mechanics of compositely crosslinked filamentous networks},}\ }\href
  {\doibase 10.1371/journal.pone.0035939} {\bibfield  {journal} {\bibinfo
  {journal} {PLOS ONE}\ }\textbf {\bibinfo {volume} {7}},\ \bibinfo {pages}
  {1--11} (\bibinfo {year} {2012})}\BibitemShut {NoStop}%
\bibitem [{\citenamefont {Landau}\ and\ \citenamefont
  {Lifshitz}(1986)}]{Landau86}%
  \BibitemOpen
  \bibfield  {author} {\bibinfo {author} {\bibfnamefont {Lev~D}\ \bibnamefont
  {Landau}}\ and\ \bibinfo {author} {\bibfnamefont {EM}~\bibnamefont
  {Lifshitz}},\ }\bibfield  {title} {\enquote {\bibinfo {title} {Theory of
  elasticity, vol. 7},}\ }\href@noop {} {\bibfield  {journal} {\bibinfo
  {journal} {Course of Theoretical Physics}\ }\textbf {\bibinfo {volume} {3}},\
  \bibinfo {pages} {109} (\bibinfo {year} {1986})}\BibitemShut {NoStop}%
\bibitem [{\citenamefont {Hutson}\ \emph {et~al.}(2003)\citenamefont {Hutson},
  \citenamefont {Tokutake}, \citenamefont {Chang}, \citenamefont {Bloor},
  \citenamefont {Venakides}, \citenamefont {Kiehart},\ and\ \citenamefont
  {Edwards}}]{Hutson145}%
  \BibitemOpen
  \bibfield  {author} {\bibinfo {author} {\bibfnamefont {M.~Shane}\
  \bibnamefont {Hutson}}, \bibinfo {author} {\bibfnamefont {Yoichiro}\
  \bibnamefont {Tokutake}}, \bibinfo {author} {\bibfnamefont {Ming-Shien}\
  \bibnamefont {Chang}}, \bibinfo {author} {\bibfnamefont {James~W.}\
  \bibnamefont {Bloor}}, \bibinfo {author} {\bibfnamefont {Stephanos}\
  \bibnamefont {Venakides}}, \bibinfo {author} {\bibfnamefont {Daniel~P.}\
  \bibnamefont {Kiehart}}, \ and\ \bibinfo {author} {\bibfnamefont {Glenn~S.}\
  \bibnamefont {Edwards}},\ }\bibfield  {title} {\enquote {\bibinfo {title}
  {Forces for morphogenesis investigated with laser microsurgery and
  quantitative modeling},}\ }\href {\doibase 10.1126/science.1079552}
  {\bibfield  {journal} {\bibinfo  {journal} {Science}\ }\textbf {\bibinfo
  {volume} {300}},\ \bibinfo {pages} {145--149} (\bibinfo {year}
  {2003})}\BibitemShut {NoStop}%
\bibitem [{\citenamefont {Brodland}\ \emph {et~al.}(2010)\citenamefont
  {Brodland}, \citenamefont {Conte}, \citenamefont {Cranston}, \citenamefont
  {Veldhuis}, \citenamefont {Narasimhan}, \citenamefont {Hutson}, \citenamefont
  {Jacinto}, \citenamefont {Ulrich}, \citenamefont {Baum},\ and\ \citenamefont
  {Miodownik}}]{Brodland22111}%
  \BibitemOpen
  \bibfield  {author} {\bibinfo {author} {\bibfnamefont {G.~Wayne}\
  \bibnamefont {Brodland}}, \bibinfo {author} {\bibfnamefont {Vito}\
  \bibnamefont {Conte}}, \bibinfo {author} {\bibfnamefont {P.~Graham}\
  \bibnamefont {Cranston}}, \bibinfo {author} {\bibfnamefont {Jim}\
  \bibnamefont {Veldhuis}}, \bibinfo {author} {\bibfnamefont {Sriram}\
  \bibnamefont {Narasimhan}}, \bibinfo {author} {\bibfnamefont {M.~Shane}\
  \bibnamefont {Hutson}}, \bibinfo {author} {\bibfnamefont {Antonio}\
  \bibnamefont {Jacinto}}, \bibinfo {author} {\bibfnamefont {Florian}\
  \bibnamefont {Ulrich}}, \bibinfo {author} {\bibfnamefont {Buzz}\ \bibnamefont
  {Baum}}, \ and\ \bibinfo {author} {\bibfnamefont {Mark}\ \bibnamefont
  {Miodownik}},\ }\bibfield  {title} {\enquote {\bibinfo {title} {Video force
  microscopy reveals the mechanics of ventral furrow invagination in
  drosophila},}\ }\href {\doibase 10.1073/pnas.1006591107} {\bibfield
  {journal} {\bibinfo  {journal} {Proceedings of the National Academy of
  Sciences}\ }\textbf {\bibinfo {volume} {107}},\ \bibinfo {pages}
  {22111--22116} (\bibinfo {year} {2010})}\BibitemShut {NoStop}%
\bibitem [{\citenamefont {Chiou}\ \emph {et~al.}(2012)\citenamefont {Chiou},
  \citenamefont {Hufnagel},\ and\ \citenamefont
  {Shraiman}}]{Chiou_Shraiman_2012}%
  \BibitemOpen
  \bibfield  {author} {\bibinfo {author} {\bibfnamefont {Kevin~K.}\
  \bibnamefont {Chiou}}, \bibinfo {author} {\bibfnamefont {Lars}\ \bibnamefont
  {Hufnagel}}, \ and\ \bibinfo {author} {\bibfnamefont {Boris~I.}\ \bibnamefont
  {Shraiman}},\ }\bibfield  {title} {\enquote {\bibinfo {title} {Mechanical
  stress inference for two dimensional cell arrays},}\ }\href {\doibase
  10.1371/journal.pcbi.1002512} {\bibfield  {journal} {\bibinfo  {journal}
  {PLOS Computational Biology}\ }\textbf {\bibinfo {volume} {8}},\ \bibinfo
  {pages} {1--9} (\bibinfo {year} {2012})}\BibitemShut {NoStop}%
\bibitem [{\citenamefont {Ishihara}\ \emph {et~al.}(2013)\citenamefont
  {Ishihara}, \citenamefont {Sugimura}, \citenamefont {Cox}, \citenamefont
  {Bonnet}, \citenamefont {Bella{\"\i}che},\ and\ \citenamefont
  {Graner}}]{Ishihara_compare_EPJE}%
  \BibitemOpen
  \bibfield  {author} {\bibinfo {author} {\bibfnamefont {S.}~\bibnamefont
  {Ishihara}}, \bibinfo {author} {\bibfnamefont {K.}~\bibnamefont {Sugimura}},
  \bibinfo {author} {\bibfnamefont {S.~J.}\ \bibnamefont {Cox}}, \bibinfo
  {author} {\bibfnamefont {I.}~\bibnamefont {Bonnet}}, \bibinfo {author}
  {\bibfnamefont {Y.}~\bibnamefont {Bella{\"\i}che}}, \ and\ \bibinfo {author}
  {\bibfnamefont {F.}~\bibnamefont {Graner}},\ }\bibfield  {title} {\enquote
  {\bibinfo {title} {Comparative study of non-invasive force and stress
  inference methods in tissue},}\ }\href {\doibase 10.1140/epje/i2013-13045-8}
  {\bibfield  {journal} {\bibinfo  {journal} {The European Physical Journal E}\
  }\textbf {\bibinfo {volume} {36}},\ \bibinfo {pages} {45} (\bibinfo {year}
  {2013})}\BibitemShut {NoStop}%
\bibitem [{\citenamefont {Ishihara}\ and\ \citenamefont
  {Sugimura}(2012)}]{ISHIHARA2012201}%
  \BibitemOpen
  \bibfield  {author} {\bibinfo {author} {\bibfnamefont {Shuji}\ \bibnamefont
  {Ishihara}}\ and\ \bibinfo {author} {\bibfnamefont {Kaoru}\ \bibnamefont
  {Sugimura}},\ }\bibfield  {title} {\enquote {\bibinfo {title} {Bayesian
  inference of force dynamics during morphogenesis},}\ }\href {\doibase
  https://doi.org/10.1016/j.jtbi.2012.08.017} {\bibfield  {journal} {\bibinfo
  {journal} {Journal of Theoretical Biology}\ }\textbf {\bibinfo {volume}
  {313}},\ \bibinfo {pages} {201 -- 211} (\bibinfo {year} {2012})}\BibitemShut
  {NoStop}%
\bibitem [{\citenamefont {Brodland}\ \emph {et~al.}(2014)\citenamefont
  {Brodland}, \citenamefont {Veldhuis}, \citenamefont {Kim}, \citenamefont
  {Perrone}, \citenamefont {Mashburn},\ and\ \citenamefont
  {Hutson}}]{Brodland_cellfit}%
  \BibitemOpen
  \bibfield  {author} {\bibinfo {author} {\bibfnamefont {G.~Wayne}\
  \bibnamefont {Brodland}}, \bibinfo {author} {\bibfnamefont {Jim~H.}\
  \bibnamefont {Veldhuis}}, \bibinfo {author} {\bibfnamefont {Steven}\
  \bibnamefont {Kim}}, \bibinfo {author} {\bibfnamefont {Matthew}\ \bibnamefont
  {Perrone}}, \bibinfo {author} {\bibfnamefont {David}\ \bibnamefont
  {Mashburn}}, \ and\ \bibinfo {author} {\bibfnamefont {M.~Shane}\ \bibnamefont
  {Hutson}},\ }\bibfield  {title} {\enquote {\bibinfo {title} {Cellfit: A
  cellular force-inference toolkit using curvilinear cell boundaries},}\ }\href
  {\doibase 10.1371/journal.pone.0099116} {\bibfield  {journal} {\bibinfo
  {journal} {PLOS ONE}\ }\textbf {\bibinfo {volume} {9}},\ \bibinfo {pages}
  {1--15} (\bibinfo {year} {2014})}\BibitemShut {NoStop}%
\bibitem [{\citenamefont {{Bi}}\ and\ \citenamefont {{Yan}}()}]{Bi_private}%
  \BibitemOpen
  \bibfield  {author} {\bibinfo {author} {\bibfnamefont {D.}~\bibnamefont
  {{Bi}}}\ and\ \bibinfo {author} {\bibfnamefont {L.}~\bibnamefont {{Yan}}},\
  }\href@noop {} {}\bibinfo {note} {Private communication}\BibitemShut
  {NoStop}%
\bibitem [{\citenamefont {Pouille}\ \emph {et~al.}(2009)\citenamefont
  {Pouille}, \citenamefont {Ahmadi}, \citenamefont {Brunet},\ and\
  \citenamefont {Farge}}]{Pouille09}%
  \BibitemOpen
  \bibfield  {author} {\bibinfo {author} {\bibfnamefont {Philippe-Alexandre}\
  \bibnamefont {Pouille}}, \bibinfo {author} {\bibfnamefont {Padra}\
  \bibnamefont {Ahmadi}}, \bibinfo {author} {\bibfnamefont {Anne-Christine}\
  \bibnamefont {Brunet}}, \ and\ \bibinfo {author} {\bibfnamefont {Emmanuel}\
  \bibnamefont {Farge}},\ }\bibfield  {title} {\enquote {\bibinfo {title}
  {Mechanical signals trigger myosin ii redistribution and mesoderm
  invagination in drosophila embryos},}\ }\href@noop {} {\bibfield  {journal}
  {\bibinfo  {journal} {Sci. Signal.}\ }\textbf {\bibinfo {volume} {2}},\
  \bibinfo {pages} {ra16--ra16} (\bibinfo {year} {2009})}\BibitemShut {NoStop}%
\bibitem [{\citenamefont {Sun}\ and\ \citenamefont
  {Irvine}(2016)}]{SUN2016694}%
  \BibitemOpen
  \bibfield  {author} {\bibinfo {author} {\bibfnamefont {Shuguo}\ \bibnamefont
  {Sun}}\ and\ \bibinfo {author} {\bibfnamefont {Kenneth~D.}\ \bibnamefont
  {Irvine}},\ }\bibfield  {title} {\enquote {\bibinfo {title} {Cellular
  organization and cytoskeletal regulation of the hippo signaling network},}\
  }\href@noop {} {\bibfield  {journal} {\bibinfo  {journal} {Trends in Cell
  Biology}\ }\textbf {\bibinfo {volume} {26}},\ \bibinfo {pages} {694 -- 704}
  (\bibinfo {year} {2016})}\BibitemShut {NoStop}%
\bibitem [{\citenamefont {Ibar}\ \emph {et~al.}(2018)\citenamefont {Ibar},
  \citenamefont {Kirichenko}, \citenamefont {Keepers}, \citenamefont {Enners},
  \citenamefont {Fleisch},\ and\ \citenamefont {Irvine}}]{Ibarjcs214700}%
  \BibitemOpen
  \bibfield  {author} {\bibinfo {author} {\bibfnamefont {Consuelo}\
  \bibnamefont {Ibar}}, \bibinfo {author} {\bibfnamefont {Elmira}\ \bibnamefont
  {Kirichenko}}, \bibinfo {author} {\bibfnamefont {Benjamin}\ \bibnamefont
  {Keepers}}, \bibinfo {author} {\bibfnamefont {Edward}\ \bibnamefont
  {Enners}}, \bibinfo {author} {\bibfnamefont {Katelyn}\ \bibnamefont
  {Fleisch}}, \ and\ \bibinfo {author} {\bibfnamefont {Kenneth~D.}\
  \bibnamefont {Irvine}},\ }\bibfield  {title} {\enquote {\bibinfo {title}
  {Tension-dependent regulation of mammalian hippo signaling through limd1},}\
  }\href {\doibase 10.1242/jcs.214700} {\bibfield  {journal} {\bibinfo
  {journal} {Journal of Cell Science}\ }\textbf {\bibinfo {volume} {131}}
  (\bibinfo {year} {2018}),\ 10.1242/jcs.214700}\BibitemShut {NoStop}%
\bibitem [{\citenamefont {Doubrovinski}\ \emph {et~al.}(2017)\citenamefont
  {Doubrovinski}, \citenamefont {Swan}, \citenamefont {Polyakov},\ and\
  \citenamefont {Wieschaus}}]{Doubrovinski17}%
  \BibitemOpen
  \bibfield  {author} {\bibinfo {author} {\bibfnamefont {Konstantin}\
  \bibnamefont {Doubrovinski}}, \bibinfo {author} {\bibfnamefont {Michael}\
  \bibnamefont {Swan}}, \bibinfo {author} {\bibfnamefont {Oleg}\ \bibnamefont
  {Polyakov}}, \ and\ \bibinfo {author} {\bibfnamefont {Eric~F}\ \bibnamefont
  {Wieschaus}},\ }\bibfield  {title} {\enquote {\bibinfo {title} {Measurement
  of cortical elasticity in drosophila melanogaster embryos using
  ferrofluids},}\ }\href@noop {} {\bibfield  {journal} {\bibinfo  {journal}
  {Proceedings of the National Academy of Sciences}\ ,\ \bibinfo {pages}
  {201616659}} (\bibinfo {year} {2017})}\BibitemShut {NoStop}%
\bibitem [{\citenamefont {{Merkel}}\ \emph {et~al.}(2018)\citenamefont
  {{Merkel}}, \citenamefont {{Baumgarten}}, \citenamefont {{Tighe}},\ and\
  \citenamefont {{Manning}}}]{2018arXiv180901586M}%
  \BibitemOpen
  \bibfield  {author} {\bibinfo {author} {\bibfnamefont {M.}~\bibnamefont
  {{Merkel}}}, \bibinfo {author} {\bibfnamefont {K.}~\bibnamefont
  {{Baumgarten}}}, \bibinfo {author} {\bibfnamefont {B.~P.}\ \bibnamefont
  {{Tighe}}}, \ and\ \bibinfo {author} {\bibfnamefont {M.~L.}\ \bibnamefont
  {{Manning}}},\ }\bibfield  {title} {\enquote {\bibinfo {title} {{A unifying
  perspective on rigidity in under-constrained materials}},}\ }\href@noop {}
  {\bibfield  {journal} {\bibinfo  {journal} {ArXiv e-prints}\ } (\bibinfo
  {year} {2018})},\ \Eprint {http://arxiv.org/abs/1809.01586} {arXiv:1809.01586
  [cond-mat.soft]} \BibitemShut {NoStop}%
\bibitem [{\citenamefont {Sharma}\ \emph {et~al.}(2016)\citenamefont {Sharma},
  \citenamefont {Licup}, \citenamefont {Jansen}, \citenamefont {Rens},
  \citenamefont {Sheinman}, \citenamefont {Koenderink},\ and\ \citenamefont
  {MacKintosh}}]{Sharma_nphys_2016}%
  \BibitemOpen
  \bibfield  {author} {\bibinfo {author} {\bibfnamefont {A.}~\bibnamefont
  {Sharma}}, \bibinfo {author} {\bibfnamefont {A.~J.}\ \bibnamefont {Licup}},
  \bibinfo {author} {\bibfnamefont {K.~A.}\ \bibnamefont {Jansen}}, \bibinfo
  {author} {\bibfnamefont {R.}~\bibnamefont {Rens}}, \bibinfo {author}
  {\bibfnamefont {M.}~\bibnamefont {Sheinman}}, \bibinfo {author}
  {\bibfnamefont {G.~H.}\ \bibnamefont {Koenderink}}, \ and\ \bibinfo {author}
  {\bibfnamefont {F.~C.}\ \bibnamefont {MacKintosh}},\ }\bibfield  {title}
  {\enquote {\bibinfo {title} {Strain-controlled criticality governs the
  nonlinear mechanics of fibre networks},}\ }\href@noop {} {\bibfield
  {journal} {\bibinfo  {journal} {Nature Physics}\ }\textbf {\bibinfo {volume}
  {12}},\ \bibinfo {pages} {584 EP --} (\bibinfo {year} {2016})}\BibitemShut
  {NoStop}%
\bibitem [{\citenamefont {Feng}\ \emph {et~al.}(2015)\citenamefont {Feng},
  \citenamefont {Levine}, \citenamefont {Mao},\ and\ \citenamefont
  {Sander}}]{feng_pre_2015}%
  \BibitemOpen
  \bibfield  {author} {\bibinfo {author} {\bibfnamefont {Jingchen}\
  \bibnamefont {Feng}}, \bibinfo {author} {\bibfnamefont {Herbert}\
  \bibnamefont {Levine}}, \bibinfo {author} {\bibfnamefont {Xiaoming}\
  \bibnamefont {Mao}}, \ and\ \bibinfo {author} {\bibfnamefont {Leonard~M.}\
  \bibnamefont {Sander}},\ }\bibfield  {title} {\enquote {\bibinfo {title}
  {Alignment and nonlinear elasticity in biopolymer gels},}\ }\href {\doibase
  10.1103/PhysRevE.91.042710} {\bibfield  {journal} {\bibinfo  {journal} {Phys.
  Rev. E}\ }\textbf {\bibinfo {volume} {91}},\ \bibinfo {pages} {042710}
  (\bibinfo {year} {2015})}\BibitemShut {NoStop}%
\bibitem [{\citenamefont {Feng}\ \emph {et~al.}(2016)\citenamefont {Feng},
  \citenamefont {Levine}, \citenamefont {Mao},\ and\ \citenamefont
  {Sander}}]{feng_2016_SM_nonlinear}%
  \BibitemOpen
  \bibfield  {author} {\bibinfo {author} {\bibfnamefont {Jingchen}\
  \bibnamefont {Feng}}, \bibinfo {author} {\bibfnamefont {Herbert}\
  \bibnamefont {Levine}}, \bibinfo {author} {\bibfnamefont {Xiaoming}\
  \bibnamefont {Mao}}, \ and\ \bibinfo {author} {\bibfnamefont {Leonard~M.}\
  \bibnamefont {Sander}},\ }\bibfield  {title} {\enquote {\bibinfo {title}
  {Nonlinear elasticity of disordered fiber networks},}\ }\href {\doibase
  10.1039/C5SM01856K} {\bibfield  {journal} {\bibinfo  {journal} {Soft Matter}\
  }\textbf {\bibinfo {volume} {12}},\ \bibinfo {pages} {1419--1424} (\bibinfo
  {year} {2016})}\BibitemShut {NoStop}%
\bibitem [{\citenamefont {Li}\ and\ \citenamefont {Sun}(2014)}]{Li2014}%
  \BibitemOpen
  \bibfield  {author} {\bibinfo {author} {\bibfnamefont {Bo}~\bibnamefont
  {Li}}\ and\ \bibinfo {author} {\bibfnamefont {Sean~X.}\ \bibnamefont {Sun}},\
  }\bibfield  {title} {\enquote {\bibinfo {title} {{Coherent motions in
  confluent cell monolayer sheets}},}\ }\href {\doibase
  10.1016/j.bpj.2014.08.006} {\bibfield  {journal} {\bibinfo  {journal}
  {Biophysical Journal}\ }\textbf {\bibinfo {volume} {107}},\ \bibinfo {pages}
  {1532--1541} (\bibinfo {year} {2014})}\BibitemShut {NoStop}%
\bibitem [{\citenamefont {Sussman}\ and\ \citenamefont
  {Merkel}(2018)}]{C7SM02127E}%
  \BibitemOpen
  \bibfield  {author} {\bibinfo {author} {\bibfnamefont {Daniel~M.}\
  \bibnamefont {Sussman}}\ and\ \bibinfo {author} {\bibfnamefont {Matthias}\
  \bibnamefont {Merkel}},\ }\bibfield  {title} {\enquote {\bibinfo {title} {No
  unjamming transition in a voronoi model of biological tissue},}\ }\href
  {\doibase 10.1039/C7SM02127E} {\bibfield  {journal} {\bibinfo  {journal}
  {Soft Matter}\ }\textbf {\bibinfo {volume} {14}},\ \bibinfo {pages}
  {3397--3403} (\bibinfo {year} {2018})}\BibitemShut {NoStop}%
\bibitem [{\citenamefont {Byrd}\ \emph {et~al.}(1995)\citenamefont {Byrd},
  \citenamefont {Lu}, \citenamefont {Nocedal},\ and\ \citenamefont
  {Zhu}}]{L-BFGS}%
  \BibitemOpen
  \bibfield  {author} {\bibinfo {author} {\bibfnamefont {Richard~H.}\
  \bibnamefont {Byrd}}, \bibinfo {author} {\bibfnamefont {Peihuang}\
  \bibnamefont {Lu}}, \bibinfo {author} {\bibfnamefont {Jorge}\ \bibnamefont
  {Nocedal}}, \ and\ \bibinfo {author} {\bibfnamefont {Ciyou}\ \bibnamefont
  {Zhu}},\ }\bibfield  {title} {\enquote {\bibinfo {title} {A limited memory
  algorithm for bound constrained optimization},}\ }\href {\doibase
  10.1137/0916069} {\bibfield  {journal} {\bibinfo  {journal} {SIAM Journal on
  Scientific Computing}\ }\textbf {\bibinfo {volume} {16}},\ \bibinfo {pages}
  {1190--1208} (\bibinfo {year} {1995})},\ \Eprint
  {http://arxiv.org/abs/http://dx.doi.org/10.1137/0916069}
  {http://dx.doi.org/10.1137/0916069} \BibitemShut {NoStop}%
\bibitem [{\citenamefont {Maloney}\ and\ \citenamefont
  {Lema\^{\i}tre}(2006)}]{Maloney_PRE_2006}%
  \BibitemOpen
  \bibfield  {author} {\bibinfo {author} {\bibfnamefont {Craig~E.}\
  \bibnamefont {Maloney}}\ and\ \bibinfo {author} {\bibfnamefont {Ana\"el}\
  \bibnamefont {Lema\^{\i}tre}},\ }\bibfield  {title} {\enquote {\bibinfo
  {title} {Amorphous systems in athermal, quasistatic shear},}\ }\href
  {\doibase 10.1103/PhysRevE.74.016118} {\bibfield  {journal} {\bibinfo
  {journal} {Phys. Rev. E}\ }\textbf {\bibinfo {volume} {74}},\ \bibinfo
  {pages} {016118} (\bibinfo {year} {2006})}\BibitemShut {NoStop}%
\bibitem [{\citenamefont {Guttman}(1946)}]{Guttman}%
  \BibitemOpen
  \bibfield  {author} {\bibinfo {author} {\bibfnamefont {Louis}\ \bibnamefont
  {Guttman}},\ }\bibfield  {title} {\enquote {\bibinfo {title} {Enlargement
  methods for computing the inverse matrix},}\ }\href@noop {} {\bibfield
  {journal} {\bibinfo  {journal} {The Annals of Mathematical Statistics}\
  }\textbf {\bibinfo {volume} {17}},\ \bibinfo {pages} {336--343} (\bibinfo
  {year} {1946})}\BibitemShut {NoStop}%
\end{thebibliography}
%

\end{document}